\documentclass[aps, prl, twocolumn, superscriptaddress]{revtex4-1}

\usepackage{amsmath}

\usepackage{xcolor}
\usepackage{graphicx}

\usepackage[T1]{fontenc}
\usepackage[colorlinks,hyperindex]{hyperref}
\hypersetup
{
   colorlinks,%
   citecolor=blue,%
   linkcolor=blue,%
   urlcolor=blue,%
}
\usepackage[capitalise]{cleveref}

\mathchardef\mhyphen="2D
\newcommand\ssp{\mathit{s.p.}}

\newcommand\logft{\mathrm{log}ft}
\newcommand\vmu{\mathrm{V_{MU}}}

\makeindex

\begin{document}

\title{$^{133}$In: A Rosetta Stone for decays of $r$-process nuclei}

\author{Z.~Y.~Xu}
\affiliation{Department of Physics and Astronomy, University of Tennessee,
Knoxville, Tennessee 37996, USA}
\author{M.~Madurga}
\affiliation{Department of Physics and Astronomy, University of Tennessee,
Knoxville, Tennessee 37996, USA}
\author{R.~Grzywacz}
\affiliation{Department of Physics and Astronomy, University of Tennessee,
Knoxville, Tennessee 37996, USA}
\affiliation{Physics Division, Oak Ridge National Laboratory, Oak Ridge,
Tennessee 37831, USA}
\author{T.~T.~King}
\affiliation{Department of Physics and Astronomy, University of Tennessee,
Knoxville, Tennessee 37996, USA}
\affiliation{Physics Division, Oak Ridge National Laboratory, Oak Ridge,
Tennessee 37831, USA}
\author{A.~Algora}
\affiliation{Instituto de F\'isica Corpuscular, CSIC-Universidad de Valencia,
E-46071, Valencia, Spain}
\affiliation{Institute for Nuclear Research (ATOMKI), H-4026 Debrecen, Bem ter
18/c, Hungary}
\author{A.~N.~Andreyev}
\affiliation{Department of Physics, University of York, North Yorkshire YO10
5DD, United Kingdom}
\affiliation{Advanced Science Research Center, Japan Atomic Energy Agency,
Tokai-mura, Japan}
\author{J.~Benito}
\affiliation{Grupo de F\'isica Nuclear and IPARCOS, Facultad de CC.\ F\'isicas,
Universidad Complutense de Madrid, E-28040 Madrid, Spain}
\author{T.~Berry}
\affiliation{Department of Physics, University of Surrey, Guildford GU2 7XH,
United Kingdom}
\author{M.~J.~G.~Borge}
\affiliation{Instituto de Estructura de la Materia, IEM-CSIC, Serrano 113 bis,
E-28006 Madrid, Spain}
\author{C.~Costache}
\affiliation{Horia Hulubei National Institute for Physics and Nuclear
Engineering, RO-077125 Bucharest, Romania}
\author{H.~De~Witte}
\affiliation{KU Leuven, Instituut voor Kern- en Stralingsfysica, B-3001 Leuven,
Belgium}
\author{A.~Fijalkowska}
\affiliation{Department of Physics and Astronomy, Rutgers University, New
Brunswick, New Jersey 08903, USA}
\affiliation{Faculty of Physics, University of Warsaw, PL 02-093 Warsaw, Poland}
\author{L.~M.~Fraile}
\affiliation{Grupo de F\'isica Nuclear and IPARCOS, Facultad de CC.\ F\'isicas,
Universidad Complutense de Madrid, E-28040 Madrid, Spain}
\author{H.~O.~U.~Fynbo}
\affiliation{Department of Physics and Astronomy, Aarhus University, DK-8000
Aarhus C, Denmark}
\author{A.~Gottardo}
\affiliation{IPN, IN2P3-CNRS, Universit\'e Paris-Sud, Universit\'e Paris Saclay,
91406 Orsay Cedex, France}
\author{C.~Halverson}
\affiliation{Department of Physics and Astronomy, University of Tennessee,
Knoxville, Tennessee 37996, USA}
\author{L.~J.~Harkness-Brennan}
\affiliation{Department of Physics, Oliver Lodge Laboratory, University of
Liverpool, Liverpool L69 7ZE, United Kingdom}
\author{J.~Heideman}
\affiliation{Department of Physics and Astronomy, University of Tennessee,
Knoxville, Tennessee 37996, USA}
\author{M.~Huyse}
\affiliation{KU Leuven, Instituut voor Kern- en Stralingsfysica, B-3001 Leuven,
Belgium}
\author{A.~Illana}
\affiliation{KU Leuven, Instituut voor Kern- en Stralingsfysica, B-3001 Leuven,
Belgium}
\affiliation{University of Jyv\"askyl\"a, Department of Physics, P.O. Box 35,
FI-40014, Jyv\"askyl\"a, Finland}
\author{\L.~Janiak}
\affiliation{Faculty of Physics, University of Warsaw, PL 02-093 Warsaw, Poland}
\affiliation{National Centre for Nuclear Research, 05-400 Otwock, \'swierk,
Poland}
\author{D.~S.~Judson}
\affiliation{Department of Physics, Oliver Lodge Laboratory, University of
Liverpool, Liverpool L69 7ZE, United Kingdom}
\author{A.~Korgul}
\affiliation{Faculty of Physics, University of Warsaw, PL 02-093 Warsaw, Poland}
\author{T.~Kurtukian-Nieto}
\affiliation{CENBG, Universit\'e de Bordeaux---UMR 5797 CNRS/IN2P3, Chemin du
Solarium, 33175 Gradignan, France}
\author{I.~Lazarus}
\affiliation{STFC Daresbury, Daresbury, Warrington WA4 4AD, United Kingdom}
\author{R.~Lic\u{a}}
\affiliation{ISOLDE, EP Department, CERN, CH-1211 Geneva, Switzerland}
\affiliation{Horia Hulubei National Institute for Physics and Nuclear
Engineering, RO-077125 Bucharest, Romania}
\author{R.~Lozeva}
\affiliation{Universit{\'e} Paris-Saclay, IJCLab, CNRS/IN2P3, F-91405 Orsay,
France}
\author{N.~Marginean}
\affiliation{Horia Hulubei National Institute for Physics and Nuclear
Engineering, RO-077125 Bucharest, Romania}
\author{R.~Marginean}
\affiliation{Horia Hulubei National Institute for Physics and Nuclear
Engineering, RO-077125 Bucharest, Romania}
\author{C.~Mazzocchi}
\affiliation{Faculty of Physics, University of Warsaw, PL 02-093 Warsaw, Poland}
\author{C.~Mihai}
\affiliation{Horia Hulubei National Institute for Physics and Nuclear
Engineering, RO-077125 Bucharest, Romania}
\author{R.~E.~Mihai}
\affiliation{Horia Hulubei National Institute for Physics and Nuclear
Engineering, RO-077125 Bucharest, Romania}
\author{A.~I.~Morales}
\affiliation{Instituto de F\'isica Corpuscular, CSIC-Universidad de Valencia,
E-46071, Valencia, Spain}
\author{R.~D.~Page}
\affiliation{Department of Physics, Oliver Lodge Laboratory, University of
Liverpool, Liverpool L69 7ZE, United Kingdom}
\author{J.~Pakarinen}
\affiliation{University of Jyv\"askyl\"a, Department of Physics, P.O. Box 35,
FI-40014, Jyv\"askyl\"a, Finland}
\affiliation{Helsinki Institute of Physics, University of Helsinki, P.O. Box 64,
FIN-00014, Helsinki, Finland}
\author{M.~Piersa-Si\l{}kowska}
\affiliation{Faculty of Physics, University of Warsaw, PL 02-093 Warsaw, Poland}
\author{Zs.~Podoly\'ak}
\affiliation{Department of Physics, University of Surrey, Guildford GU2 7XH,
United Kingdom}
\author{P.~Sarriguren}
\affiliation{Instituto de Estructura de la Materia, IEM-CSIC, Serrano 113 bis,
E-28006 Madrid, Spain}
\author{M.~Singh}
\affiliation{Department of Physics and Astronomy, University of Tennessee,
Knoxville, Tennessee 37996, USA}
\author{Ch.~Sotty}
\affiliation{Horia Hulubei National Institute for Physics and Nuclear
Engineering, RO-077125 Bucharest, Romania}
\author{M.~Stepaniuk}
\affiliation{Faculty of Physics, University of Warsaw, PL 02-093 Warsaw, Poland}
\author{O.~Tengblad}
\affiliation{Instituto de Estructura de la Materia, IEM-CSIC, Serrano 113 bis,
E-28006 Madrid, Spain}
\author{A.~Turturica}
\affiliation{Horia Hulubei National Institute for Physics and Nuclear
Engineering, RO-077125 Bucharest, Romania}
\author{P.~Van~Duppen}
\affiliation{KU Leuven, Instituut voor Kern- en Stralingsfysica, B-3001 Leuven,
Belgium}
\author{V.~Vedia}
\affiliation{Grupo de F\'isica Nuclear and IPARCOS, Facultad de CC.\ F\'isicas,
Universidad Complutense de Madrid, E-28040 Madrid, Spain}
\author{S.~Vi\~nals}
\affiliation{Instituto de Estructura de la Materia, IEM-CSIC, Serrano 113 bis,
E-28006 Madrid, Spain}
\author{N.~Warr}
\affiliation{Institut f\"ur Kernphysik, Universit\"at zu K\"oln, 50937 K\"oln,
Germany}
\author{R.~Yokoyama}
\affiliation{Department of Physics and Astronomy, University of Tennessee,
Knoxville, Tennessee 37996, USA}
\author{C.~X.~Yuan}
\affiliation{Sino-French Institute of Nuclear Engineering and Technology, Sun
Yat-Sen University, Zhuhai, 519082, Guangdong, China}

\date{today}

\begin{abstract}
   The $\beta$ decays from both the ground state and a long-lived isomer of
   $^{133}$In were studied at the ISOLDE Decay Station (IDS). With a hybrid
   detection system sensitive to $\beta$, $\gamma$, and neutron spectroscopy,
   the comparative partial half-lives ($\logft$) have been measured for all
   their dominant $\beta$-decay channels for the first time, including a
   low-energy Gamow-Teller transition and several First-Forbidden (FF)
   transitions. Uniquely for such a heavy neutron-rich nucleus, their $\beta$
   decays selectively populate only a few isolated neutron unbound states in
   $^{133}$Sn. Precise energy and branching-ratio measurements of those
   resonances allow us to benchmark $\beta$-decay theories at an unprecedented
   level in this region of the nuclear chart. The results show good agreement
   with the newly developed large-scale shell model (LSSM) calculations. The
   experimental findings establish an archetype for the $\beta$ decay of
   neutron-rich nuclei southeast of $^{132}$Sn and will serve as a guide for
   future theoretical development aiming to describe accurately the key $\beta$
   decays in the rapid-neutron capture ($r$-) process.
\end{abstract}

\maketitle

\begin{figure}[htbp]
   \centering
   \includegraphics[width=0.48\textwidth]{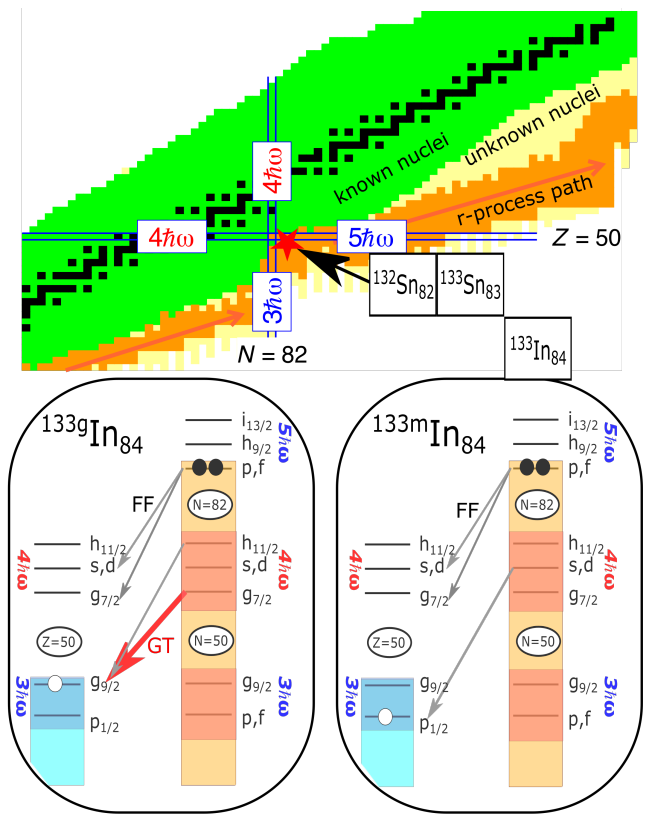}
   \caption{(Top) Chart of nuclei centered on $^{133}$In (red star). The label
   $\hbar\omega$ refers to the harmonic-oscillator shells around $^{132}$Sn. The
   $r$-process path is taken from Ref.\ \cite{skynet}. (Bottom) Proton and
   neutron single-particle ($\ssp$) diagram with dominant $\beta$-decay channels
   in $^{133g}$In and $^{133m}$In. Red and gray arrows represent GT and FF
   transitions respectively.
   }\label{nchart}
\end{figure}

\noindent\textbf{\textit{Introduction}---}
The rapid-neutron capture ($r$-) process is responsible for the creation of half
of the heavy elements in the universe \cite{rprocess1957-1, rprocess1957-2}.
Many stable nuclei present today are decay products of the very short-lived
nuclei produced in extreme environments such as neutron star mergers or
supernovae \cite{nsm-rprocess, hypernovae}. Most of these progenitor nuclei have
large neutron-to-proton ratios, and state-of-the-art nuclear research facilities
cannot produce samples in sufficient quantities for experimental work. Yet,
measured elemental abundance in stars cannot be explained without knowing their
decay properties including half-lives $T_{1/2}$ and $\beta$-delayed
neutron-emission probabilities $P_n$ \cite{rProcessReview, mumpower16,
GorielyReview}. Modern nuclear theories were developed to predict these
quantities for radioactive isotopes far from their stable counterparts
\cite{MollerPRC, Marketin2016, KouraGT, Moller2019, Engel20}. To verify those
models, experimental efforts were carried out continuously pursuing those gross
decay properties of isotopes close to the $r$-process path \cite{Nishimu11,
Morales14, 78Ni_Xu, LorussoPRL, Wu17, briken1, briken2}. Due to the complicated
nature of those decays far off stability, the agreement with model predictions
can be ambiguous, i.e., theories may arrive at a similar gross property for a
single isotope using different footing. In addition, it is generally hard
to find conclusive answers on how to improve the theories when a discrepancy
emergies. Thus, it is desirable to measure the observables capable of
benchmarking $\beta$-decay calculations on a more fundamental level. In this
Letter, we report a $\beta$-decay strength measurement of $^{133}$In ($Z=49$,
$N=84$), a nucleus close to many $r$-process nuclei southeast of $^{132}$Sn
($Z=50$, $N=82$), see \cref{nchart}. We examined decays from both the ground
state ($^{133g}$In) and the isomer ($^{133m}$In) via $\beta$-delayed $\gamma$
and neutron spectroscopy, demonstrating as a textbook example the interplay
between allowed Gamow-Teller (GT) and First-Forbidden (FF) transitions in
extremely neutron-rich nuclei near the $r$-process path. Thus, our measurement
must be accounted for by the models used to predict the decay properties of the
$r$-process nuclei.

In the nuclear shell model \cite{shell1,shell2}, the doubly magic $^{132}$Sn
arranges protons ($\pi$) and neutrons ($\nu$) respectively into the closed
$3\hbar\omega$ and $4\hbar\omega$ major shells, see \cref{nchart}. To the
southeast of $^{132}$Sn, where $^{133}$In resides, the proton Fermi surface is
near the $\pi g_{9/2}$ orbital ($3\hbar\omega$) whereas neutrons start filling
the $5\hbar\omega$ shell above $N=82$, generating large $2\hbar\omega$ asymmetry
between the proton and neutron Fermi surfaces. Since $\pi g_{9/2}$ is partially
occupied, the GT transformation $\nu g_{7/2}\rightarrow\pi g_{9/2}$ (the red
arrow in \cref{nchart}) is expected to be strong. Other competing GT channels
have to induce proton excitation across the $Z=50$ shell (e.g., $\nu
g_{7/2}\rightarrow \pi g_{7/2}$) and thus are much less favorable energetically.
Consequently, the $\nu g_{7/2}\rightarrow\pi g_{9/2}$ transformation is the
single dominant decay channel in the majority of nuclei in this region. Besides,
a few FF transitions contribute significantly to the $\beta$-decay rates by
involving neutron and proton orbitals with opposite parities near the Fermi
surface (the gray arrows in \cref{nchart}, e.g., $\nu h_{11/2}\rightarrow\pi
g_{9/2}$). The proximity of $^{133}$In to the $^{132}$Sn core reduces the number
of active nucleons and the degrees of freedom in the decay process, making it an
ideal ground to validate nuclear theories. On the other hand, the extreme
neutron excess ($N-Z$=35) and large $Q_{\beta}$ energy window (>13 MeV) give
$^{133}$In more complete access than nearby nuclei, such as $^{131}$In ($Z=49$,
$N=82$) and $^{133}$Sn ($Z=50$, $N=83$), to the dominant $\beta$-decay channels
that are responsible for the gross decay properties in the region. Overall, the
unique combination of a large variety of decay modes and simple representation
makes $^{133}$In a perfect study-case nucleus, or a Rosetta Stone, to understand
how the $r$-process nuclei decay near the neutron $N=82$ shell closure.

We studied the $\beta$ decay of $^{133}$In using the neutron time-of-flight
(TOF) technique in combination with high-resolution $\gamma$-ray spectroscopic
system. The $\beta$ decay mostly populated neutron-unbound states in $^{133}$Sn,
which promptly decayed to $^{132}$Sn via neutron emission \cite{hoff, monika,
benito}. If the neutron emission feeds an excited state in $^{132}$Sn, the
nucleus will also undergo $\gamma$ decay(s) to the ground state. Although
several groups have conducted spectroscopic studies of $^{133}$Sn in the past
\cite{hoff, kate, allmond, vaquero17, monika}, the knowledge of states above the
neutron separation energy was scarce due to either the weak production rate or
inefficient neutron detection. By taking advantage of neutron and $\gamma$
spectroscopy measured in coincidence with $\beta$ decay, we revealed for the
first time all the dominant $\beta$-decay transitions in $^{133}$In above the
neutron separation energy. Owing to selective laser ionization of the $^{133}$In
samples \cite{monika}, the decays from the $9/2^+$ ground state ($^{133g}$In)
and the $1/2^-$ isomer ($^{133m}$In) were separated unambiguously. The simple
structure of $^{133}$Sn, the $\beta$-decay selection rules, and the laser
ionization all together allowed us to achieve a superior precision measurement.
In addition, we used the new observation to benchmark large-scale shell-model
(LSSM) calculations. The new measurement provides valuable insights into
understanding the $\beta$ decays of $r$-process nuclei.\\

\begin{figure}[htbp]
   \centering
   \includegraphics[width=0.5\textwidth]{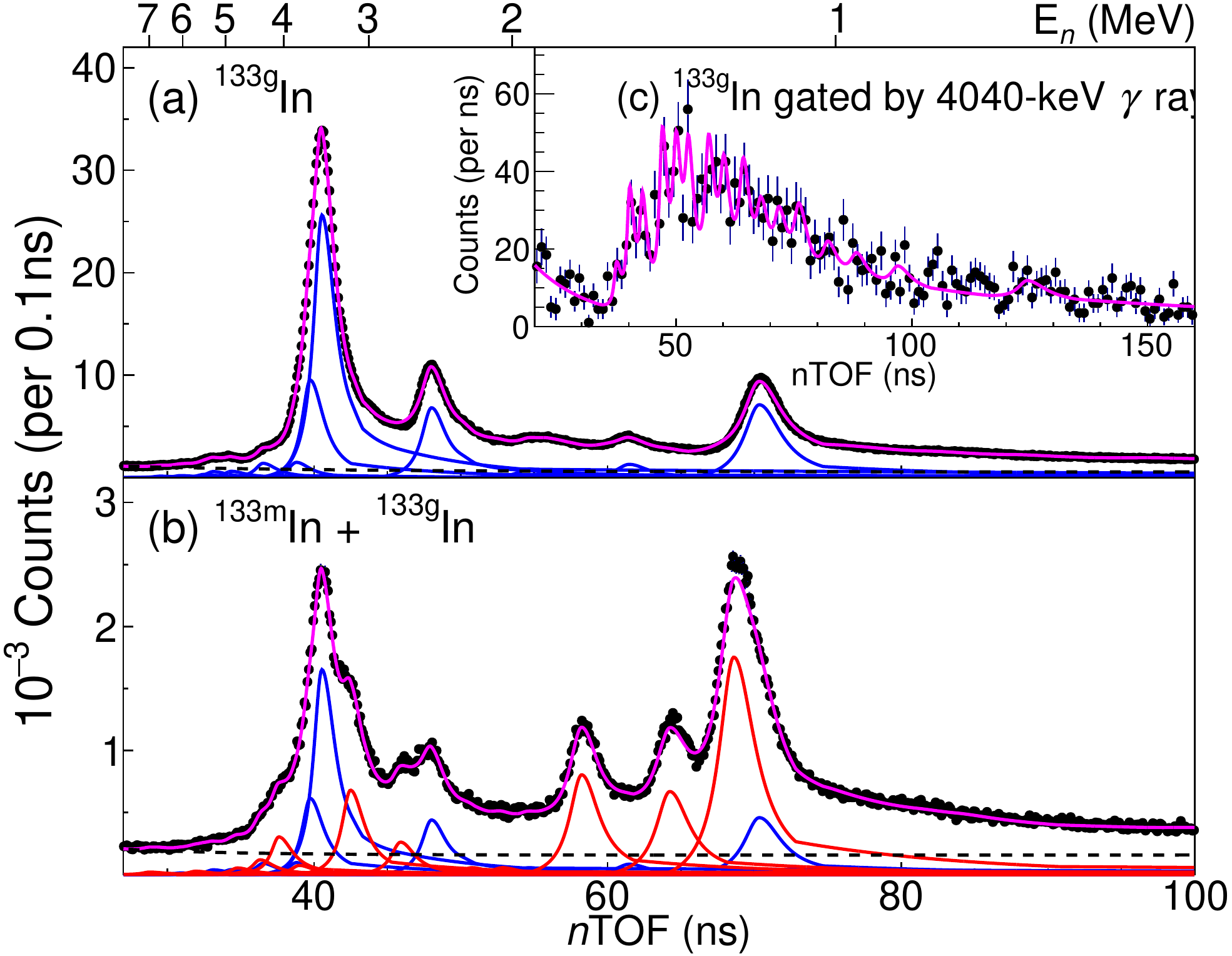}
   \caption{Neutron TOF spectra taken in coincidence with the $^{133}$In $\beta$
   decays, with (a) corresponding to the pure ground-state decay and (b) to an
   admixture of ground-state (40\%) and isomeric decays (60\%). The inset (c)
   shows the ground-state spectrum in coincidence with the 4041-keV $\gamma$
   decay in $^{132}$Sn. On top of the background (dashed line), the spectra are
   fitted by the neutron response functions (magenta) consisting of 18 (blue)
   and 13 (red) peaks in the ground-state and isomeric decays, respectively.
   }\label{ntof}
\end{figure}

\noindent\textbf{\textit{Experiment and result}---}
The Isotope Separator On-Line (ISOLDE) facility at CERN \cite{isolde} and
Resonance Ionization Laser Ion Source \cite{rilis} produced the isotopes of
interest. Through the General Purpose Separator (GPS) \cite{isolde}, the
beams were brought to the ISOLDE Decay Station for $\beta$-decay measurements.
The neutron TOF spectra measured in coincidence with the $\beta$ decay of
$^{133}$In are presented in \cref{ntof}, with \cref{ntof}(a) corresponding to
the pure ground-state decay and \cref{ntof}(b) to an admixture of ground-state
(40\%) and isomeric decays (60\%). Those neutrons were emitted from the
neutron-unbound states in $^{133}$Sn after being populated in the $\beta$ decay.
Neutron emissions may leave the residual $^{132}$Sn nucleus in an excited state.
However, we did not observe any of the strong neutron peaks in \cref{ntof}
coinciding with the $^{132}$Sn $\gamma$ decay, see \cref{ntof}(c), implying
strong direct ground-state feedings in the neutron emissions. The spectra are
fitted by a neutron response function (magenta) consisting of 18 and 13 peaks in
$^{133g}$In (blue) and $^{133m}$In (red) decays, respectively. We extracted the
excitation energies ($E_{ex}$) and decay probabilities ($I_{\beta}$) of
individual states from the fitting result. The full details of the experimental
setup, data analysis, and the list of neutron unbound states identified in
$^{133}$Sn are presented in Ref.\ \cite{133In-prc}.

The main achievement of this work is the observation and quantification of the
$\beta$-decay channels in $^{133g,m}$In. The strongest transitions are mediated
by transforming a neutron from inside the $N=82$ core to a proton on either $\pi
g_{9/2}$ (ground-state decay) or $\pi p_{1/2}$ (isomeric decay), leaving the
proton $Z=50$ shell closed and two neutrons outside $N=82$ coupled to a
spin-zero pair, see \cref{nchart}. We refer to the $^{133}$Sn states so
populated as $\nu$2p-1h (neutron two particle one hole) states hereafter. Using
the analysis methodology detailed in Ref.\ \cite{133In-prc}, we identified four
such states, including the $11/2^-\ (\nu h^{-1}_{11/2})$ state at 3.564(1) MeV
\cite{monika}, the $3/2^+(\nu d^{-1}_{3/2})$ state at 3.62(2) MeV, the
$1/2^+(\nu s^{-1}_{1/2})$ state at 3.79(2) MeV, and the $7/2^+(\nu
g^{-1}_{7/2})$ state at 5.93(9) MeV (the superscript of an orbital indicates
occupation number, being positive for particles and negative for holes). Our
experiment observed most of these states for the first time, the sole exception
being the $11/2^-$ state \cite{hoff, vaquero17, monika}. We extracted
comparative partial half-lives ($\logft$) for those transitions. The $\logft$
values quantify the strength of a given $\beta$-decay transition and correlate
to the $\beta$-decay strength as $S_{\beta}=1/ft$ \cite{sbeta}, where $f$ is the
Fermi function \cite{fermi} for the electron distribution feeding a given state
and $t=T_{1/2}/I_{\beta}$ is the partial half-life of a transition with
$I_{\beta}$ probability. From the $9/2^+$ ground state, the $\logft$ to the
$11/2^-$ and $7/2^+$ states are 5.7(1) and 4.7(1), respectively. From the
$1/2^-$ isomer, the $\logft$ values to the $3/2^+$ and $1/2^+$ states are 5.4(1)
and 5.8(1), respectively. Based on the constraints imposed by $\beta$-decay
selection rules, the $7/2^+$ state was populated via a GT transition, whereas
the other three states were fed by FF transitions. These assignments are in line
with the systematics gleaned from the $\logft$ values mentioned above
\cite{logftreview}.\\

\noindent\textbf{\textit{Comparison with LSSM}---}
We carried out LSSM calculations to interpret our results quantitatively. A
model space containing multiple complete proton and neutron major shells around
$^{132}$Sn exceeds current computational capability. To focus on the strong
decay channels in $^{133}$In, e.g. $\nu g_{7/2}\rightarrow\pi g_{9/2}$, we built
the model space on a $^{88}$Sr core ($Z=38$, $N=50$), including the $0g_{7/2}$,
$1d_{5/2}$, $1d_{3/2}$, $2s_{1/2}$, $0h_{11/2}$, $1f_{7/2}$ orbitals for valence
neutrons and the $1p_{1/2}$, $0g_{9/2}$, $0g_{7/2}$, $1d_{5/2}$, $1d_{3/2}$,
$2s_{1/2}$ orbitals for valence protons. This choice retains important orbital
partners relevant for $\beta$ decay, see \cref{nchart}. We truncated the number
of allowed p-h excitations across $^{132}$Sn to 2p-2h as the first-order
approximation. We used three sets of two-body interactions constructed from the
effective nucleon-nucleon ($NN$) potentials of (i) N$^3$LO \cite{n3lo}, (ii)
Argonne V18 \cite{v18}, and (iii) $\vmu$ plus M3Y \cite{vmu, m3y}. N$^3$LO and
V18 were derived using the many-body perturbation theory \cite{cens}, with the
procedure outlined in Ref.\ \cite{jj77}. $\vmu$ was obtained by computing the
matrix elements directly within our model space. We determined the
single-particle ($\ssp$) energies from the spectroscopic data in the vicinity of
$^{132}$Sn. The GT and FF operators were defined in Ref.\ \cite{yoshida}, and
their effective scaling factors were listed as follows that best reproduce our
data.
\begin{eqnarray*}
   &&q(\mathrm{GT}) = 0.6,\ q(M_0^T)=1.5,\ q(M_0^S)=0.6,\\
   &&q(x)=0.5,\ q(u)=0.4,\ q(z)=0.8.
\end{eqnarray*}

\begin{figure*}[htbp]
   \centering
   \includegraphics[width=0.8\textwidth]{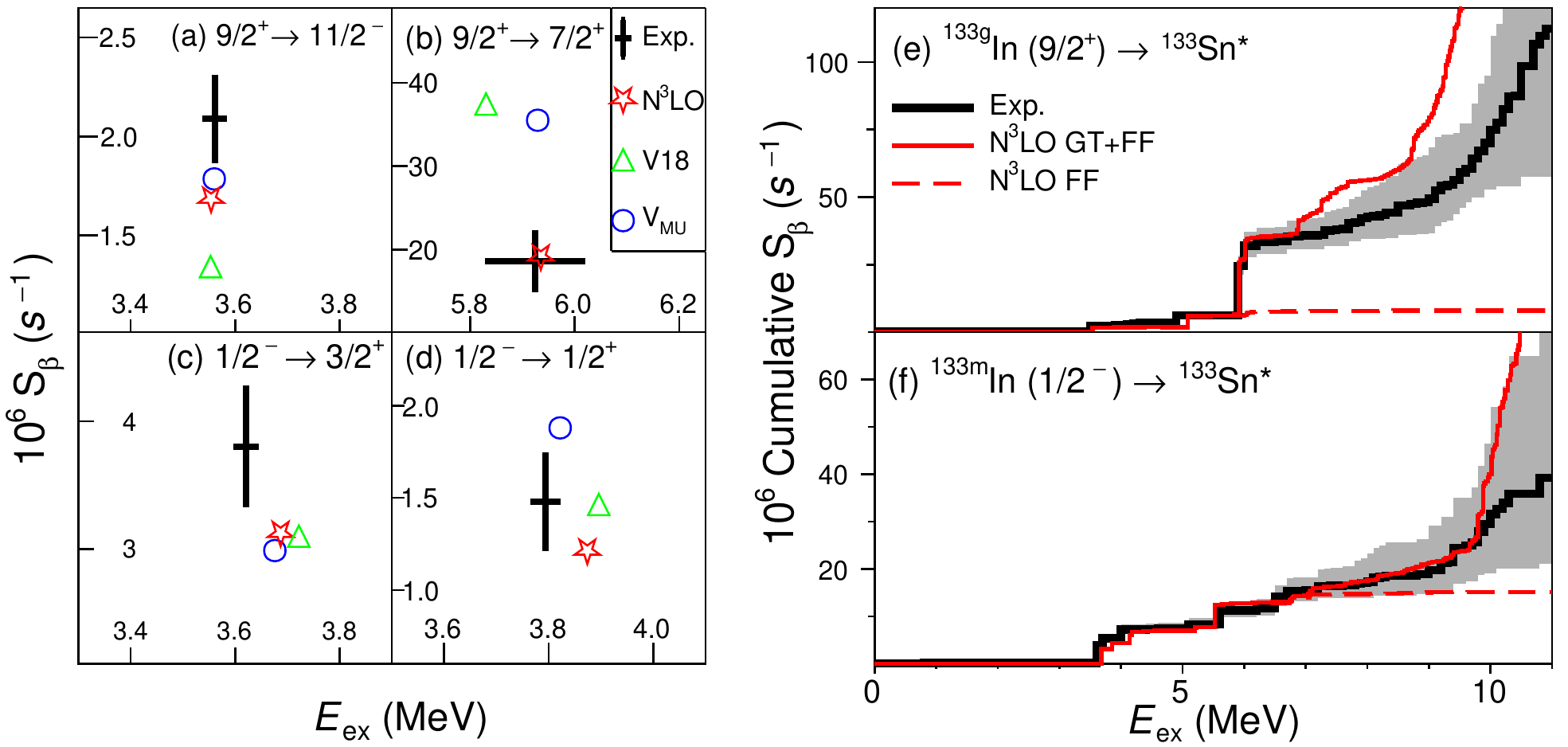}
   \caption{Comparisons of excitation energy and decay strength between LSSM and
   experimental data. Figures (a)-(d) show the results of four individual
   transitions populating the $\nu$2p-1h states. Figures (e) and (f) present
   cumulative strength distribution up to $E_{ex}=11$ MeV for $^{133g,m}$In,
   respectively. The calculation only includes the results from N$^3$LO because
   of its better agreement in the GT strength. The theoretical FF contribution
   is drawn explicitly in dashed lines.
   }\label{bgt}
\end{figure*}

We first examined the individual transitions populating the four $\nu$2p-1h
states, see Figs.\ \ref{bgt} (a)--(d). All three nuclear potentials reproduced
the experimental FF strengths feeding the $11/2^-$, $3/2^+$, and $1/2^+$ states
at lower excitation energy. Additionally, they gave consistent microscopic
compositions of those states: the greatest fractions in the $11/2^-$ and $3/2^+$
wavefunctions were $\nu h^{-1}_{11/2}\times f^{2}_{7/2}$ and $\nu
d^{-1}_{3/2}\times f^{2}_{7/2}$, respectively ($>85\%$). The $1/2^+$ state was
somewhat mixed, with the leading order term $\nu s^{-1}_{1/2}\times f^{2}_{7/2}$
being less than 55\%. Regarding the $7/2^+$ state, the calculations diverged in
the GT strength, giving $36\times10^{-6}\ s^{-1}$ ($\vmu$), $37\times10^{-6}\
s^{-1}$ (V18), and $19\times10^{-6}\ s^{-1}$ (N$^3$LO) respectively. Although
all models predicted a similar fraction of $\nu g^{-1}_{7/2}\times f^{2}_{7/2}$
($\sim45\%$) in their wavefunctions, they differed in the amounts of proton
excitation across $Z=50$, 0.4 in N$^3$LO, and 0.1 in V18 and $\vmu$. The
experimental GT strength, $20(4)\times10^{-6}\ s^{-1}$, was as quenched as the
N$^3$LO prediction, suggesting sizeable proton core excitation contributing to
the state. The comparison reveals the sensitivity of this particular GT decay
strength to the employed $NN$ interactions. Considering this $\nu
g_{7/2}\rightarrow\pi g_{9/2}$ transition dominates the decay rate (and
half-life) in not only $^{133}$In but also a large number of neutron-rich nuclei
southeast of $^{132}$Sn, it is of paramount importance to reproduce this decay
in $^{133}$In in any theoretical calculations aiming to provide reliable
nuclear-decay input to astrophysical applications.

Next, we presented in Figs.\ \ref{bgt} (e, f) the cumulative $\beta$-strength
distribution from the experiment and LSSM with N$^3$LO. The calculations
reproduced the experimental distribution of both states below 9 MeV, giving
half-lives of 145 ms for the ground state and 169 ms for the isomer, in good
agreement with the literature values (162 and 167 ms) \cite{monika}. Towards
higher excitation energy, a sharp kink emerged in the calculations and drove the
distributions up over the experimental ones. Because FF decays are extremely
weak there, see Figs.\ \ref{bgt} (e, f), those strengths are ascribed to the GT
decays involving both the neutron and proton orbitals in the 50--82 shell, or
the $4\hbar\omega$ shell, in \cref{nchart}. The disagreement is most likely
caused by the truncation of 2p-2h excitation across $^{132}$Sn, which is not
sufficient to describe fully the $NN$ correlations and strength distribution at
such high energy. Even though it has a relatively minor impact on the calculated
half-lives and thus the $r$-process, the problem will have to be addressed with
more advanced theoretical treatment in the future.\\

\noindent\textbf{\textit{Feedback to global calculations}---}
Although the LSSM calculations achieved a satisfactory agreement with our data,
it is impractical to make systematic calculations across the nuclear chart due
to the large model spaces. Therefore, global nuclear models are indispensable
for modeling the $r$-process. Our new measurements can serve as constraints and
validation points to improve the accuracy of those global models beyond what was
previously achievable. The measured branching ratios from this work allowed the
extraction of partial half-lives of GT and FF transitions of an $r$-process
nucleus. According to our LSSM calculations in \cref{bgt}, FF transitions
dominate the strength below the GT peak at 6 MeV, whereas those above 6 MeV are
mostly GT transitions. Therefore, the partial half-life of FF transitions is
obtained by summing $\beta$-decay probabilities below the $7/2^+$ state at 5.93
MeV, including the bound states \cite{hoff, monika, benito}. The GT transitions
contain the rest of the feeding intensities from 5.93 MeV onward. To accommodate
the model dependency, we estimated a systematic uncertainty of attributing 50\%
of the strength above 6 MeV to FF transitions. The resultant partial half-lives
are $t^\mathrm{GT}=260(40)$ and $t^\mathrm{FF}=435(60)$ ms for $^{133g}$In, and
$t^\mathrm{GT}=1130(500)$ and $t^\mathrm{FF}=195(10)$ ms for $^{133m}$In.
Although the two states have similar half-lives, the ground-state decay is
dominated by GT transitions, whereas the isomeric decay is mostly carried by FF
transitions.

\begin{figure}[htbp]
   \centering
   \includegraphics[width=0.45\textwidth]{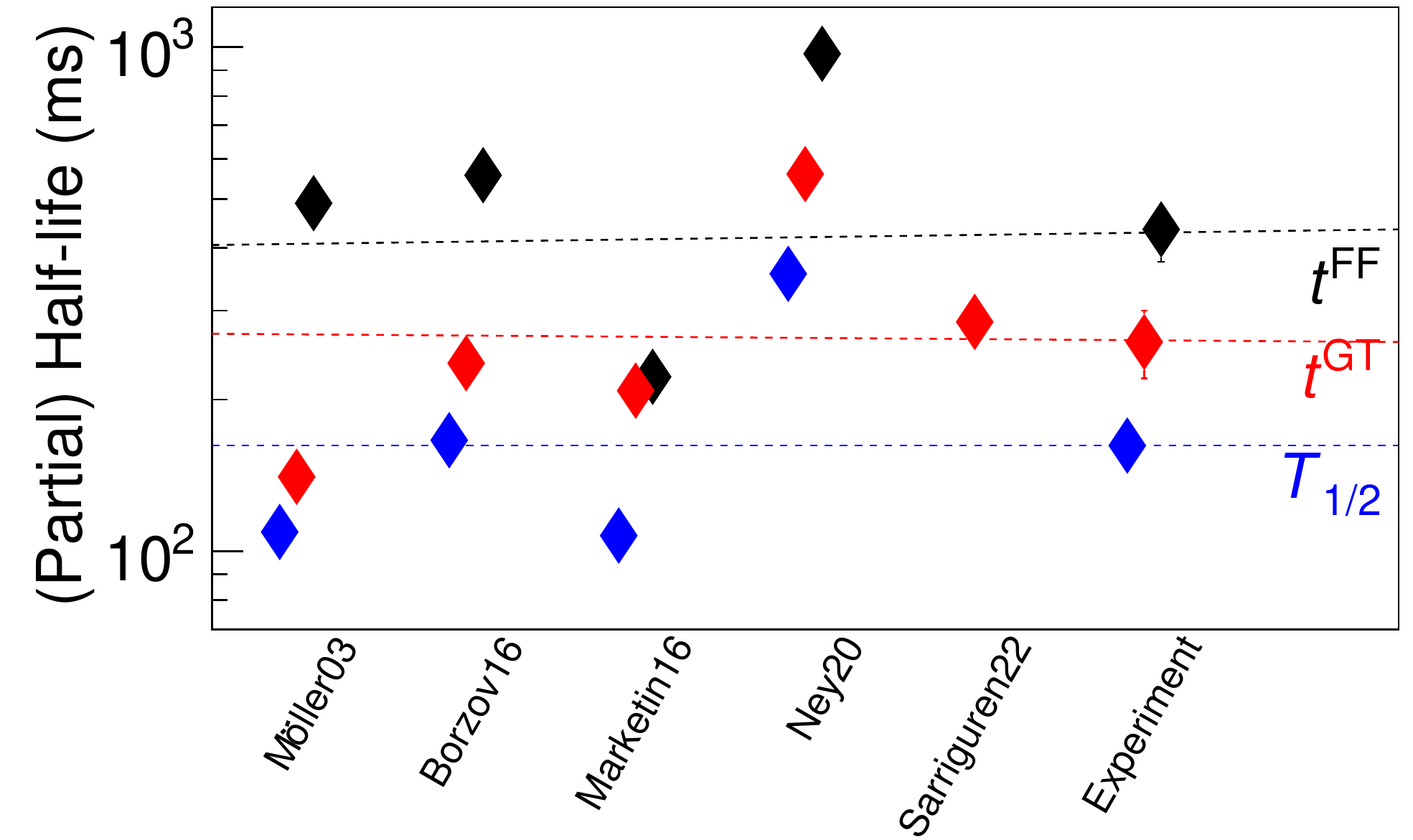}
   \caption{Comparison between several global calculations and the experimental
   half-life (guided by dashed lines) of $^{133g}$In. The total half-life (blue)
   is divided into GT (red) and FF (black) partial half-lives. Calculations are
   explained in the main text.
   }\label{phalf}
\end{figure}

Because global models only predict ground-state decays to date, the comparison
in \cref{phalf} is presented for $^{133g}$In exclusively. The global models
include M\"oller03 (FRDM+QRPA) \cite{MollerPRC}, Borzov16 (DF+CQRPA)
\cite{borzov16}, Marketin16 (RHB+$pn$-RQRPA) \cite{Marketin2016}, Ney20
(EFA-pnFAM) \cite{Engel20}, and Sarriguren22 (HF+BCS+QRPA) \cite{PedroNPA,
*PedroPriv}. All five are the QRPA calculations that differ in their degree of
self-consistency, density functional, or calculation method. In the results
of Moller03, the discrepancy is mainly driven by the GT decays, while in
Marketin16, it is caused by FF transitions with overestimated strength. Although
Ney20 finds a reasonable ratio between the GT and FF strengths, its absolute
decay rates are underestimated by more than a factor of two. The deviations
suggest the strength distributions of those models need to be revised for
$^{133}$In to improve their prediction power for other $r$-process nuclei
further away from $^{132}$Sn. Borzov16 achieves the best agreement overall with
the experimental data. Even though Sarriguren22 does not include FF decays, it
provides a reasonable partial GT half-life for $^{133g}$In.\\

\noindent\textbf{\textit{Summary and prospects}---}
In conclusion, we established with high precision the $\beta$-decay strength
distribution of $^{133g,m}$In. Its ground-state decay is dominated by a GT
transformation, while the isomer almost exclusively decays through FF
transitions. The experimental findings were used to benchmark LSSM calculations
with effective interactions. For the GT transformation $9/2^+\rightarrow7/2^+$,
only N$^3$LO produced a good agreement with the data. In contrast, all the
models agreed with the FF decays at lower excitation energy. The comparison of
several existing global models shows a wide range of competition between GT and
FF transitions in this simple nucleus, with only Borzov16 estimating their
relative contributions and absolute decay rates correctly. It is noteworthy that
the novel $ab$-initio theories developed rapidly in nuclear physics during the
last decade. While not yet available for global predictions, they have already
given essential advancement in understanding nuclear $\beta$-decay probabilities
\cite{BGTnature}. The measurements from this work will serve as an anchor point
on the neutron-rich side of the nuclear chart, where the strengths are more
fragmented and quenched than those in the $^{100}$Sn region along the $Z=N$ line
\cite{sn100nature, sn100prl}.\\


\begin{acknowledgments}
We acknowledge the support of the ISOLDE Collaboration and technical teams.
The authors thank Dr.\ Soda Yoshida, Dr.\ Yutaka Utsuno, Dr.\ Noritaka Shimizu,
Dr.\ Kate L Jones, and Dr.\ Ivan N Borzov for valuable discussions.
This project was supported by the European Unions Horizon 2020 research and
innovation programme Grant Agreements No.\ 654002 (ENSAR2),
by the Office of Nuclear Physics, U.S. Department of Energy under Award No.\
DE-FG02-96ER40983 (UTK) and DE-AC05-00OR22725 (ORNL),
by the National Nuclear Security Administration under the Stewardship Science
Academic Alliances program through DOE Award No.\ DE-NA0002132,
by the Romanian IFA project CERN-RO/ISOLDE,
by the Research Foundation Flanders (FWO, Belgium),
by the Interuniversity Attraction Poles Programme initiated by the Belgian
Science Policy Office (BriX network P7/12),
by the German BMBF under contracts 05P18PKCIA and 05P21PKCI1 in Verbundprojekte
05P2018 and 05P2021,
by the UK Science and Technology Facilities Research Council (STFC) of the UK
Grant No.\ ST/R004056/1, ST/P004598/1, ST/P003885/1, ST/V001027/1, and
ST/V001035/1,
by National Natural Science Foundation of China under Grant No.\ 11775316,
by the Polish National Science Center under Grants No.\ 2019/33/N/ST2/03023,
No.\ 2020/36/T/ST2/00547, and No.\ 2020/39/B/ST2/02346,
by Spanish MCIN/AEI FPA2015-65035-P, PGC2018-093636-B-I00, RTI2018-098868-B-I00,
PID2019-104390GB-I00, PID2019-104714GB-C21, and IJCI-2014-19172 grants,
by Universidad Complutense de Madrid (Spain) through Grupo de F\'isica Nuclear
(910059) and Predoctoral Grant No.\ CT27/16-CT28/16.
The LSSM calculations were carried out by KSHELL \cite{kshell}.
\end{acknowledgments}


\begin{thebibliography}{48}%
\makeatletter
\providecommand \@ifxundefined [1]{%
 \@ifx{#1\undefined}
}%
\providecommand \@ifnum [1]{%
 \ifnum #1\expandafter \@firstoftwo
 \else \expandafter \@secondoftwo
 \fi
}%
\providecommand \@ifx [1]{%
 \ifx #1\expandafter \@firstoftwo
 \else \expandafter \@secondoftwo
 \fi
}%
\providecommand \natexlab [1]{#1}%
\providecommand \enquote  [1]{``#1''}%
\providecommand \bibnamefont  [1]{#1}%
\providecommand \bibfnamefont [1]{#1}%
\providecommand \citenamefont [1]{#1}%
\providecommand \href@noop [0]{\@secondoftwo}%
\providecommand \href [0]{\begingroup \@sanitize@url \@href}%
\providecommand \@href[1]{\@@startlink{#1}\@@href}%
\providecommand \@@href[1]{\endgroup#1\@@endlink}%
\providecommand \@sanitize@url [0]{\catcode `\\12\catcode `\$12\catcode
  `\&12\catcode `\#12\catcode `\^12\catcode `\_12\catcode `\%12\relax}%
\providecommand \@@startlink[1]{}%
\providecommand \@@endlink[0]{}%
\providecommand \url  [0]{\begingroup\@sanitize@url \@url }%
\providecommand \@url [1]{\endgroup\@href {#1}{\urlprefix }}%
\providecommand \urlprefix  [0]{URL }%
\providecommand \Eprint [0]{\href }%
\providecommand \doibase [0]{https://doi.org/}%
\providecommand \selectlanguage [0]{\@gobble}%
\providecommand \bibinfo  [0]{\@secondoftwo}%
\providecommand \bibfield  [0]{\@secondoftwo}%
\providecommand \translation [1]{[#1]}%
\providecommand \BibitemOpen [0]{}%
\providecommand \bibitemStop [0]{}%
\providecommand \bibitemNoStop [0]{.\EOS\space}%
\providecommand \EOS [0]{\spacefactor3000\relax}%
\providecommand \BibitemShut  [1]{\csname bibitem#1\endcsname}%
\let\auto@bib@innerbib\@empty
\bibitem [{\citenamefont {Burbidge}\ \emph {et~al.}(1957)\citenamefont
  {Burbidge}, \citenamefont {Burbidge}, \citenamefont {Fowler},\ and\
  \citenamefont {Hoyle}}]{rprocess1957-1}%
  \BibitemOpen
  \bibfield  {author} {\bibinfo {author} {\bibfnamefont {E.~M.}\ \bibnamefont
  {Burbidge}}, \bibinfo {author} {\bibfnamefont {G.~R.}\ \bibnamefont
  {Burbidge}}, \bibinfo {author} {\bibfnamefont {W.~A.}\ \bibnamefont
  {Fowler}},\ and\ \bibinfo {author} {\bibfnamefont {F.}~\bibnamefont
  {Hoyle}},\ }\href {https://doi.org/10.1103/RevModPhys.29.547} {\bibfield
  {journal} {\bibinfo  {journal} {Rev. Mod. Phys.}\ }\textbf {\bibinfo {volume}
  {29}},\ \bibinfo {pages} {547} (\bibinfo {year} {1957})}\BibitemShut
  {NoStop}%
\bibitem [{\citenamefont {Cameron}(1957)}]{rprocess1957-2}%
  \BibitemOpen
  \bibfield  {author} {\bibinfo {author} {\bibfnamefont {A.~G.}\ \bibnamefont
  {Cameron}},\ }\href@noop {} {\emph {\bibinfo {title} {{Stellar evolution,
  nuclear astrophysics, and nucleogenesis. Second edition}}}},\ \bibinfo
  {series} {Technical Report}, Vol.\ \bibinfo {volume} {CRL-41}\ (\bibinfo
  {publisher} {Atomic Energy of Canada Ltd},\ \bibinfo {address} {Chalk River,
  Ontario},\ \bibinfo {year} {1957})\BibitemShut {NoStop}%
\bibitem [{\citenamefont {Pian}\ \emph {et~al.}(2017)\citenamefont {Pian},
  \citenamefont {D'Avanzo}, \citenamefont {Benetti}, \citenamefont {Branchesi},
  \citenamefont {Brocato}, \citenamefont {Campana}, \citenamefont {Cappellaro},
  \citenamefont {Covino}, \citenamefont {D'Elia}, \citenamefont {Fynbo},
  \citenamefont {Getman}, \citenamefont {Ghirlanda}, \citenamefont
  {Ghisellini}, \citenamefont {Grado}, \citenamefont {Greco}, \citenamefont
  {Hjorth}, \citenamefont {Kouveliotou}, \citenamefont {Levan}, \citenamefont
  {Limatola}, \citenamefont {Malesani}, \citenamefont {Mazzali}, \citenamefont
  {Melandri}, \citenamefont {M{\o}ller}, \citenamefont {Nicastro},
  \citenamefont {Palazzi}, \citenamefont {Piranomonte}, \citenamefont {Rossi},
  \citenamefont {Salafia}, \citenamefont {Selsing}, \citenamefont {Stratta},
  \citenamefont {Tanaka}, \citenamefont {Tanvir}, \citenamefont {Tomasella},
  \citenamefont {Watson}, \citenamefont {Yang}, \citenamefont {Amati},
  \citenamefont {Antonelli}, \citenamefont {Ascenzi}, \citenamefont
  {Bernardini}, \citenamefont {Bo{\"e}r}, \citenamefont {Bufano}, \citenamefont
  {Bulgarelli}, \citenamefont {Capaccioli}, \citenamefont {Casella},
  \citenamefont {Castro-Tirado}, \citenamefont {Chassande-Mottin},
  \citenamefont {Ciolfi}, \citenamefont {Copperwheat}, \citenamefont {Dadina},
  \citenamefont {De~Cesare}, \citenamefont {Di~Paola}, \citenamefont {Fan},
  \citenamefont {Gendre}, \citenamefont {Giuffrida}, \citenamefont {Giunta},
  \citenamefont {Hunt}, \citenamefont {Israel}, \citenamefont {Jin},
  \citenamefont {Kasliwal}, \citenamefont {Klose}, \citenamefont {Lisi},
  \citenamefont {Longo}, \citenamefont {Maiorano}, \citenamefont {Mapelli},
  \citenamefont {Masetti}, \citenamefont {Nava}, \citenamefont {Patricelli},
  \citenamefont {Perley}, \citenamefont {Pescalli}, \citenamefont {Piran},
  \citenamefont {Possenti}, \citenamefont {Pulone}, \citenamefont {Razzano},
  \citenamefont {Salvaterra}, \citenamefont {Schipani}, \citenamefont {Spera},
  \citenamefont {Stamerra}, \citenamefont {Stella}, \citenamefont
  {Tagliaferri}, \citenamefont {Testa}, \citenamefont {Troja}, \citenamefont
  {Turatto}, \citenamefont {Vergani},\ and\ \citenamefont
  {Vergani}}]{nsm-rprocess}%
  \BibitemOpen
  \bibfield  {author} {\bibinfo {author} {\bibfnamefont {E.}~\bibnamefont
  {Pian}}, \bibinfo {author} {\bibfnamefont {P.}~\bibnamefont {D'Avanzo}},
  \bibinfo {author} {\bibfnamefont {S.}~\bibnamefont {Benetti}}, \bibinfo
  {author} {\bibfnamefont {M.}~\bibnamefont {Branchesi}}, \bibinfo {author}
  {\bibfnamefont {E.}~\bibnamefont {Brocato}}, \bibinfo {author} {\bibfnamefont
  {S.}~\bibnamefont {Campana}}, \bibinfo {author} {\bibfnamefont
  {E.}~\bibnamefont {Cappellaro}}, \bibinfo {author} {\bibfnamefont
  {S.}~\bibnamefont {Covino}}, \bibinfo {author} {\bibfnamefont
  {V.}~\bibnamefont {D'Elia}}, \bibinfo {author} {\bibfnamefont {J.~P.~U.}\
  \bibnamefont {Fynbo}}, \bibinfo {author} {\bibfnamefont {F.}~\bibnamefont
  {Getman}}, \bibinfo {author} {\bibfnamefont {G.}~\bibnamefont {Ghirlanda}},
  \bibinfo {author} {\bibfnamefont {G.}~\bibnamefont {Ghisellini}}, \bibinfo
  {author} {\bibfnamefont {A.}~\bibnamefont {Grado}}, \bibinfo {author}
  {\bibfnamefont {G.}~\bibnamefont {Greco}}, \bibinfo {author} {\bibfnamefont
  {J.}~\bibnamefont {Hjorth}}, \bibinfo {author} {\bibfnamefont
  {C.}~\bibnamefont {Kouveliotou}}, \bibinfo {author} {\bibfnamefont
  {A.}~\bibnamefont {Levan}}, \bibinfo {author} {\bibfnamefont
  {L.}~\bibnamefont {Limatola}}, \bibinfo {author} {\bibfnamefont
  {D.}~\bibnamefont {Malesani}}, \bibinfo {author} {\bibfnamefont {P.~A.}\
  \bibnamefont {Mazzali}}, \bibinfo {author} {\bibfnamefont {A.}~\bibnamefont
  {Melandri}}, \bibinfo {author} {\bibfnamefont {P.}~\bibnamefont {M{\o}ller}},
  \bibinfo {author} {\bibfnamefont {L.}~\bibnamefont {Nicastro}}, \bibinfo
  {author} {\bibfnamefont {E.}~\bibnamefont {Palazzi}}, \bibinfo {author}
  {\bibfnamefont {S.}~\bibnamefont {Piranomonte}}, \bibinfo {author}
  {\bibfnamefont {A.}~\bibnamefont {Rossi}}, \bibinfo {author} {\bibfnamefont
  {O.~S.}\ \bibnamefont {Salafia}}, \bibinfo {author} {\bibfnamefont
  {J.}~\bibnamefont {Selsing}}, \bibinfo {author} {\bibfnamefont
  {G.}~\bibnamefont {Stratta}}, \bibinfo {author} {\bibfnamefont
  {M.}~\bibnamefont {Tanaka}}, \bibinfo {author} {\bibfnamefont {N.~R.}\
  \bibnamefont {Tanvir}}, \bibinfo {author} {\bibfnamefont {L.}~\bibnamefont
  {Tomasella}}, \bibinfo {author} {\bibfnamefont {D.}~\bibnamefont {Watson}},
  \bibinfo {author} {\bibfnamefont {S.}~\bibnamefont {Yang}}, \bibinfo {author}
  {\bibfnamefont {L.}~\bibnamefont {Amati}}, \bibinfo {author} {\bibfnamefont
  {L.~A.}\ \bibnamefont {Antonelli}}, \bibinfo {author} {\bibfnamefont
  {S.}~\bibnamefont {Ascenzi}}, \bibinfo {author} {\bibfnamefont {M.~G.}\
  \bibnamefont {Bernardini}}, \bibinfo {author} {\bibfnamefont
  {M.}~\bibnamefont {Bo{\"e}r}}, \bibinfo {author} {\bibfnamefont
  {F.}~\bibnamefont {Bufano}}, \bibinfo {author} {\bibfnamefont
  {A.}~\bibnamefont {Bulgarelli}}, \bibinfo {author} {\bibfnamefont
  {M.}~\bibnamefont {Capaccioli}}, \bibinfo {author} {\bibfnamefont
  {P.}~\bibnamefont {Casella}}, \bibinfo {author} {\bibfnamefont {A.~J.}\
  \bibnamefont {Castro-Tirado}}, \bibinfo {author} {\bibfnamefont
  {E.}~\bibnamefont {Chassande-Mottin}}, \bibinfo {author} {\bibfnamefont
  {R.}~\bibnamefont {Ciolfi}}, \bibinfo {author} {\bibfnamefont {C.~M.}\
  \bibnamefont {Copperwheat}}, \bibinfo {author} {\bibfnamefont
  {M.}~\bibnamefont {Dadina}}, \bibinfo {author} {\bibfnamefont
  {G.}~\bibnamefont {De~Cesare}}, \bibinfo {author} {\bibfnamefont
  {A.}~\bibnamefont {Di~Paola}}, \bibinfo {author} {\bibfnamefont {Y.~Z.}\
  \bibnamefont {Fan}}, \bibinfo {author} {\bibfnamefont {B.}~\bibnamefont
  {Gendre}}, \bibinfo {author} {\bibfnamefont {G.}~\bibnamefont {Giuffrida}},
  \bibinfo {author} {\bibfnamefont {A.}~\bibnamefont {Giunta}}, \bibinfo
  {author} {\bibfnamefont {L.~K.}\ \bibnamefont {Hunt}}, \bibinfo {author}
  {\bibfnamefont {G.~L.}\ \bibnamefont {Israel}}, \bibinfo {author}
  {\bibfnamefont {Z.~P.}\ \bibnamefont {Jin}}, \bibinfo {author} {\bibfnamefont
  {M.~M.}\ \bibnamefont {Kasliwal}}, \bibinfo {author} {\bibfnamefont
  {S.}~\bibnamefont {Klose}}, \bibinfo {author} {\bibfnamefont
  {M.}~\bibnamefont {Lisi}}, \bibinfo {author} {\bibfnamefont {F.}~\bibnamefont
  {Longo}}, \bibinfo {author} {\bibfnamefont {E.}~\bibnamefont {Maiorano}},
  \bibinfo {author} {\bibfnamefont {M.}~\bibnamefont {Mapelli}}, \bibinfo
  {author} {\bibfnamefont {N.}~\bibnamefont {Masetti}}, \bibinfo {author}
  {\bibfnamefont {L.}~\bibnamefont {Nava}}, \bibinfo {author} {\bibfnamefont
  {B.}~\bibnamefont {Patricelli}}, \bibinfo {author} {\bibfnamefont
  {D.}~\bibnamefont {Perley}}, \bibinfo {author} {\bibfnamefont
  {A.}~\bibnamefont {Pescalli}}, \bibinfo {author} {\bibfnamefont
  {T.}~\bibnamefont {Piran}}, \bibinfo {author} {\bibfnamefont
  {A.}~\bibnamefont {Possenti}}, \bibinfo {author} {\bibfnamefont
  {L.}~\bibnamefont {Pulone}}, \bibinfo {author} {\bibfnamefont
  {M.}~\bibnamefont {Razzano}}, \bibinfo {author} {\bibfnamefont
  {R.}~\bibnamefont {Salvaterra}}, \bibinfo {author} {\bibfnamefont
  {P.}~\bibnamefont {Schipani}}, \bibinfo {author} {\bibfnamefont
  {M.}~\bibnamefont {Spera}}, \bibinfo {author} {\bibfnamefont
  {A.}~\bibnamefont {Stamerra}}, \bibinfo {author} {\bibfnamefont
  {L.}~\bibnamefont {Stella}}, \bibinfo {author} {\bibfnamefont
  {G.}~\bibnamefont {Tagliaferri}}, \bibinfo {author} {\bibfnamefont
  {V.}~\bibnamefont {Testa}}, \bibinfo {author} {\bibfnamefont
  {E.}~\bibnamefont {Troja}}, \bibinfo {author} {\bibfnamefont
  {M.}~\bibnamefont {Turatto}}, \bibinfo {author} {\bibfnamefont {S.~D.}\
  \bibnamefont {Vergani}},\ and\ \bibinfo {author} {\bibfnamefont
  {D.}~\bibnamefont {Vergani}},\ }\href {https://doi.org/10.1038/nature24298}
  {\bibfield  {journal} {\bibinfo  {journal} {Nature}\ }\textbf {\bibinfo
  {volume} {551}},\ \bibinfo {pages} {67} (\bibinfo {year} {2017})}\BibitemShut
  {NoStop}%
\bibitem [{\citenamefont {Yong}\ \emph {et~al.}(2021)\citenamefont {Yong},
  \citenamefont {Kobayashi}, \citenamefont {Da~Costa}, \citenamefont {Bessell},
  \citenamefont {Chiti}, \citenamefont {Frebel}, \citenamefont {Lind},
  \citenamefont {Mackey}, \citenamefont {Nordlander}, \citenamefont {Asplund},
  \citenamefont {Casey}, \citenamefont {Marino}, \citenamefont {Murphy},\ and\
  \citenamefont {Schmidt}}]{hypernovae}%
  \BibitemOpen
  \bibfield  {author} {\bibinfo {author} {\bibfnamefont {D.}~\bibnamefont
  {Yong}}, \bibinfo {author} {\bibfnamefont {C.}~\bibnamefont {Kobayashi}},
  \bibinfo {author} {\bibfnamefont {G.~S.}\ \bibnamefont {Da~Costa}}, \bibinfo
  {author} {\bibfnamefont {M.~S.}\ \bibnamefont {Bessell}}, \bibinfo {author}
  {\bibfnamefont {A.}~\bibnamefont {Chiti}}, \bibinfo {author} {\bibfnamefont
  {A.}~\bibnamefont {Frebel}}, \bibinfo {author} {\bibfnamefont
  {K.}~\bibnamefont {Lind}}, \bibinfo {author} {\bibfnamefont {A.~D.}\
  \bibnamefont {Mackey}}, \bibinfo {author} {\bibfnamefont {T.}~\bibnamefont
  {Nordlander}}, \bibinfo {author} {\bibfnamefont {M.}~\bibnamefont {Asplund}},
  \bibinfo {author} {\bibfnamefont {A.~R.}\ \bibnamefont {Casey}}, \bibinfo
  {author} {\bibfnamefont {A.~F.}\ \bibnamefont {Marino}}, \bibinfo {author}
  {\bibfnamefont {S.~J.}\ \bibnamefont {Murphy}},\ and\ \bibinfo {author}
  {\bibfnamefont {B.~P.}\ \bibnamefont {Schmidt}},\ }\href
  {https://doi.org/10.1038/s41586-021-03611-2} {\bibfield  {journal} {\bibinfo
  {journal} {Nature}\ }\textbf {\bibinfo {volume} {595}},\ \bibinfo {pages}
  {223} (\bibinfo {year} {2021})}\BibitemShut {NoStop}%
\bibitem [{\citenamefont {Cowan}\ \emph {et~al.}(2021)\citenamefont {Cowan},
  \citenamefont {Sneden}, \citenamefont {Lawler}, \citenamefont {Aprahamian},
  \citenamefont {Wiescher}, \citenamefont {Langanke}, \citenamefont
  {Mart\'{\i}nez-Pinedo},\ and\ \citenamefont {Thielemann}}]{rProcessReview}%
  \BibitemOpen
  \bibfield  {author} {\bibinfo {author} {\bibfnamefont {J.~J.}\ \bibnamefont
  {Cowan}}, \bibinfo {author} {\bibfnamefont {C.}~\bibnamefont {Sneden}},
  \bibinfo {author} {\bibfnamefont {J.~E.}\ \bibnamefont {Lawler}}, \bibinfo
  {author} {\bibfnamefont {A.}~\bibnamefont {Aprahamian}}, \bibinfo {author}
  {\bibfnamefont {M.}~\bibnamefont {Wiescher}}, \bibinfo {author}
  {\bibfnamefont {K.}~\bibnamefont {Langanke}}, \bibinfo {author}
  {\bibfnamefont {G.}~\bibnamefont {Mart\'{\i}nez-Pinedo}},\ and\ \bibinfo
  {author} {\bibfnamefont {F.-K.}\ \bibnamefont {Thielemann}},\ }\href
  {https://doi.org/10.1103/RevModPhys.93.015002} {\bibfield  {journal}
  {\bibinfo  {journal} {Rev. Mod. Phys.}\ }\textbf {\bibinfo {volume} {93}},\
  \bibinfo {pages} {015002} (\bibinfo {year} {2021})}\BibitemShut {NoStop}%
\bibitem [{\citenamefont {Mumpower}\ \emph {et~al.}(2016)\citenamefont
  {Mumpower}, \citenamefont {Surman}, \citenamefont {McLaughlin},\ and\
  \citenamefont {Aprahamian}}]{mumpower16}%
  \BibitemOpen
  \bibfield  {author} {\bibinfo {author} {\bibfnamefont {M.}~\bibnamefont
  {Mumpower}}, \bibinfo {author} {\bibfnamefont {R.}~\bibnamefont {Surman}},
  \bibinfo {author} {\bibfnamefont {G.}~\bibnamefont {McLaughlin}},\ and\
  \bibinfo {author} {\bibfnamefont {A.}~\bibnamefont {Aprahamian}},\ }\href
  {https://doi.org/https://doi.org/10.1016/j.ppnp.2015.09.001} {\bibfield
  {journal} {\bibinfo  {journal} {Progress in Particle and Nuclear Physics}\
  }\textbf {\bibinfo {volume} {86}},\ \bibinfo {pages} {86} (\bibinfo {year}
  {2016})}\BibitemShut {NoStop}%
\bibitem [{\citenamefont {Arnould}\ and\ \citenamefont
  {Goriely}(2020)}]{GorielyReview}%
  \BibitemOpen
  \bibfield  {author} {\bibinfo {author} {\bibfnamefont {M.}~\bibnamefont
  {Arnould}}\ and\ \bibinfo {author} {\bibfnamefont {S.}~\bibnamefont
  {Goriely}},\ }\href
  {https://doi.org/https://doi.org/10.1016/j.ppnp.2020.103766} {\bibfield
  {journal} {\bibinfo  {journal} {Progress in Particle and Nuclear Physics}\
  }\textbf {\bibinfo {volume} {112}},\ \bibinfo {pages} {103766} (\bibinfo
  {year} {2020})}\BibitemShut {NoStop}%
\bibitem [{\citenamefont {M\"oller}\ \emph {et~al.}(2003)\citenamefont
  {M\"oller}, \citenamefont {Pfeiffer},\ and\ \citenamefont
  {Kratz}}]{MollerPRC}%
  \BibitemOpen
  \bibfield  {author} {\bibinfo {author} {\bibfnamefont {P.}~\bibnamefont
  {M\"oller}}, \bibinfo {author} {\bibfnamefont {B.}~\bibnamefont {Pfeiffer}},\
  and\ \bibinfo {author} {\bibfnamefont {K.-L.}\ \bibnamefont {Kratz}},\ }\href
  {https://doi.org/10.1103/PhysRevC.67.055802} {\bibfield  {journal} {\bibinfo
  {journal} {Phys. Rev. C}\ }\textbf {\bibinfo {volume} {67}},\ \bibinfo
  {pages} {055802} (\bibinfo {year} {2003})}\BibitemShut {NoStop}%
\bibitem [{\citenamefont {Marketin}\ \emph {et~al.}(2016)\citenamefont
  {Marketin}, \citenamefont {Huther},\ and\ \citenamefont
  {Mart\'{\i}nez-Pinedo}}]{Marketin2016}%
  \BibitemOpen
  \bibfield  {author} {\bibinfo {author} {\bibfnamefont {T.}~\bibnamefont
  {Marketin}}, \bibinfo {author} {\bibfnamefont {L.}~\bibnamefont {Huther}},\
  and\ \bibinfo {author} {\bibfnamefont {G.}~\bibnamefont
  {Mart\'{\i}nez-Pinedo}},\ }\href {https://doi.org/10.1103/PhysRevC.93.025805}
  {\bibfield  {journal} {\bibinfo  {journal} {Phys. Rev. C}\ }\textbf {\bibinfo
  {volume} {93}},\ \bibinfo {pages} {025805} (\bibinfo {year}
  {2016})}\BibitemShut {NoStop}%
\bibitem [{\citenamefont {Koura}\ \emph {et~al.}(2017)\citenamefont {Koura},
  \citenamefont {Yoshida}, \citenamefont {Tachibana},\ and\ \citenamefont
  {Chiba}}]{KouraGT}%
  \BibitemOpen
  \bibfield  {author} {\bibinfo {author} {\bibfnamefont {H.}~\bibnamefont
  {Koura}}, \bibinfo {author} {\bibfnamefont {T.}~\bibnamefont {Yoshida}},
  \bibinfo {author} {\bibfnamefont {T.}~\bibnamefont {Tachibana}},\ and\
  \bibinfo {author} {\bibfnamefont {S.}~\bibnamefont {Chiba}},\ }\href
  {https://doi.org/10.1051/epjconf/201714612003} {\bibfield  {journal}
  {\bibinfo  {journal} {EPJ Web Conf.}\ }\textbf {\bibinfo {volume} {146}},\
  \bibinfo {pages} {12003} (\bibinfo {year} {2017})}\BibitemShut {NoStop}%
\bibitem [{\citenamefont {M\"oller}\ \emph {et~al.}(2019)\citenamefont
  {M\"oller}, \citenamefont {Mumpower}, \citenamefont {Kawano},\ and\
  \citenamefont {Myers}}]{Moller2019}%
  \BibitemOpen
  \bibfield  {author} {\bibinfo {author} {\bibfnamefont {P.}~\bibnamefont
  {M\"oller}}, \bibinfo {author} {\bibfnamefont {M.}~\bibnamefont {Mumpower}},
  \bibinfo {author} {\bibfnamefont {T.}~\bibnamefont {Kawano}},\ and\ \bibinfo
  {author} {\bibfnamefont {W.}~\bibnamefont {Myers}},\ }\href
  {https://doi.org/https://doi.org/10.1016/j.adt.2018.03.003} {\bibfield
  {journal} {\bibinfo  {journal} {Atomic Data and Nuclear Data Tables}\
  }\textbf {\bibinfo {volume} {125}},\ \bibinfo {pages} {1} (\bibinfo {year}
  {2019})}\BibitemShut {NoStop}%
\bibitem [{\citenamefont {Ney}\ \emph {et~al.}(2020)\citenamefont {Ney},
  \citenamefont {Engel}, \citenamefont {Li},\ and\ \citenamefont
  {Schunck}}]{Engel20}%
  \BibitemOpen
  \bibfield  {author} {\bibinfo {author} {\bibfnamefont {E.~M.}\ \bibnamefont
  {Ney}}, \bibinfo {author} {\bibfnamefont {J.}~\bibnamefont {Engel}}, \bibinfo
  {author} {\bibfnamefont {T.}~\bibnamefont {Li}},\ and\ \bibinfo {author}
  {\bibfnamefont {N.}~\bibnamefont {Schunck}},\ }\href
  {https://doi.org/10.1103/PhysRevC.102.034326} {\bibfield  {journal} {\bibinfo
   {journal} {Phys. Rev. C}\ }\textbf {\bibinfo {volume} {102}},\ \bibinfo
  {pages} {034326} (\bibinfo {year} {2020})}\BibitemShut {NoStop}%
\bibitem [{\citenamefont {Nishimura}\ \emph {et~al.}(2011)\citenamefont
  {Nishimura}, \citenamefont {Li}, \citenamefont {Watanabe}, \citenamefont
  {Yoshinaga}, \citenamefont {Sumikama}, \citenamefont {Tachibana},
  \citenamefont {Yamaguchi}, \citenamefont {Kurata-Nishimura}, \citenamefont
  {Lorusso}, \citenamefont {Miyashita}, \citenamefont {Odahara}, \citenamefont
  {Baba}, \citenamefont {Berryman}, \citenamefont {Blasi}, \citenamefont
  {Bracco}, \citenamefont {Camera}, \citenamefont {Chiba}, \citenamefont
  {Doornenbal}, \citenamefont {Go}, \citenamefont {Hashimoto}, \citenamefont
  {Hayakawa}, \citenamefont {Hinke}, \citenamefont {Ideguchi}, \citenamefont
  {Isobe}, \citenamefont {Ito}, \citenamefont {Jenkins}, \citenamefont
  {Kawada}, \citenamefont {Kobayashi}, \citenamefont {Kondo}, \citenamefont
  {Kr\"ucken}, \citenamefont {Kubono}, \citenamefont {Nakano}, \citenamefont
  {Ong}, \citenamefont {Ota}, \citenamefont {Podoly\'ak}, \citenamefont
  {Sakurai}, \citenamefont {Scheit}, \citenamefont {Steiger}, \citenamefont
  {Steppenbeck}, \citenamefont {Sugimoto}, \citenamefont {Takano},
  \citenamefont {Takashima}, \citenamefont {Tajiri}, \citenamefont {Teranishi},
  \citenamefont {Wakabayashi}, \citenamefont {Walker}, \citenamefont
  {Wieland},\ and\ \citenamefont {Yamaguchi}}]{Nishimu11}%
  \BibitemOpen
  \bibfield  {author} {\bibinfo {author} {\bibfnamefont {S.}~\bibnamefont
  {Nishimura}}, \bibinfo {author} {\bibfnamefont {Z.}~\bibnamefont {Li}},
  \bibinfo {author} {\bibfnamefont {H.}~\bibnamefont {Watanabe}}, \bibinfo
  {author} {\bibfnamefont {K.}~\bibnamefont {Yoshinaga}}, \bibinfo {author}
  {\bibfnamefont {T.}~\bibnamefont {Sumikama}}, \bibinfo {author}
  {\bibfnamefont {T.}~\bibnamefont {Tachibana}}, \bibinfo {author}
  {\bibfnamefont {K.}~\bibnamefont {Yamaguchi}}, \bibinfo {author}
  {\bibfnamefont {M.}~\bibnamefont {Kurata-Nishimura}}, \bibinfo {author}
  {\bibfnamefont {G.}~\bibnamefont {Lorusso}}, \bibinfo {author} {\bibfnamefont
  {Y.}~\bibnamefont {Miyashita}}, \bibinfo {author} {\bibfnamefont
  {A.}~\bibnamefont {Odahara}}, \bibinfo {author} {\bibfnamefont
  {H.}~\bibnamefont {Baba}}, \bibinfo {author} {\bibfnamefont {J.~S.}\
  \bibnamefont {Berryman}}, \bibinfo {author} {\bibfnamefont {N.}~\bibnamefont
  {Blasi}}, \bibinfo {author} {\bibfnamefont {A.}~\bibnamefont {Bracco}},
  \bibinfo {author} {\bibfnamefont {F.}~\bibnamefont {Camera}}, \bibinfo
  {author} {\bibfnamefont {J.}~\bibnamefont {Chiba}}, \bibinfo {author}
  {\bibfnamefont {P.}~\bibnamefont {Doornenbal}}, \bibinfo {author}
  {\bibfnamefont {S.}~\bibnamefont {Go}}, \bibinfo {author} {\bibfnamefont
  {T.}~\bibnamefont {Hashimoto}}, \bibinfo {author} {\bibfnamefont
  {S.}~\bibnamefont {Hayakawa}}, \bibinfo {author} {\bibfnamefont
  {C.}~\bibnamefont {Hinke}}, \bibinfo {author} {\bibfnamefont
  {E.}~\bibnamefont {Ideguchi}}, \bibinfo {author} {\bibfnamefont
  {T.}~\bibnamefont {Isobe}}, \bibinfo {author} {\bibfnamefont
  {Y.}~\bibnamefont {Ito}}, \bibinfo {author} {\bibfnamefont {D.~G.}\
  \bibnamefont {Jenkins}}, \bibinfo {author} {\bibfnamefont {Y.}~\bibnamefont
  {Kawada}}, \bibinfo {author} {\bibfnamefont {N.}~\bibnamefont {Kobayashi}},
  \bibinfo {author} {\bibfnamefont {Y.}~\bibnamefont {Kondo}}, \bibinfo
  {author} {\bibfnamefont {R.}~\bibnamefont {Kr\"ucken}}, \bibinfo {author}
  {\bibfnamefont {S.}~\bibnamefont {Kubono}}, \bibinfo {author} {\bibfnamefont
  {T.}~\bibnamefont {Nakano}}, \bibinfo {author} {\bibfnamefont {H.~J.}\
  \bibnamefont {Ong}}, \bibinfo {author} {\bibfnamefont {S.}~\bibnamefont
  {Ota}}, \bibinfo {author} {\bibfnamefont {Z.}~\bibnamefont {Podoly\'ak}},
  \bibinfo {author} {\bibfnamefont {H.}~\bibnamefont {Sakurai}}, \bibinfo
  {author} {\bibfnamefont {H.}~\bibnamefont {Scheit}}, \bibinfo {author}
  {\bibfnamefont {K.}~\bibnamefont {Steiger}}, \bibinfo {author} {\bibfnamefont
  {D.}~\bibnamefont {Steppenbeck}}, \bibinfo {author} {\bibfnamefont
  {K.}~\bibnamefont {Sugimoto}}, \bibinfo {author} {\bibfnamefont
  {S.}~\bibnamefont {Takano}}, \bibinfo {author} {\bibfnamefont
  {A.}~\bibnamefont {Takashima}}, \bibinfo {author} {\bibfnamefont
  {K.}~\bibnamefont {Tajiri}}, \bibinfo {author} {\bibfnamefont
  {T.}~\bibnamefont {Teranishi}}, \bibinfo {author} {\bibfnamefont
  {Y.}~\bibnamefont {Wakabayashi}}, \bibinfo {author} {\bibfnamefont {P.~M.}\
  \bibnamefont {Walker}}, \bibinfo {author} {\bibfnamefont {O.}~\bibnamefont
  {Wieland}},\ and\ \bibinfo {author} {\bibfnamefont {H.}~\bibnamefont
  {Yamaguchi}},\ }\href {https://doi.org/10.1103/PhysRevLett.106.052502}
  {\bibfield  {journal} {\bibinfo  {journal} {Phys. Rev. Lett.}\ }\textbf
  {\bibinfo {volume} {106}},\ \bibinfo {pages} {052502} (\bibinfo {year}
  {2011})}\BibitemShut {NoStop}%
\bibitem [{\citenamefont {Morales}\ \emph {et~al.}(2014)\citenamefont
  {Morales}, \citenamefont {Benlliure}, \citenamefont {Kurtuki\'an-Nieto},
  \citenamefont {Schmidt}, \citenamefont {Verma}, \citenamefont {Regan},
  \citenamefont {Podoly\'ak}, \citenamefont {G\'orska}, \citenamefont {Pietri},
  \citenamefont {Kumar}, \citenamefont {Casarejos}, \citenamefont {Al-Dahan},
  \citenamefont {Algora}, \citenamefont {Alkhomashi}, \citenamefont
  {\'Alvarez-Pol}, \citenamefont {Benzoni}, \citenamefont {Blazhev},
  \citenamefont {Boutachkov}, \citenamefont {Bruce}, \citenamefont {C\'aceres},
  \citenamefont {Cullen}, \citenamefont {Denis~Bacelar}, \citenamefont
  {Doornenbal}, \citenamefont {Est\'evez-Aguado}, \citenamefont {Farrelly},
  \citenamefont {Fujita}, \citenamefont {Garnsworthy}, \citenamefont
  {Gelletly}, \citenamefont {Gerl}, \citenamefont {Grebosz}, \citenamefont
  {Hoischen}, \citenamefont {Kojouharov}, \citenamefont {Kurz}, \citenamefont
  {Lalkovski}, \citenamefont {Liu}, \citenamefont {Mihai}, \citenamefont
  {Molina}, \citenamefont {M\"ucher}, \citenamefont {Rubio}, \citenamefont
  {Shaffner}, \citenamefont {Steer}, \citenamefont {Tamii}, \citenamefont
  {Tashenov}, \citenamefont {Valiente-Dob\'on}, \citenamefont {Walker},
  \citenamefont {Wollersheim},\ and\ \citenamefont {Woods}}]{Morales14}%
  \BibitemOpen
  \bibfield  {author} {\bibinfo {author} {\bibfnamefont {A.~I.}\ \bibnamefont
  {Morales}}, \bibinfo {author} {\bibfnamefont {J.}~\bibnamefont {Benlliure}},
  \bibinfo {author} {\bibfnamefont {T.}~\bibnamefont {Kurtuki\'an-Nieto}},
  \bibinfo {author} {\bibfnamefont {K.-H.}\ \bibnamefont {Schmidt}}, \bibinfo
  {author} {\bibfnamefont {S.}~\bibnamefont {Verma}}, \bibinfo {author}
  {\bibfnamefont {P.~H.}\ \bibnamefont {Regan}}, \bibinfo {author}
  {\bibfnamefont {Z.}~\bibnamefont {Podoly\'ak}}, \bibinfo {author}
  {\bibfnamefont {M.}~\bibnamefont {G\'orska}}, \bibinfo {author}
  {\bibfnamefont {S.}~\bibnamefont {Pietri}}, \bibinfo {author} {\bibfnamefont
  {R.}~\bibnamefont {Kumar}}, \bibinfo {author} {\bibfnamefont
  {E.}~\bibnamefont {Casarejos}}, \bibinfo {author} {\bibfnamefont
  {N.}~\bibnamefont {Al-Dahan}}, \bibinfo {author} {\bibfnamefont
  {A.}~\bibnamefont {Algora}}, \bibinfo {author} {\bibfnamefont
  {N.}~\bibnamefont {Alkhomashi}}, \bibinfo {author} {\bibfnamefont
  {H.}~\bibnamefont {\'Alvarez-Pol}}, \bibinfo {author} {\bibfnamefont
  {G.}~\bibnamefont {Benzoni}}, \bibinfo {author} {\bibfnamefont
  {A.}~\bibnamefont {Blazhev}}, \bibinfo {author} {\bibfnamefont
  {P.}~\bibnamefont {Boutachkov}}, \bibinfo {author} {\bibfnamefont {A.~M.}\
  \bibnamefont {Bruce}}, \bibinfo {author} {\bibfnamefont {L.~S.}\ \bibnamefont
  {C\'aceres}}, \bibinfo {author} {\bibfnamefont {I.~J.}\ \bibnamefont
  {Cullen}}, \bibinfo {author} {\bibfnamefont {A.~M.}\ \bibnamefont
  {Denis~Bacelar}}, \bibinfo {author} {\bibfnamefont {P.}~\bibnamefont
  {Doornenbal}}, \bibinfo {author} {\bibfnamefont {M.~E.}\ \bibnamefont
  {Est\'evez-Aguado}}, \bibinfo {author} {\bibfnamefont {G.}~\bibnamefont
  {Farrelly}}, \bibinfo {author} {\bibfnamefont {Y.}~\bibnamefont {Fujita}},
  \bibinfo {author} {\bibfnamefont {A.~B.}\ \bibnamefont {Garnsworthy}},
  \bibinfo {author} {\bibfnamefont {W.}~\bibnamefont {Gelletly}}, \bibinfo
  {author} {\bibfnamefont {J.}~\bibnamefont {Gerl}}, \bibinfo {author}
  {\bibfnamefont {J.}~\bibnamefont {Grebosz}}, \bibinfo {author} {\bibfnamefont
  {R.}~\bibnamefont {Hoischen}}, \bibinfo {author} {\bibfnamefont
  {I.}~\bibnamefont {Kojouharov}}, \bibinfo {author} {\bibfnamefont
  {N.}~\bibnamefont {Kurz}}, \bibinfo {author} {\bibfnamefont {S.}~\bibnamefont
  {Lalkovski}}, \bibinfo {author} {\bibfnamefont {Z.}~\bibnamefont {Liu}},
  \bibinfo {author} {\bibfnamefont {C.}~\bibnamefont {Mihai}}, \bibinfo
  {author} {\bibfnamefont {F.}~\bibnamefont {Molina}}, \bibinfo {author}
  {\bibfnamefont {D.}~\bibnamefont {M\"ucher}}, \bibinfo {author}
  {\bibfnamefont {B.}~\bibnamefont {Rubio}}, \bibinfo {author} {\bibfnamefont
  {H.}~\bibnamefont {Shaffner}}, \bibinfo {author} {\bibfnamefont {S.~J.}\
  \bibnamefont {Steer}}, \bibinfo {author} {\bibfnamefont {A.}~\bibnamefont
  {Tamii}}, \bibinfo {author} {\bibfnamefont {S.}~\bibnamefont {Tashenov}},
  \bibinfo {author} {\bibfnamefont {J.~J.}\ \bibnamefont {Valiente-Dob\'on}},
  \bibinfo {author} {\bibfnamefont {P.~M.}\ \bibnamefont {Walker}}, \bibinfo
  {author} {\bibfnamefont {H.~J.}\ \bibnamefont {Wollersheim}},\ and\ \bibinfo
  {author} {\bibfnamefont {P.~J.}\ \bibnamefont {Woods}},\ }\href
  {https://doi.org/10.1103/PhysRevLett.113.022702} {\bibfield  {journal}
  {\bibinfo  {journal} {Phys. Rev. Lett.}\ }\textbf {\bibinfo {volume} {113}},\
  \bibinfo {pages} {022702} (\bibinfo {year} {2014})}\BibitemShut {NoStop}%
\bibitem [{\citenamefont {Xu}\ \emph {et~al.}(2014)\citenamefont {Xu},
  \citenamefont {Nishimura}, \citenamefont {Lorusso}, \citenamefont {Browne},
  \citenamefont {Doornenbal}, \citenamefont {Gey}, \citenamefont {Jung},
  \citenamefont {Li}, \citenamefont {Niikura}, \citenamefont {S\"oderstr\"om},
  \citenamefont {Sumikama}, \citenamefont {Taprogge}, \citenamefont {Vajta},
  \citenamefont {Watanabe}, \citenamefont {Wu}, \citenamefont {Yagi},
  \citenamefont {Yoshinaga}, \citenamefont {Baba}, \citenamefont {Franchoo},
  \citenamefont {Isobe}, \citenamefont {John}, \citenamefont {Kojouharov},
  \citenamefont {Kubono}, \citenamefont {Kurz}, \citenamefont {Matea},
  \citenamefont {Matsui}, \citenamefont {Mengoni}, \citenamefont {Morfouace},
  \citenamefont {Napoli}, \citenamefont {Naqvi}, \citenamefont {Nishibata},
  \citenamefont {Odahara}, \citenamefont {\ifmmode~\mbox{\c{S}}\else
  \c{S}\fi{}ahin}, \citenamefont {Sakurai}, \citenamefont {Schaffner},
  \citenamefont {Stefan}, \citenamefont {Suzuki}, \citenamefont {Taniuchi},\
  and\ \citenamefont {Werner}}]{78Ni_Xu}%
  \BibitemOpen
  \bibfield  {author} {\bibinfo {author} {\bibfnamefont {Z.~Y.}\ \bibnamefont
  {Xu}}, \bibinfo {author} {\bibfnamefont {S.}~\bibnamefont {Nishimura}},
  \bibinfo {author} {\bibfnamefont {G.}~\bibnamefont {Lorusso}}, \bibinfo
  {author} {\bibfnamefont {F.}~\bibnamefont {Browne}}, \bibinfo {author}
  {\bibfnamefont {P.}~\bibnamefont {Doornenbal}}, \bibinfo {author}
  {\bibfnamefont {G.}~\bibnamefont {Gey}}, \bibinfo {author} {\bibfnamefont
  {H.-S.}\ \bibnamefont {Jung}}, \bibinfo {author} {\bibfnamefont
  {Z.}~\bibnamefont {Li}}, \bibinfo {author} {\bibfnamefont {M.}~\bibnamefont
  {Niikura}}, \bibinfo {author} {\bibfnamefont {P.-A.}\ \bibnamefont
  {S\"oderstr\"om}}, \bibinfo {author} {\bibfnamefont {T.}~\bibnamefont
  {Sumikama}}, \bibinfo {author} {\bibfnamefont {J.}~\bibnamefont {Taprogge}},
  \bibinfo {author} {\bibfnamefont {Z.}~\bibnamefont {Vajta}}, \bibinfo
  {author} {\bibfnamefont {H.}~\bibnamefont {Watanabe}}, \bibinfo {author}
  {\bibfnamefont {J.}~\bibnamefont {Wu}}, \bibinfo {author} {\bibfnamefont
  {A.}~\bibnamefont {Yagi}}, \bibinfo {author} {\bibfnamefont {K.}~\bibnamefont
  {Yoshinaga}}, \bibinfo {author} {\bibfnamefont {H.}~\bibnamefont {Baba}},
  \bibinfo {author} {\bibfnamefont {S.}~\bibnamefont {Franchoo}}, \bibinfo
  {author} {\bibfnamefont {T.}~\bibnamefont {Isobe}}, \bibinfo {author}
  {\bibfnamefont {P.~R.}\ \bibnamefont {John}}, \bibinfo {author}
  {\bibfnamefont {I.}~\bibnamefont {Kojouharov}}, \bibinfo {author}
  {\bibfnamefont {S.}~\bibnamefont {Kubono}}, \bibinfo {author} {\bibfnamefont
  {N.}~\bibnamefont {Kurz}}, \bibinfo {author} {\bibfnamefont {I.}~\bibnamefont
  {Matea}}, \bibinfo {author} {\bibfnamefont {K.}~\bibnamefont {Matsui}},
  \bibinfo {author} {\bibfnamefont {D.}~\bibnamefont {Mengoni}}, \bibinfo
  {author} {\bibfnamefont {P.}~\bibnamefont {Morfouace}}, \bibinfo {author}
  {\bibfnamefont {D.~R.}\ \bibnamefont {Napoli}}, \bibinfo {author}
  {\bibfnamefont {F.}~\bibnamefont {Naqvi}}, \bibinfo {author} {\bibfnamefont
  {H.}~\bibnamefont {Nishibata}}, \bibinfo {author} {\bibfnamefont
  {A.}~\bibnamefont {Odahara}}, \bibinfo {author} {\bibfnamefont
  {E.}~\bibnamefont {\ifmmode~\mbox{\c{S}}\else \c{S}\fi{}ahin}}, \bibinfo
  {author} {\bibfnamefont {H.}~\bibnamefont {Sakurai}}, \bibinfo {author}
  {\bibfnamefont {H.}~\bibnamefont {Schaffner}}, \bibinfo {author}
  {\bibfnamefont {I.~G.}\ \bibnamefont {Stefan}}, \bibinfo {author}
  {\bibfnamefont {D.}~\bibnamefont {Suzuki}}, \bibinfo {author} {\bibfnamefont
  {R.}~\bibnamefont {Taniuchi}},\ and\ \bibinfo {author} {\bibfnamefont
  {V.}~\bibnamefont {Werner}},\ }\href
  {https://doi.org/10.1103/PhysRevLett.113.032505} {\bibfield  {journal}
  {\bibinfo  {journal} {Phys. Rev. Lett.}\ }\textbf {\bibinfo {volume} {113}},\
  \bibinfo {pages} {032505} (\bibinfo {year} {2014})}\BibitemShut {NoStop}%
\bibitem [{\citenamefont {Lorusso}\ \emph {et~al.}(2015)\citenamefont
  {Lorusso}, \citenamefont {Nishimura}, \citenamefont {Xu}, \citenamefont
  {Jungclaus}, \citenamefont {Shimizu}, \citenamefont {Simpson}, \citenamefont
  {S\"oderstr\"om}, \citenamefont {Watanabe}, \citenamefont {Browne},
  \citenamefont {Doornenbal}, \citenamefont {Gey}, \citenamefont {Jung},
  \citenamefont {Meyer}, \citenamefont {Sumikama}, \citenamefont {Taprogge},
  \citenamefont {Vajta}, \citenamefont {Wu}, \citenamefont {Baba},
  \citenamefont {Benzoni}, \citenamefont {Chae}, \citenamefont {Crespi},
  \citenamefont {Fukuda}, \citenamefont {Gernh\"auser}, \citenamefont {Inabe},
  \citenamefont {Isobe}, \citenamefont {Kajino}, \citenamefont {Kameda},
  \citenamefont {Kim}, \citenamefont {Kim}, \citenamefont {Kojouharov},
  \citenamefont {Kondev}, \citenamefont {Kubo}, \citenamefont {Kurz},
  \citenamefont {Kwon}, \citenamefont {Lane}, \citenamefont {Li}, \citenamefont
  {Montaner-Piz\'a}, \citenamefont {Moschner}, \citenamefont {Naqvi},
  \citenamefont {Niikura}, \citenamefont {Nishibata}, \citenamefont {Odahara},
  \citenamefont {Orlandi}, \citenamefont {Patel}, \citenamefont {Podoly\'ak},
  \citenamefont {Sakurai}, \citenamefont {Schaffner}, \citenamefont {Schury},
  \citenamefont {Shibagaki}, \citenamefont {Steiger}, \citenamefont {Suzuki},
  \citenamefont {Takeda}, \citenamefont {Wendt}, \citenamefont {Yagi},\ and\
  \citenamefont {Yoshinaga}}]{LorussoPRL}%
  \BibitemOpen
  \bibfield  {author} {\bibinfo {author} {\bibfnamefont {G.}~\bibnamefont
  {Lorusso}}, \bibinfo {author} {\bibfnamefont {S.}~\bibnamefont {Nishimura}},
  \bibinfo {author} {\bibfnamefont {Z.~Y.}\ \bibnamefont {Xu}}, \bibinfo
  {author} {\bibfnamefont {A.}~\bibnamefont {Jungclaus}}, \bibinfo {author}
  {\bibfnamefont {Y.}~\bibnamefont {Shimizu}}, \bibinfo {author} {\bibfnamefont
  {G.~S.}\ \bibnamefont {Simpson}}, \bibinfo {author} {\bibfnamefont {P.-A.}\
  \bibnamefont {S\"oderstr\"om}}, \bibinfo {author} {\bibfnamefont
  {H.}~\bibnamefont {Watanabe}}, \bibinfo {author} {\bibfnamefont
  {F.}~\bibnamefont {Browne}}, \bibinfo {author} {\bibfnamefont
  {P.}~\bibnamefont {Doornenbal}}, \bibinfo {author} {\bibfnamefont
  {G.}~\bibnamefont {Gey}}, \bibinfo {author} {\bibfnamefont {H.~S.}\
  \bibnamefont {Jung}}, \bibinfo {author} {\bibfnamefont {B.}~\bibnamefont
  {Meyer}}, \bibinfo {author} {\bibfnamefont {T.}~\bibnamefont {Sumikama}},
  \bibinfo {author} {\bibfnamefont {J.}~\bibnamefont {Taprogge}}, \bibinfo
  {author} {\bibfnamefont {Z.}~\bibnamefont {Vajta}}, \bibinfo {author}
  {\bibfnamefont {J.}~\bibnamefont {Wu}}, \bibinfo {author} {\bibfnamefont
  {H.}~\bibnamefont {Baba}}, \bibinfo {author} {\bibfnamefont {G.}~\bibnamefont
  {Benzoni}}, \bibinfo {author} {\bibfnamefont {K.~Y.}\ \bibnamefont {Chae}},
  \bibinfo {author} {\bibfnamefont {F.~C.~L.}\ \bibnamefont {Crespi}}, \bibinfo
  {author} {\bibfnamefont {N.}~\bibnamefont {Fukuda}}, \bibinfo {author}
  {\bibfnamefont {R.}~\bibnamefont {Gernh\"auser}}, \bibinfo {author}
  {\bibfnamefont {N.}~\bibnamefont {Inabe}}, \bibinfo {author} {\bibfnamefont
  {T.}~\bibnamefont {Isobe}}, \bibinfo {author} {\bibfnamefont
  {T.}~\bibnamefont {Kajino}}, \bibinfo {author} {\bibfnamefont
  {D.}~\bibnamefont {Kameda}}, \bibinfo {author} {\bibfnamefont {G.~D.}\
  \bibnamefont {Kim}}, \bibinfo {author} {\bibfnamefont {Y.-K.}\ \bibnamefont
  {Kim}}, \bibinfo {author} {\bibfnamefont {I.}~\bibnamefont {Kojouharov}},
  \bibinfo {author} {\bibfnamefont {F.~G.}\ \bibnamefont {Kondev}}, \bibinfo
  {author} {\bibfnamefont {T.}~\bibnamefont {Kubo}}, \bibinfo {author}
  {\bibfnamefont {N.}~\bibnamefont {Kurz}}, \bibinfo {author} {\bibfnamefont
  {Y.~K.}\ \bibnamefont {Kwon}}, \bibinfo {author} {\bibfnamefont {G.~J.}\
  \bibnamefont {Lane}}, \bibinfo {author} {\bibfnamefont {Z.}~\bibnamefont
  {Li}}, \bibinfo {author} {\bibfnamefont {A.}~\bibnamefont {Montaner-Piz\'a}},
  \bibinfo {author} {\bibfnamefont {K.}~\bibnamefont {Moschner}}, \bibinfo
  {author} {\bibfnamefont {F.}~\bibnamefont {Naqvi}}, \bibinfo {author}
  {\bibfnamefont {M.}~\bibnamefont {Niikura}}, \bibinfo {author} {\bibfnamefont
  {H.}~\bibnamefont {Nishibata}}, \bibinfo {author} {\bibfnamefont
  {A.}~\bibnamefont {Odahara}}, \bibinfo {author} {\bibfnamefont
  {R.}~\bibnamefont {Orlandi}}, \bibinfo {author} {\bibfnamefont
  {Z.}~\bibnamefont {Patel}}, \bibinfo {author} {\bibfnamefont
  {Z.}~\bibnamefont {Podoly\'ak}}, \bibinfo {author} {\bibfnamefont
  {H.}~\bibnamefont {Sakurai}}, \bibinfo {author} {\bibfnamefont
  {H.}~\bibnamefont {Schaffner}}, \bibinfo {author} {\bibfnamefont
  {P.}~\bibnamefont {Schury}}, \bibinfo {author} {\bibfnamefont
  {S.}~\bibnamefont {Shibagaki}}, \bibinfo {author} {\bibfnamefont
  {K.}~\bibnamefont {Steiger}}, \bibinfo {author} {\bibfnamefont
  {H.}~\bibnamefont {Suzuki}}, \bibinfo {author} {\bibfnamefont
  {H.}~\bibnamefont {Takeda}}, \bibinfo {author} {\bibfnamefont
  {A.}~\bibnamefont {Wendt}}, \bibinfo {author} {\bibfnamefont
  {A.}~\bibnamefont {Yagi}},\ and\ \bibinfo {author} {\bibfnamefont
  {K.}~\bibnamefont {Yoshinaga}},\ }\href
  {https://doi.org/10.1103/PhysRevLett.114.192501} {\bibfield  {journal}
  {\bibinfo  {journal} {Phys. Rev. Lett.}\ }\textbf {\bibinfo {volume} {114}},\
  \bibinfo {pages} {192501} (\bibinfo {year} {2015})}\BibitemShut {NoStop}%
\bibitem [{\citenamefont {Wu}\ \emph {et~al.}(2017)\citenamefont {Wu},
  \citenamefont {Nishimura}, \citenamefont {Lorusso}, \citenamefont {M\"oller},
  \citenamefont {Ideguchi}, \citenamefont {Regan}, \citenamefont {Simpson},
  \citenamefont {S\"oderstr\"om}, \citenamefont {Walker}, \citenamefont
  {Watanabe}, \citenamefont {Xu}, \citenamefont {Baba}, \citenamefont {Browne},
  \citenamefont {Daido}, \citenamefont {Doornenbal}, \citenamefont {Fang},
  \citenamefont {Fukuda}, \citenamefont {Gey}, \citenamefont {Isobe},
  \citenamefont {Korkulu}, \citenamefont {Lee}, \citenamefont {Liu},
  \citenamefont {Li}, \citenamefont {Patel}, \citenamefont {Phong},
  \citenamefont {Rice}, \citenamefont {Sakurai}, \citenamefont {Sinclair},
  \citenamefont {Sumikama}, \citenamefont {Tanaka}, \citenamefont {Yagi},
  \citenamefont {Ye}, \citenamefont {Yokoyama}, \citenamefont {Zhang},
  \citenamefont {Ahn}, \citenamefont {Alharbi}, \citenamefont {Aoi},
  \citenamefont {Bello~Garrote}, \citenamefont {Benzoni}, \citenamefont
  {Bruce}, \citenamefont {Carroll}, \citenamefont {Chae}, \citenamefont
  {Dombradi}, \citenamefont {Estrade}, \citenamefont {Gottardo}, \citenamefont
  {Griffin}, \citenamefont {Inabe}, \citenamefont {Kameda}, \citenamefont
  {Kanaoka}, \citenamefont {Kojouharov}, \citenamefont {Kondev}, \citenamefont
  {Kubo}, \citenamefont {Kubono}, \citenamefont {Kurz}, \citenamefont {Kuti},
  \citenamefont {Lalkovski}, \citenamefont {Lane}, \citenamefont {Lee},
  \citenamefont {Lokotko}, \citenamefont {Lotay}, \citenamefont {Moon},
  \citenamefont {Murai}, \citenamefont {Nishibata}, \citenamefont {Nishizuka},
  \citenamefont {Nita}, \citenamefont {Odahara}, \citenamefont {Podoly\'ak},
  \citenamefont {Roberts}, \citenamefont {Schaffner}, \citenamefont {Shand},
  \citenamefont {Shimizu}, \citenamefont {Suzuki}, \citenamefont {Takeda},
  \citenamefont {Taprogge}, \citenamefont {Terashima}, \citenamefont {Vajta},\
  and\ \citenamefont {Yoshida}}]{Wu17}%
  \BibitemOpen
  \bibfield  {author} {\bibinfo {author} {\bibfnamefont {J.}~\bibnamefont
  {Wu}}, \bibinfo {author} {\bibfnamefont {S.}~\bibnamefont {Nishimura}},
  \bibinfo {author} {\bibfnamefont {G.}~\bibnamefont {Lorusso}}, \bibinfo
  {author} {\bibfnamefont {P.}~\bibnamefont {M\"oller}}, \bibinfo {author}
  {\bibfnamefont {E.}~\bibnamefont {Ideguchi}}, \bibinfo {author}
  {\bibfnamefont {P.-H.}\ \bibnamefont {Regan}}, \bibinfo {author}
  {\bibfnamefont {G.~S.}\ \bibnamefont {Simpson}}, \bibinfo {author}
  {\bibfnamefont {P.-A.}\ \bibnamefont {S\"oderstr\"om}}, \bibinfo {author}
  {\bibfnamefont {P.~M.}\ \bibnamefont {Walker}}, \bibinfo {author}
  {\bibfnamefont {H.}~\bibnamefont {Watanabe}}, \bibinfo {author}
  {\bibfnamefont {Z.~Y.}\ \bibnamefont {Xu}}, \bibinfo {author} {\bibfnamefont
  {H.}~\bibnamefont {Baba}}, \bibinfo {author} {\bibfnamefont {F.}~\bibnamefont
  {Browne}}, \bibinfo {author} {\bibfnamefont {R.}~\bibnamefont {Daido}},
  \bibinfo {author} {\bibfnamefont {P.}~\bibnamefont {Doornenbal}}, \bibinfo
  {author} {\bibfnamefont {Y.~F.}\ \bibnamefont {Fang}}, \bibinfo {author}
  {\bibfnamefont {N.}~\bibnamefont {Fukuda}}, \bibinfo {author} {\bibfnamefont
  {G.}~\bibnamefont {Gey}}, \bibinfo {author} {\bibfnamefont {T.}~\bibnamefont
  {Isobe}}, \bibinfo {author} {\bibfnamefont {Z.}~\bibnamefont {Korkulu}},
  \bibinfo {author} {\bibfnamefont {P.~S.}\ \bibnamefont {Lee}}, \bibinfo
  {author} {\bibfnamefont {J.~J.}\ \bibnamefont {Liu}}, \bibinfo {author}
  {\bibfnamefont {Z.}~\bibnamefont {Li}}, \bibinfo {author} {\bibfnamefont
  {Z.}~\bibnamefont {Patel}}, \bibinfo {author} {\bibfnamefont
  {V.}~\bibnamefont {Phong}}, \bibinfo {author} {\bibfnamefont
  {S.}~\bibnamefont {Rice}}, \bibinfo {author} {\bibfnamefont {H.}~\bibnamefont
  {Sakurai}}, \bibinfo {author} {\bibfnamefont {L.}~\bibnamefont {Sinclair}},
  \bibinfo {author} {\bibfnamefont {T.}~\bibnamefont {Sumikama}}, \bibinfo
  {author} {\bibfnamefont {M.}~\bibnamefont {Tanaka}}, \bibinfo {author}
  {\bibfnamefont {A.}~\bibnamefont {Yagi}}, \bibinfo {author} {\bibfnamefont
  {Y.~L.}\ \bibnamefont {Ye}}, \bibinfo {author} {\bibfnamefont
  {R.}~\bibnamefont {Yokoyama}}, \bibinfo {author} {\bibfnamefont {G.~X.}\
  \bibnamefont {Zhang}}, \bibinfo {author} {\bibfnamefont {D.~S.}\ \bibnamefont
  {Ahn}}, \bibinfo {author} {\bibfnamefont {T.}~\bibnamefont {Alharbi}},
  \bibinfo {author} {\bibfnamefont {N.}~\bibnamefont {Aoi}}, \bibinfo {author}
  {\bibfnamefont {F.~L.}\ \bibnamefont {Bello~Garrote}}, \bibinfo {author}
  {\bibfnamefont {G.}~\bibnamefont {Benzoni}}, \bibinfo {author} {\bibfnamefont
  {A.~M.}\ \bibnamefont {Bruce}}, \bibinfo {author} {\bibfnamefont {R.~J.}\
  \bibnamefont {Carroll}}, \bibinfo {author} {\bibfnamefont {K.~Y.}\
  \bibnamefont {Chae}}, \bibinfo {author} {\bibfnamefont {Z.}~\bibnamefont
  {Dombradi}}, \bibinfo {author} {\bibfnamefont {A.}~\bibnamefont {Estrade}},
  \bibinfo {author} {\bibfnamefont {A.}~\bibnamefont {Gottardo}}, \bibinfo
  {author} {\bibfnamefont {C.~J.}\ \bibnamefont {Griffin}}, \bibinfo {author}
  {\bibfnamefont {N.}~\bibnamefont {Inabe}}, \bibinfo {author} {\bibfnamefont
  {D.}~\bibnamefont {Kameda}}, \bibinfo {author} {\bibfnamefont
  {H.}~\bibnamefont {Kanaoka}}, \bibinfo {author} {\bibfnamefont
  {I.}~\bibnamefont {Kojouharov}}, \bibinfo {author} {\bibfnamefont {F.~G.}\
  \bibnamefont {Kondev}}, \bibinfo {author} {\bibfnamefont {T.}~\bibnamefont
  {Kubo}}, \bibinfo {author} {\bibfnamefont {S.}~\bibnamefont {Kubono}},
  \bibinfo {author} {\bibfnamefont {N.}~\bibnamefont {Kurz}}, \bibinfo {author}
  {\bibfnamefont {I.}~\bibnamefont {Kuti}}, \bibinfo {author} {\bibfnamefont
  {S.}~\bibnamefont {Lalkovski}}, \bibinfo {author} {\bibfnamefont {G.~J.}\
  \bibnamefont {Lane}}, \bibinfo {author} {\bibfnamefont {E.~J.}\ \bibnamefont
  {Lee}}, \bibinfo {author} {\bibfnamefont {T.}~\bibnamefont {Lokotko}},
  \bibinfo {author} {\bibfnamefont {G.}~\bibnamefont {Lotay}}, \bibinfo
  {author} {\bibfnamefont {C.-B.}\ \bibnamefont {Moon}}, \bibinfo {author}
  {\bibfnamefont {D.}~\bibnamefont {Murai}}, \bibinfo {author} {\bibfnamefont
  {H.}~\bibnamefont {Nishibata}}, \bibinfo {author} {\bibfnamefont
  {I.}~\bibnamefont {Nishizuka}}, \bibinfo {author} {\bibfnamefont {C.~R.}\
  \bibnamefont {Nita}}, \bibinfo {author} {\bibfnamefont {A.}~\bibnamefont
  {Odahara}}, \bibinfo {author} {\bibfnamefont {Z.}~\bibnamefont {Podoly\'ak}},
  \bibinfo {author} {\bibfnamefont {O.~J.}\ \bibnamefont {Roberts}}, \bibinfo
  {author} {\bibfnamefont {H.}~\bibnamefont {Schaffner}}, \bibinfo {author}
  {\bibfnamefont {C.}~\bibnamefont {Shand}}, \bibinfo {author} {\bibfnamefont
  {Y.}~\bibnamefont {Shimizu}}, \bibinfo {author} {\bibfnamefont
  {H.}~\bibnamefont {Suzuki}}, \bibinfo {author} {\bibfnamefont
  {H.}~\bibnamefont {Takeda}}, \bibinfo {author} {\bibfnamefont
  {J.}~\bibnamefont {Taprogge}}, \bibinfo {author} {\bibfnamefont
  {S.}~\bibnamefont {Terashima}}, \bibinfo {author} {\bibfnamefont
  {Z.}~\bibnamefont {Vajta}},\ and\ \bibinfo {author} {\bibfnamefont
  {S.}~\bibnamefont {Yoshida}},\ }\href
  {https://doi.org/10.1103/PhysRevLett.118.072701} {\bibfield  {journal}
  {\bibinfo  {journal} {Phys. Rev. Lett.}\ }\textbf {\bibinfo {volume} {118}},\
  \bibinfo {pages} {072701} (\bibinfo {year} {2017})}\BibitemShut {NoStop}%
\bibitem [{\citenamefont {Hall}\ \emph {et~al.}(2021)\citenamefont {Hall},
  \citenamefont {Davinson}, \citenamefont {Estrade}, \citenamefont {Liu},
  \citenamefont {Lorusso}, \citenamefont {Montes}, \citenamefont {Nishimura},
  \citenamefont {Phong}, \citenamefont {Woods}, \citenamefont {Agramunt},
  \citenamefont {Ahn}, \citenamefont {Algora}, \citenamefont {Allmond},
  \citenamefont {Baba}, \citenamefont {Bae}, \citenamefont {Brewer},
  \citenamefont {Bruno}, \citenamefont {Caballero-Folch}, \citenamefont
  {Calvi{\~n}o}, \citenamefont {Coleman-Smith}, \citenamefont {Cortes},
  \citenamefont {Dillmann}, \citenamefont {Domingo-Pardo}, \citenamefont
  {Fijalkowska}, \citenamefont {Fukuda}, \citenamefont {Go}, \citenamefont
  {Griffin}, \citenamefont {Grzywacz}, \citenamefont {Ha}, \citenamefont
  {Harkness-Brennan}, \citenamefont {Isobe}, \citenamefont {Kahl},
  \citenamefont {Khiem}, \citenamefont {Kiss}, \citenamefont {Korgul},
  \citenamefont {Kubono}, \citenamefont {Labiche}, \citenamefont {Lazarus},
  \citenamefont {Liang}, \citenamefont {Liu}, \citenamefont {Matsui},
  \citenamefont {Miernik}, \citenamefont {Moon}, \citenamefont {Morales},
  \citenamefont {Morrall}, \citenamefont {Mumpower}, \citenamefont {Nepal},
  \citenamefont {Page}, \citenamefont {Piersa}, \citenamefont {Pucknell},
  \citenamefont {Rasco}, \citenamefont {Rubio}, \citenamefont {Rykaczewski},
  \citenamefont {Sakurai}, \citenamefont {Shimizu}, \citenamefont {Stracener},
  \citenamefont {Sumikama}, \citenamefont {Suzuki}, \citenamefont {Tain},
  \citenamefont {Takeda}, \citenamefont {Tarife{\~n}o-Saldivia}, \citenamefont
  {Tolosa-Delgado}, \citenamefont {Woli{\'n}ska-Cichocka},\ and\ \citenamefont
  {Yokoyama}}]{briken1}%
  \BibitemOpen
  \bibfield  {author} {\bibinfo {author} {\bibfnamefont {O.}~\bibnamefont
  {Hall}}, \bibinfo {author} {\bibfnamefont {T.}~\bibnamefont {Davinson}},
  \bibinfo {author} {\bibfnamefont {A.}~\bibnamefont {Estrade}}, \bibinfo
  {author} {\bibfnamefont {J.}~\bibnamefont {Liu}}, \bibinfo {author}
  {\bibfnamefont {G.}~\bibnamefont {Lorusso}}, \bibinfo {author} {\bibfnamefont
  {F.}~\bibnamefont {Montes}}, \bibinfo {author} {\bibfnamefont
  {S.}~\bibnamefont {Nishimura}}, \bibinfo {author} {\bibfnamefont
  {V.}~\bibnamefont {Phong}}, \bibinfo {author} {\bibfnamefont
  {P.}~\bibnamefont {Woods}}, \bibinfo {author} {\bibfnamefont
  {J.}~\bibnamefont {Agramunt}}, \bibinfo {author} {\bibfnamefont
  {D.}~\bibnamefont {Ahn}}, \bibinfo {author} {\bibfnamefont {A.}~\bibnamefont
  {Algora}}, \bibinfo {author} {\bibfnamefont {J.}~\bibnamefont {Allmond}},
  \bibinfo {author} {\bibfnamefont {H.}~\bibnamefont {Baba}}, \bibinfo {author}
  {\bibfnamefont {S.}~\bibnamefont {Bae}}, \bibinfo {author} {\bibfnamefont
  {N.}~\bibnamefont {Brewer}}, \bibinfo {author} {\bibfnamefont
  {C.}~\bibnamefont {Bruno}}, \bibinfo {author} {\bibfnamefont
  {R.}~\bibnamefont {Caballero-Folch}}, \bibinfo {author} {\bibfnamefont
  {F.}~\bibnamefont {Calvi{\~n}o}}, \bibinfo {author} {\bibfnamefont
  {P.}~\bibnamefont {Coleman-Smith}}, \bibinfo {author} {\bibfnamefont
  {G.}~\bibnamefont {Cortes}}, \bibinfo {author} {\bibfnamefont
  {I.}~\bibnamefont {Dillmann}}, \bibinfo {author} {\bibfnamefont
  {C.}~\bibnamefont {Domingo-Pardo}}, \bibinfo {author} {\bibfnamefont
  {A.}~\bibnamefont {Fijalkowska}}, \bibinfo {author} {\bibfnamefont
  {N.}~\bibnamefont {Fukuda}}, \bibinfo {author} {\bibfnamefont
  {S.}~\bibnamefont {Go}}, \bibinfo {author} {\bibfnamefont {C.}~\bibnamefont
  {Griffin}}, \bibinfo {author} {\bibfnamefont {R.}~\bibnamefont {Grzywacz}},
  \bibinfo {author} {\bibfnamefont {J.}~\bibnamefont {Ha}}, \bibinfo {author}
  {\bibfnamefont {L.}~\bibnamefont {Harkness-Brennan}}, \bibinfo {author}
  {\bibfnamefont {T.}~\bibnamefont {Isobe}}, \bibinfo {author} {\bibfnamefont
  {D.}~\bibnamefont {Kahl}}, \bibinfo {author} {\bibfnamefont {L.}~\bibnamefont
  {Khiem}}, \bibinfo {author} {\bibfnamefont {G.}~\bibnamefont {Kiss}},
  \bibinfo {author} {\bibfnamefont {A.}~\bibnamefont {Korgul}}, \bibinfo
  {author} {\bibfnamefont {S.}~\bibnamefont {Kubono}}, \bibinfo {author}
  {\bibfnamefont {M.}~\bibnamefont {Labiche}}, \bibinfo {author} {\bibfnamefont
  {I.}~\bibnamefont {Lazarus}}, \bibinfo {author} {\bibfnamefont
  {J.}~\bibnamefont {Liang}}, \bibinfo {author} {\bibfnamefont
  {Z.}~\bibnamefont {Liu}}, \bibinfo {author} {\bibfnamefont {K.}~\bibnamefont
  {Matsui}}, \bibinfo {author} {\bibfnamefont {K.}~\bibnamefont {Miernik}},
  \bibinfo {author} {\bibfnamefont {B.}~\bibnamefont {Moon}}, \bibinfo {author}
  {\bibfnamefont {A.}~\bibnamefont {Morales}}, \bibinfo {author} {\bibfnamefont
  {P.}~\bibnamefont {Morrall}}, \bibinfo {author} {\bibfnamefont
  {M.}~\bibnamefont {Mumpower}}, \bibinfo {author} {\bibfnamefont
  {N.}~\bibnamefont {Nepal}}, \bibinfo {author} {\bibfnamefont
  {R.}~\bibnamefont {Page}}, \bibinfo {author} {\bibfnamefont {M.}~\bibnamefont
  {Piersa}}, \bibinfo {author} {\bibfnamefont {V.}~\bibnamefont {Pucknell}},
  \bibinfo {author} {\bibfnamefont {B.}~\bibnamefont {Rasco}}, \bibinfo
  {author} {\bibfnamefont {B.}~\bibnamefont {Rubio}}, \bibinfo {author}
  {\bibfnamefont {K.}~\bibnamefont {Rykaczewski}}, \bibinfo {author}
  {\bibfnamefont {H.}~\bibnamefont {Sakurai}}, \bibinfo {author} {\bibfnamefont
  {Y.}~\bibnamefont {Shimizu}}, \bibinfo {author} {\bibfnamefont
  {D.}~\bibnamefont {Stracener}}, \bibinfo {author} {\bibfnamefont
  {T.}~\bibnamefont {Sumikama}}, \bibinfo {author} {\bibfnamefont
  {H.}~\bibnamefont {Suzuki}}, \bibinfo {author} {\bibfnamefont
  {J.}~\bibnamefont {Tain}}, \bibinfo {author} {\bibfnamefont {H.}~\bibnamefont
  {Takeda}}, \bibinfo {author} {\bibfnamefont {A.}~\bibnamefont
  {Tarife{\~n}o-Saldivia}}, \bibinfo {author} {\bibfnamefont {A.}~\bibnamefont
  {Tolosa-Delgado}}, \bibinfo {author} {\bibfnamefont {M.}~\bibnamefont
  {Woli{\'n}ska-Cichocka}},\ and\ \bibinfo {author} {\bibfnamefont
  {R.}~\bibnamefont {Yokoyama}},\ }\href
  {https://doi.org/https://doi.org/10.1016/j.physletb.2021.136266} {\bibfield
  {journal} {\bibinfo  {journal} {Physics Letters B}\ }\textbf {\bibinfo
  {volume} {816}},\ \bibinfo {pages} {136266} (\bibinfo {year}
  {2021})}\BibitemShut {NoStop}%
\bibitem [{\citenamefont {Phong}\ \emph {et~al.}(2022)\citenamefont {Phong},
  \citenamefont {Nishimura}, \citenamefont {Lorusso}, \citenamefont {Davinson},
  \citenamefont {Estrade}, \citenamefont {Hall}, \citenamefont {Kawano},
  \citenamefont {Liu}, \citenamefont {Montes}, \citenamefont {Nishimura},
  \citenamefont {Grzywacz}, \citenamefont {Rykaczewski}, \citenamefont
  {Agramunt}, \citenamefont {Ahn}, \citenamefont {Algora}, \citenamefont
  {Allmond}, \citenamefont {Baba}, \citenamefont {Bae}, \citenamefont {Brewer},
  \citenamefont {Bruno}, \citenamefont {Caballero-Folch}, \citenamefont
  {Calvi\~no}, \citenamefont {Coleman-Smith}, \citenamefont {Cortes},
  \citenamefont {Dillmann}, \citenamefont {Domingo-Pardo}, \citenamefont
  {Fijalkowska}, \citenamefont {Fukuda}, \citenamefont {Go}, \citenamefont
  {Griffin}, \citenamefont {Ha}, \citenamefont {Harkness-Brennan},
  \citenamefont {Isobe}, \citenamefont {Kahl}, \citenamefont {Khiem},
  \citenamefont {Kiss}, \citenamefont {Korgul}, \citenamefont {Kubono},
  \citenamefont {Labiche}, \citenamefont {Lazarus}, \citenamefont {Liang},
  \citenamefont {Liu}, \citenamefont {Matsui}, \citenamefont {Miernik},
  \citenamefont {Moon}, \citenamefont {Morales}, \citenamefont {Morrall},
  \citenamefont {Nepal}, \citenamefont {Page}, \citenamefont
  {Piersa-Si\l{}kowska}, \citenamefont {Pucknell}, \citenamefont {Rasco},
  \citenamefont {Rubio}, \citenamefont {Sakurai}, \citenamefont {Shimizu},
  \citenamefont {Stracener}, \citenamefont {Sumikama}, \citenamefont {Suzuki},
  \citenamefont {Tain}, \citenamefont {Takeda}, \citenamefont {Tarife\~no
  Saldivia}, \citenamefont {Tolosa-Delgado}, \citenamefont {Woli\ifmmode
  \acute{n}\else~\'{n}\fi{}ska Cichocka}, \citenamefont {Woods},\ and\
  \citenamefont {Yokoyama}}]{briken2}%
  \BibitemOpen
  \bibfield  {author} {\bibinfo {author} {\bibfnamefont {V.~H.}\ \bibnamefont
  {Phong}}, \bibinfo {author} {\bibfnamefont {S.}~\bibnamefont {Nishimura}},
  \bibinfo {author} {\bibfnamefont {G.}~\bibnamefont {Lorusso}}, \bibinfo
  {author} {\bibfnamefont {T.}~\bibnamefont {Davinson}}, \bibinfo {author}
  {\bibfnamefont {A.}~\bibnamefont {Estrade}}, \bibinfo {author} {\bibfnamefont
  {O.}~\bibnamefont {Hall}}, \bibinfo {author} {\bibfnamefont {T.}~\bibnamefont
  {Kawano}}, \bibinfo {author} {\bibfnamefont {J.}~\bibnamefont {Liu}},
  \bibinfo {author} {\bibfnamefont {F.}~\bibnamefont {Montes}}, \bibinfo
  {author} {\bibfnamefont {N.}~\bibnamefont {Nishimura}}, \bibinfo {author}
  {\bibfnamefont {R.}~\bibnamefont {Grzywacz}}, \bibinfo {author}
  {\bibfnamefont {K.~P.}\ \bibnamefont {Rykaczewski}}, \bibinfo {author}
  {\bibfnamefont {J.}~\bibnamefont {Agramunt}}, \bibinfo {author}
  {\bibfnamefont {D.~S.}\ \bibnamefont {Ahn}}, \bibinfo {author} {\bibfnamefont
  {A.}~\bibnamefont {Algora}}, \bibinfo {author} {\bibfnamefont {J.~M.}\
  \bibnamefont {Allmond}}, \bibinfo {author} {\bibfnamefont {H.}~\bibnamefont
  {Baba}}, \bibinfo {author} {\bibfnamefont {S.}~\bibnamefont {Bae}}, \bibinfo
  {author} {\bibfnamefont {N.~T.}\ \bibnamefont {Brewer}}, \bibinfo {author}
  {\bibfnamefont {C.~G.}\ \bibnamefont {Bruno}}, \bibinfo {author}
  {\bibfnamefont {R.}~\bibnamefont {Caballero-Folch}}, \bibinfo {author}
  {\bibfnamefont {F.}~\bibnamefont {Calvi\~no}}, \bibinfo {author}
  {\bibfnamefont {P.~J.}\ \bibnamefont {Coleman-Smith}}, \bibinfo {author}
  {\bibfnamefont {G.}~\bibnamefont {Cortes}}, \bibinfo {author} {\bibfnamefont
  {I.}~\bibnamefont {Dillmann}}, \bibinfo {author} {\bibfnamefont
  {C.}~\bibnamefont {Domingo-Pardo}}, \bibinfo {author} {\bibfnamefont
  {A.}~\bibnamefont {Fijalkowska}}, \bibinfo {author} {\bibfnamefont
  {N.}~\bibnamefont {Fukuda}}, \bibinfo {author} {\bibfnamefont
  {S.}~\bibnamefont {Go}}, \bibinfo {author} {\bibfnamefont {C.~J.}\
  \bibnamefont {Griffin}}, \bibinfo {author} {\bibfnamefont {J.}~\bibnamefont
  {Ha}}, \bibinfo {author} {\bibfnamefont {L.~J.}\ \bibnamefont
  {Harkness-Brennan}}, \bibinfo {author} {\bibfnamefont {T.}~\bibnamefont
  {Isobe}}, \bibinfo {author} {\bibfnamefont {D.}~\bibnamefont {Kahl}},
  \bibinfo {author} {\bibfnamefont {L.~H.}\ \bibnamefont {Khiem}}, \bibinfo
  {author} {\bibfnamefont {G.~G.}\ \bibnamefont {Kiss}}, \bibinfo {author}
  {\bibfnamefont {A.}~\bibnamefont {Korgul}}, \bibinfo {author} {\bibfnamefont
  {S.}~\bibnamefont {Kubono}}, \bibinfo {author} {\bibfnamefont
  {M.}~\bibnamefont {Labiche}}, \bibinfo {author} {\bibfnamefont
  {I.}~\bibnamefont {Lazarus}}, \bibinfo {author} {\bibfnamefont
  {J.}~\bibnamefont {Liang}}, \bibinfo {author} {\bibfnamefont
  {Z.}~\bibnamefont {Liu}}, \bibinfo {author} {\bibfnamefont {K.}~\bibnamefont
  {Matsui}}, \bibinfo {author} {\bibfnamefont {K.}~\bibnamefont {Miernik}},
  \bibinfo {author} {\bibfnamefont {B.}~\bibnamefont {Moon}}, \bibinfo {author}
  {\bibfnamefont {A.~I.}\ \bibnamefont {Morales}}, \bibinfo {author}
  {\bibfnamefont {P.}~\bibnamefont {Morrall}}, \bibinfo {author} {\bibfnamefont
  {N.}~\bibnamefont {Nepal}}, \bibinfo {author} {\bibfnamefont {R.~D.}\
  \bibnamefont {Page}}, \bibinfo {author} {\bibfnamefont {M.}~\bibnamefont
  {Piersa-Si\l{}kowska}}, \bibinfo {author} {\bibfnamefont {V.~F.~E.}\
  \bibnamefont {Pucknell}}, \bibinfo {author} {\bibfnamefont {B.~C.}\
  \bibnamefont {Rasco}}, \bibinfo {author} {\bibfnamefont {B.}~\bibnamefont
  {Rubio}}, \bibinfo {author} {\bibfnamefont {H.}~\bibnamefont {Sakurai}},
  \bibinfo {author} {\bibfnamefont {Y.}~\bibnamefont {Shimizu}}, \bibinfo
  {author} {\bibfnamefont {D.~W.}\ \bibnamefont {Stracener}}, \bibinfo {author}
  {\bibfnamefont {T.}~\bibnamefont {Sumikama}}, \bibinfo {author}
  {\bibfnamefont {H.}~\bibnamefont {Suzuki}}, \bibinfo {author} {\bibfnamefont
  {J.~L.}\ \bibnamefont {Tain}}, \bibinfo {author} {\bibfnamefont
  {H.}~\bibnamefont {Takeda}}, \bibinfo {author} {\bibfnamefont
  {A.}~\bibnamefont {Tarife\~no Saldivia}}, \bibinfo {author} {\bibfnamefont
  {A.}~\bibnamefont {Tolosa-Delgado}}, \bibinfo {author} {\bibfnamefont
  {M.}~\bibnamefont {Woli\ifmmode \acute{n}\else~\'{n}\fi{}ska Cichocka}},
  \bibinfo {author} {\bibfnamefont {P.~J.}\ \bibnamefont {Woods}},\ and\
  \bibinfo {author} {\bibfnamefont {R.}~\bibnamefont {Yokoyama}},\ }\href
  {https://doi.org/10.1103/PhysRevLett.129.172701} {\bibfield  {journal}
  {\bibinfo  {journal} {Phys. Rev. Lett.}\ }\textbf {\bibinfo {volume} {129}},\
  \bibinfo {pages} {172701} (\bibinfo {year} {2022})}\BibitemShut {NoStop}%
\bibitem [{\citenamefont {Lippuner}\ and\ \citenamefont
  {Roberts}(2015)}]{skynet}%
  \BibitemOpen
  \bibfield  {author} {\bibinfo {author} {\bibfnamefont {J.}~\bibnamefont
  {Lippuner}}\ and\ \bibinfo {author} {\bibfnamefont {L.~F.}\ \bibnamefont
  {Roberts}},\ }\href {https://doi.org/10.1088/0004-637x/815/2/82} {\bibfield
  {journal} {\bibinfo  {journal} {The Astrophysical Journal}\ }\textbf
  {\bibinfo {volume} {815}},\ \bibinfo {pages} {82} (\bibinfo {year}
  {2015})}\BibitemShut {NoStop}%
\bibitem [{\citenamefont {Mayer}(1949)}]{shell1}%
  \BibitemOpen
  \bibfield  {author} {\bibinfo {author} {\bibfnamefont {M.~G.}\ \bibnamefont
  {Mayer}},\ }\href {https://doi.org/10.1103/PhysRev.75.1969} {\bibfield
  {journal} {\bibinfo  {journal} {Phys. Rev.}\ }\textbf {\bibinfo {volume}
  {75}},\ \bibinfo {pages} {1969} (\bibinfo {year} {1949})}\BibitemShut
  {NoStop}%
\bibitem [{\citenamefont {Haxel}\ \emph {et~al.}(1949)\citenamefont {Haxel},
  \citenamefont {Jensen},\ and\ \citenamefont {Suess}}]{shell2}%
  \BibitemOpen
  \bibfield  {author} {\bibinfo {author} {\bibfnamefont {O.}~\bibnamefont
  {Haxel}}, \bibinfo {author} {\bibfnamefont {J.~H.~D.}\ \bibnamefont
  {Jensen}},\ and\ \bibinfo {author} {\bibfnamefont {H.~E.}\ \bibnamefont
  {Suess}},\ }\href {https://doi.org/10.1103/PhysRev.75.1766.2} {\bibfield
  {journal} {\bibinfo  {journal} {Phys. Rev.}\ }\textbf {\bibinfo {volume}
  {75}},\ \bibinfo {pages} {1766} (\bibinfo {year} {1949})}\BibitemShut
  {NoStop}%
\bibitem [{\citenamefont {Hoff}\ \emph {et~al.}(1996)\citenamefont {Hoff},
  \citenamefont {Baumann}, \citenamefont {Huck}, \citenamefont {Knipper},
  \citenamefont {Walter}, \citenamefont {Marguier}, \citenamefont {Fogelberg},
  \citenamefont {Lindroth}, \citenamefont {Mach}, \citenamefont {Sanchez-Vega},
  \citenamefont {Taylor}, \citenamefont {Van~Duppen}, \citenamefont {Jokinen},
  \citenamefont {Lindroos}, \citenamefont {Ramdhane}, \citenamefont
  {Kurcewicz}, \citenamefont {Jonson}, \citenamefont {Nyman}, \citenamefont
  {Jading}, \citenamefont {Kratz}, \citenamefont {W\"ohr}, \citenamefont
  {L\o{}vh\o{}iden}, \citenamefont {Thorsteinsen},\ and\ \citenamefont
  {Blomqvist}}]{hoff}%
  \BibitemOpen
  \bibfield  {author} {\bibinfo {author} {\bibfnamefont {P.}~\bibnamefont
  {Hoff}}, \bibinfo {author} {\bibfnamefont {P.}~\bibnamefont {Baumann}},
  \bibinfo {author} {\bibfnamefont {A.}~\bibnamefont {Huck}}, \bibinfo {author}
  {\bibfnamefont {A.}~\bibnamefont {Knipper}}, \bibinfo {author} {\bibfnamefont
  {G.}~\bibnamefont {Walter}}, \bibinfo {author} {\bibfnamefont
  {G.}~\bibnamefont {Marguier}}, \bibinfo {author} {\bibfnamefont
  {B.}~\bibnamefont {Fogelberg}}, \bibinfo {author} {\bibfnamefont
  {A.}~\bibnamefont {Lindroth}}, \bibinfo {author} {\bibfnamefont
  {H.}~\bibnamefont {Mach}}, \bibinfo {author} {\bibfnamefont {M.}~\bibnamefont
  {Sanchez-Vega}}, \bibinfo {author} {\bibfnamefont {R.~B.~E.}\ \bibnamefont
  {Taylor}}, \bibinfo {author} {\bibfnamefont {P.}~\bibnamefont {Van~Duppen}},
  \bibinfo {author} {\bibfnamefont {A.}~\bibnamefont {Jokinen}}, \bibinfo
  {author} {\bibfnamefont {M.}~\bibnamefont {Lindroos}}, \bibinfo {author}
  {\bibfnamefont {M.}~\bibnamefont {Ramdhane}}, \bibinfo {author}
  {\bibfnamefont {W.}~\bibnamefont {Kurcewicz}}, \bibinfo {author}
  {\bibfnamefont {B.}~\bibnamefont {Jonson}}, \bibinfo {author} {\bibfnamefont
  {G.}~\bibnamefont {Nyman}}, \bibinfo {author} {\bibfnamefont
  {Y.}~\bibnamefont {Jading}}, \bibinfo {author} {\bibfnamefont {K.-L.}\
  \bibnamefont {Kratz}}, \bibinfo {author} {\bibfnamefont {A.}~\bibnamefont
  {W\"ohr}}, \bibinfo {author} {\bibfnamefont {G.}~\bibnamefont
  {L\o{}vh\o{}iden}}, \bibinfo {author} {\bibfnamefont {T.~F.}\ \bibnamefont
  {Thorsteinsen}},\ and\ \bibinfo {author} {\bibfnamefont {J.}~\bibnamefont
  {Blomqvist}} (\bibinfo {collaboration} {ISOLDE Collaboration}),\ }\href
  {https://doi.org/10.1103/PhysRevLett.77.1020} {\bibfield  {journal} {\bibinfo
   {journal} {Phys. Rev. Lett.}\ }\textbf {\bibinfo {volume} {77}},\ \bibinfo
  {pages} {1020} (\bibinfo {year} {1996})}\BibitemShut {NoStop}%
\bibitem [{\citenamefont {Piersa}\ \emph {et~al.}(2019)\citenamefont {Piersa},
  \citenamefont {Korgul}, \citenamefont {Fraile}, \citenamefont {Benito},
  \citenamefont {Adamska}, \citenamefont {Andreyev}, \citenamefont
  {\'Alvarez-Rodr\'{\i}guez}, \citenamefont {Barzakh}, \citenamefont {Benzoni},
  \citenamefont {Berry}, \citenamefont {Borge}, \citenamefont {Carmona},
  \citenamefont {Chrysalidis}, \citenamefont {Correia}, \citenamefont
  {Costache}, \citenamefont {Cubiss}, \citenamefont {Day~Goodacre},
  \citenamefont {De~Witte}, \citenamefont {Fedorov}, \citenamefont {Fedosseev},
  \citenamefont {Fern\'andez-Mart\'{\i}nez}, \citenamefont {Fija\l{}kowska},
  \citenamefont {Fila}, \citenamefont {Fynbo}, \citenamefont {Galaviz},
  \citenamefont {Greenlees}, \citenamefont {Grzywacz}, \citenamefont
  {Harkness-Brennan}, \citenamefont {Henrich}, \citenamefont {Huyse},
  \citenamefont {Illana}, \citenamefont {Janas}, \citenamefont {Johnston},
  \citenamefont {Judson}, \citenamefont {Karanyonchev}, \citenamefont
  {Kici\'nska-Habior}, \citenamefont {Konki}, \citenamefont {Kurcewicz},
  \citenamefont {Lazarus}, \citenamefont {Lic\ifmmode~\u{a}\else \u{a}\fi{}},
  \citenamefont {Mach}, \citenamefont {Madurga}, \citenamefont
  {Marroqu\'{\i}n}, \citenamefont {Marsh}, \citenamefont {Mart\'{\i}nez},
  \citenamefont {Mazzocchi}, \citenamefont {M\ifmmode~\u{a}\else
  \u{a}\fi{}rginean}, \citenamefont {M\ifmmode~\u{a}\else \u{a}\fi{}rginean},
  \citenamefont {Miernik}, \citenamefont {Mihai}, \citenamefont {N\'acher},
  \citenamefont {Negret}, \citenamefont {Olaizola}, \citenamefont {Page},
  \citenamefont {Paulaskalas}, \citenamefont {Pascu}, \citenamefont {Perea},
  \citenamefont {Pucknell}, \citenamefont {Rahkila}, \citenamefont {Rapisarda},
  \citenamefont {R\'egis}, \citenamefont {Rotaru}, \citenamefont {Rothe},
  \citenamefont {S\'anchez-Tembleque}, \citenamefont {Simpson}, \citenamefont
  {Sotty}, \citenamefont {Stan}, \citenamefont {St\ifmmode~\u{a}\else
  \u{a}\fi{}noiu}, \citenamefont {Stryjczyk}, \citenamefont {Tengblad},
  \citenamefont {Turturica}, \citenamefont {Ud\'{\i}as}, \citenamefont
  {Van~Duppen}, \citenamefont {Vedia}, \citenamefont {Villa}, \citenamefont
  {Vi\~nals}, \citenamefont {Wadsworth}, \citenamefont {Walters},\ and\
  \citenamefont {Warr}}]{monika}%
  \BibitemOpen
  \bibfield  {author} {\bibinfo {author} {\bibfnamefont {M.}~\bibnamefont
  {Piersa}}, \bibinfo {author} {\bibfnamefont {A.}~\bibnamefont {Korgul}},
  \bibinfo {author} {\bibfnamefont {L.~M.}\ \bibnamefont {Fraile}}, \bibinfo
  {author} {\bibfnamefont {J.}~\bibnamefont {Benito}}, \bibinfo {author}
  {\bibfnamefont {E.}~\bibnamefont {Adamska}}, \bibinfo {author} {\bibfnamefont
  {A.~N.}\ \bibnamefont {Andreyev}}, \bibinfo {author} {\bibfnamefont
  {R.}~\bibnamefont {\'Alvarez-Rodr\'{\i}guez}}, \bibinfo {author}
  {\bibfnamefont {A.~E.}\ \bibnamefont {Barzakh}}, \bibinfo {author}
  {\bibfnamefont {G.}~\bibnamefont {Benzoni}}, \bibinfo {author} {\bibfnamefont
  {T.}~\bibnamefont {Berry}}, \bibinfo {author} {\bibfnamefont {M.~J.~G.}\
  \bibnamefont {Borge}}, \bibinfo {author} {\bibfnamefont {M.}~\bibnamefont
  {Carmona}}, \bibinfo {author} {\bibfnamefont {K.}~\bibnamefont
  {Chrysalidis}}, \bibinfo {author} {\bibfnamefont {J.~G.}\ \bibnamefont
  {Correia}}, \bibinfo {author} {\bibfnamefont {C.}~\bibnamefont {Costache}},
  \bibinfo {author} {\bibfnamefont {J.~G.}\ \bibnamefont {Cubiss}}, \bibinfo
  {author} {\bibfnamefont {T.}~\bibnamefont {Day~Goodacre}}, \bibinfo {author}
  {\bibfnamefont {H.}~\bibnamefont {De~Witte}}, \bibinfo {author}
  {\bibfnamefont {D.~V.}\ \bibnamefont {Fedorov}}, \bibinfo {author}
  {\bibfnamefont {V.~N.}\ \bibnamefont {Fedosseev}}, \bibinfo {author}
  {\bibfnamefont {G.}~\bibnamefont {Fern\'andez-Mart\'{\i}nez}}, \bibinfo
  {author} {\bibfnamefont {A.}~\bibnamefont {Fija\l{}kowska}}, \bibinfo
  {author} {\bibfnamefont {M.}~\bibnamefont {Fila}}, \bibinfo {author}
  {\bibfnamefont {H.}~\bibnamefont {Fynbo}}, \bibinfo {author} {\bibfnamefont
  {D.}~\bibnamefont {Galaviz}}, \bibinfo {author} {\bibfnamefont {P.~T.}\
  \bibnamefont {Greenlees}}, \bibinfo {author} {\bibfnamefont {R.}~\bibnamefont
  {Grzywacz}}, \bibinfo {author} {\bibfnamefont {L.~J.}\ \bibnamefont
  {Harkness-Brennan}}, \bibinfo {author} {\bibfnamefont {C.}~\bibnamefont
  {Henrich}}, \bibinfo {author} {\bibfnamefont {M.}~\bibnamefont {Huyse}},
  \bibinfo {author} {\bibfnamefont {A.}~\bibnamefont {Illana}}, \bibinfo
  {author} {\bibfnamefont {Z.}~\bibnamefont {Janas}}, \bibinfo {author}
  {\bibfnamefont {K.}~\bibnamefont {Johnston}}, \bibinfo {author}
  {\bibfnamefont {D.~S.}\ \bibnamefont {Judson}}, \bibinfo {author}
  {\bibfnamefont {V.}~\bibnamefont {Karanyonchev}}, \bibinfo {author}
  {\bibfnamefont {M.}~\bibnamefont {Kici\'nska-Habior}}, \bibinfo {author}
  {\bibfnamefont {J.}~\bibnamefont {Konki}}, \bibinfo {author} {\bibfnamefont
  {J.}~\bibnamefont {Kurcewicz}}, \bibinfo {author} {\bibfnamefont
  {I.}~\bibnamefont {Lazarus}}, \bibinfo {author} {\bibfnamefont
  {R.}~\bibnamefont {Lic\ifmmode~\u{a}\else \u{a}\fi{}}}, \bibinfo {author}
  {\bibfnamefont {H.}~\bibnamefont {Mach}}, \bibinfo {author} {\bibfnamefont
  {M.}~\bibnamefont {Madurga}}, \bibinfo {author} {\bibfnamefont
  {I.}~\bibnamefont {Marroqu\'{\i}n}}, \bibinfo {author} {\bibfnamefont
  {B.}~\bibnamefont {Marsh}}, \bibinfo {author} {\bibfnamefont {M.~C.}\
  \bibnamefont {Mart\'{\i}nez}}, \bibinfo {author} {\bibfnamefont
  {C.}~\bibnamefont {Mazzocchi}}, \bibinfo {author} {\bibfnamefont
  {N.}~\bibnamefont {M\ifmmode~\u{a}\else \u{a}\fi{}rginean}}, \bibinfo
  {author} {\bibfnamefont {R.}~\bibnamefont {M\ifmmode~\u{a}\else
  \u{a}\fi{}rginean}}, \bibinfo {author} {\bibfnamefont {K.}~\bibnamefont
  {Miernik}}, \bibinfo {author} {\bibfnamefont {C.}~\bibnamefont {Mihai}},
  \bibinfo {author} {\bibfnamefont {E.}~\bibnamefont {N\'acher}}, \bibinfo
  {author} {\bibfnamefont {A.}~\bibnamefont {Negret}}, \bibinfo {author}
  {\bibfnamefont {B.}~\bibnamefont {Olaizola}}, \bibinfo {author}
  {\bibfnamefont {R.~D.}\ \bibnamefont {Page}}, \bibinfo {author}
  {\bibfnamefont {S.}~\bibnamefont {Paulaskalas}}, \bibinfo {author}
  {\bibfnamefont {S.}~\bibnamefont {Pascu}}, \bibinfo {author} {\bibfnamefont
  {A.}~\bibnamefont {Perea}}, \bibinfo {author} {\bibfnamefont
  {V.}~\bibnamefont {Pucknell}}, \bibinfo {author} {\bibfnamefont
  {P.}~\bibnamefont {Rahkila}}, \bibinfo {author} {\bibfnamefont
  {E.}~\bibnamefont {Rapisarda}}, \bibinfo {author} {\bibfnamefont {J.-M.}\
  \bibnamefont {R\'egis}}, \bibinfo {author} {\bibfnamefont {F.}~\bibnamefont
  {Rotaru}}, \bibinfo {author} {\bibfnamefont {S.}~\bibnamefont {Rothe}},
  \bibinfo {author} {\bibfnamefont {V.}~\bibnamefont {S\'anchez-Tembleque}},
  \bibinfo {author} {\bibfnamefont {G.}~\bibnamefont {Simpson}}, \bibinfo
  {author} {\bibfnamefont {C.}~\bibnamefont {Sotty}}, \bibinfo {author}
  {\bibfnamefont {L.}~\bibnamefont {Stan}}, \bibinfo {author} {\bibfnamefont
  {M.}~\bibnamefont {St\ifmmode~\u{a}\else \u{a}\fi{}noiu}}, \bibinfo {author}
  {\bibfnamefont {M.}~\bibnamefont {Stryjczyk}}, \bibinfo {author}
  {\bibfnamefont {O.}~\bibnamefont {Tengblad}}, \bibinfo {author}
  {\bibfnamefont {A.}~\bibnamefont {Turturica}}, \bibinfo {author}
  {\bibfnamefont {J.~M.}\ \bibnamefont {Ud\'{\i}as}}, \bibinfo {author}
  {\bibfnamefont {P.}~\bibnamefont {Van~Duppen}}, \bibinfo {author}
  {\bibfnamefont {V.}~\bibnamefont {Vedia}}, \bibinfo {author} {\bibfnamefont
  {A.}~\bibnamefont {Villa}}, \bibinfo {author} {\bibfnamefont
  {S.}~\bibnamefont {Vi\~nals}}, \bibinfo {author} {\bibfnamefont
  {R.}~\bibnamefont {Wadsworth}}, \bibinfo {author} {\bibfnamefont {W.~B.}\
  \bibnamefont {Walters}},\ and\ \bibinfo {author} {\bibfnamefont
  {N.}~\bibnamefont {Warr}} (\bibinfo {collaboration} {IDS Collaboration}),\
  }\href {https://doi.org/10.1103/PhysRevC.99.024304} {\bibfield  {journal}
  {\bibinfo  {journal} {Phys. Rev. C}\ }\textbf {\bibinfo {volume} {99}},\
  \bibinfo {pages} {024304} (\bibinfo {year} {2019})}\BibitemShut {NoStop}%
\bibitem [{\citenamefont {Benito}\ \emph {et~al.}(2020)\citenamefont {Benito},
  \citenamefont {Fraile}, \citenamefont {Korgul}, \citenamefont {Piersa},
  \citenamefont {Adamska}, \citenamefont {Andreyev}, \citenamefont
  {\'Alvarez-Rodr\'{\i}guez}, \citenamefont {Barzakh}, \citenamefont {Benzoni},
  \citenamefont {Berry}, \citenamefont {Borge}, \citenamefont {Carmona},
  \citenamefont {Chrysalidis}, \citenamefont {Costache}, \citenamefont
  {Cubiss}, \citenamefont {Day~Goodacre}, \citenamefont {De~Witte},
  \citenamefont {Fedorov}, \citenamefont {Fedosseev}, \citenamefont
  {Fern\'andez-Mart\'{\i}nez}, \citenamefont {Fija\l{}kowska}, \citenamefont
  {Fila}, \citenamefont {Fynbo}, \citenamefont {Galaviz}, \citenamefont
  {Galve}, \citenamefont {Garc\'{\i}a-D\'{\i}ez}, \citenamefont {Greenlees},
  \citenamefont {Grzywacz}, \citenamefont {Harkness-Brennan}, \citenamefont
  {Henrich}, \citenamefont {Huyse}, \citenamefont {Ib\'a\~nez}, \citenamefont
  {Illana}, \citenamefont {Janas}, \citenamefont {Jolie}, \citenamefont
  {Judson}, \citenamefont {Karayonchev}, \citenamefont {Kici\'nska-Habior},
  \citenamefont {Konki}, \citenamefont {Kurcewicz}, \citenamefont {Lazarus},
  \citenamefont {Lic\ifmmode~\u{a}\else \u{a}\fi{}}, \citenamefont
  {L\'opez-Montes}, \citenamefont {Lund}, \citenamefont {Mach}, \citenamefont
  {Madurga}, \citenamefont {Marroqu\'{\i}n}, \citenamefont {Marsh},
  \citenamefont {Mart\'{\i}nez}, \citenamefont {Mazzocchi}, \citenamefont
  {M\ifmmode~\u{a}\else \u{a}\fi{}rginean}, \citenamefont {M\ifmmode~\u{a}\else
  \u{a}\fi{}rginean}, \citenamefont {Miernik}, \citenamefont {Mihai},
  \citenamefont {Mihai}, \citenamefont {N\'acher}, \citenamefont {Negret},
  \citenamefont {Olaizola}, \citenamefont {Page}, \citenamefont {Paulauskas},
  \citenamefont {Pascu}, \citenamefont {Perea}, \citenamefont {Pucknell},
  \citenamefont {Rahkila}, \citenamefont {Raison}, \citenamefont {Rapisarda},
  \citenamefont {R\'egis}, \citenamefont {Rezynkina}, \citenamefont {Rotaru},
  \citenamefont {Rothe}, \citenamefont {S\'anchez-Parcerisa}, \citenamefont
  {S\'anchez-Tembleque}, \citenamefont {Schomacker}, \citenamefont {Simpson},
  \citenamefont {Sotty}, \citenamefont {Stan}, \citenamefont
  {St\ifmmode~\u{a}\else \u{a}\fi{}noiu}, \citenamefont {Stryjczyk},
  \citenamefont {Tengblad}, \citenamefont {Turturica}, \citenamefont
  {Ud\'{\i}as}, \citenamefont {Van~Duppen}, \citenamefont {Vedia},
  \citenamefont {Villa-Abaunza}, \citenamefont {Vi\~nals}, \citenamefont
  {Walters}, \citenamefont {Wadsworth},\ and\ \citenamefont {Warr}}]{benito}%
  \BibitemOpen
  \bibfield  {author} {\bibinfo {author} {\bibfnamefont {J.}~\bibnamefont
  {Benito}}, \bibinfo {author} {\bibfnamefont {L.~M.}\ \bibnamefont {Fraile}},
  \bibinfo {author} {\bibfnamefont {A.}~\bibnamefont {Korgul}}, \bibinfo
  {author} {\bibfnamefont {M.}~\bibnamefont {Piersa}}, \bibinfo {author}
  {\bibfnamefont {E.}~\bibnamefont {Adamska}}, \bibinfo {author} {\bibfnamefont
  {A.~N.}\ \bibnamefont {Andreyev}}, \bibinfo {author} {\bibfnamefont
  {R.}~\bibnamefont {\'Alvarez-Rodr\'{\i}guez}}, \bibinfo {author}
  {\bibfnamefont {A.~E.}\ \bibnamefont {Barzakh}}, \bibinfo {author}
  {\bibfnamefont {G.}~\bibnamefont {Benzoni}}, \bibinfo {author} {\bibfnamefont
  {T.}~\bibnamefont {Berry}}, \bibinfo {author} {\bibfnamefont {M.~J.~G.}\
  \bibnamefont {Borge}}, \bibinfo {author} {\bibfnamefont {M.}~\bibnamefont
  {Carmona}}, \bibinfo {author} {\bibfnamefont {K.}~\bibnamefont
  {Chrysalidis}}, \bibinfo {author} {\bibfnamefont {C.}~\bibnamefont
  {Costache}}, \bibinfo {author} {\bibfnamefont {J.~G.}\ \bibnamefont
  {Cubiss}}, \bibinfo {author} {\bibfnamefont {T.}~\bibnamefont
  {Day~Goodacre}}, \bibinfo {author} {\bibfnamefont {H.}~\bibnamefont
  {De~Witte}}, \bibinfo {author} {\bibfnamefont {D.~V.}\ \bibnamefont
  {Fedorov}}, \bibinfo {author} {\bibfnamefont {V.~N.}\ \bibnamefont
  {Fedosseev}}, \bibinfo {author} {\bibfnamefont {G.}~\bibnamefont
  {Fern\'andez-Mart\'{\i}nez}}, \bibinfo {author} {\bibfnamefont
  {A.}~\bibnamefont {Fija\l{}kowska}}, \bibinfo {author} {\bibfnamefont
  {M.}~\bibnamefont {Fila}}, \bibinfo {author} {\bibfnamefont {H.}~\bibnamefont
  {Fynbo}}, \bibinfo {author} {\bibfnamefont {D.}~\bibnamefont {Galaviz}},
  \bibinfo {author} {\bibfnamefont {P.}~\bibnamefont {Galve}}, \bibinfo
  {author} {\bibfnamefont {M.}~\bibnamefont {Garc\'{\i}a-D\'{\i}ez}}, \bibinfo
  {author} {\bibfnamefont {P.~T.}\ \bibnamefont {Greenlees}}, \bibinfo {author}
  {\bibfnamefont {R.}~\bibnamefont {Grzywacz}}, \bibinfo {author}
  {\bibfnamefont {L.~J.}\ \bibnamefont {Harkness-Brennan}}, \bibinfo {author}
  {\bibfnamefont {C.}~\bibnamefont {Henrich}}, \bibinfo {author} {\bibfnamefont
  {M.}~\bibnamefont {Huyse}}, \bibinfo {author} {\bibfnamefont
  {P.}~\bibnamefont {Ib\'a\~nez}}, \bibinfo {author} {\bibfnamefont
  {A.}~\bibnamefont {Illana}}, \bibinfo {author} {\bibfnamefont
  {Z.}~\bibnamefont {Janas}}, \bibinfo {author} {\bibfnamefont
  {J.}~\bibnamefont {Jolie}}, \bibinfo {author} {\bibfnamefont {D.~S.}\
  \bibnamefont {Judson}}, \bibinfo {author} {\bibfnamefont {V.}~\bibnamefont
  {Karayonchev}}, \bibinfo {author} {\bibfnamefont {M.}~\bibnamefont
  {Kici\'nska-Habior}}, \bibinfo {author} {\bibfnamefont {J.}~\bibnamefont
  {Konki}}, \bibinfo {author} {\bibfnamefont {J.}~\bibnamefont {Kurcewicz}},
  \bibinfo {author} {\bibfnamefont {I.}~\bibnamefont {Lazarus}}, \bibinfo
  {author} {\bibfnamefont {R.}~\bibnamefont {Lic\ifmmode~\u{a}\else
  \u{a}\fi{}}}, \bibinfo {author} {\bibfnamefont {A.}~\bibnamefont
  {L\'opez-Montes}}, \bibinfo {author} {\bibfnamefont {M.}~\bibnamefont
  {Lund}}, \bibinfo {author} {\bibfnamefont {H.}~\bibnamefont {Mach}}, \bibinfo
  {author} {\bibfnamefont {M.}~\bibnamefont {Madurga}}, \bibinfo {author}
  {\bibfnamefont {I.}~\bibnamefont {Marroqu\'{\i}n}}, \bibinfo {author}
  {\bibfnamefont {B.}~\bibnamefont {Marsh}}, \bibinfo {author} {\bibfnamefont
  {M.~C.}\ \bibnamefont {Mart\'{\i}nez}}, \bibinfo {author} {\bibfnamefont
  {C.}~\bibnamefont {Mazzocchi}}, \bibinfo {author} {\bibfnamefont
  {N.}~\bibnamefont {M\ifmmode~\u{a}\else \u{a}\fi{}rginean}}, \bibinfo
  {author} {\bibfnamefont {R.}~\bibnamefont {M\ifmmode~\u{a}\else
  \u{a}\fi{}rginean}}, \bibinfo {author} {\bibfnamefont {K.}~\bibnamefont
  {Miernik}}, \bibinfo {author} {\bibfnamefont {C.}~\bibnamefont {Mihai}},
  \bibinfo {author} {\bibfnamefont {R.~E.}\ \bibnamefont {Mihai}}, \bibinfo
  {author} {\bibfnamefont {E.}~\bibnamefont {N\'acher}}, \bibinfo {author}
  {\bibfnamefont {A.}~\bibnamefont {Negret}}, \bibinfo {author} {\bibfnamefont
  {B.}~\bibnamefont {Olaizola}}, \bibinfo {author} {\bibfnamefont {R.~D.}\
  \bibnamefont {Page}}, \bibinfo {author} {\bibfnamefont {S.~V.}\ \bibnamefont
  {Paulauskas}}, \bibinfo {author} {\bibfnamefont {S.}~\bibnamefont {Pascu}},
  \bibinfo {author} {\bibfnamefont {A.}~\bibnamefont {Perea}}, \bibinfo
  {author} {\bibfnamefont {V.}~\bibnamefont {Pucknell}}, \bibinfo {author}
  {\bibfnamefont {P.}~\bibnamefont {Rahkila}}, \bibinfo {author} {\bibfnamefont
  {C.}~\bibnamefont {Raison}}, \bibinfo {author} {\bibfnamefont
  {E.}~\bibnamefont {Rapisarda}}, \bibinfo {author} {\bibfnamefont {J.-M.}\
  \bibnamefont {R\'egis}}, \bibinfo {author} {\bibfnamefont {K.}~\bibnamefont
  {Rezynkina}}, \bibinfo {author} {\bibfnamefont {F.}~\bibnamefont {Rotaru}},
  \bibinfo {author} {\bibfnamefont {S.}~\bibnamefont {Rothe}}, \bibinfo
  {author} {\bibfnamefont {D.}~\bibnamefont {S\'anchez-Parcerisa}}, \bibinfo
  {author} {\bibfnamefont {V.}~\bibnamefont {S\'anchez-Tembleque}}, \bibinfo
  {author} {\bibfnamefont {K.}~\bibnamefont {Schomacker}}, \bibinfo {author}
  {\bibfnamefont {G.~S.}\ \bibnamefont {Simpson}}, \bibinfo {author}
  {\bibfnamefont {C.}~\bibnamefont {Sotty}}, \bibinfo {author} {\bibfnamefont
  {L.}~\bibnamefont {Stan}}, \bibinfo {author} {\bibfnamefont {M.}~\bibnamefont
  {St\ifmmode~\u{a}\else \u{a}\fi{}noiu}}, \bibinfo {author} {\bibfnamefont
  {M.}~\bibnamefont {Stryjczyk}}, \bibinfo {author} {\bibfnamefont
  {O.}~\bibnamefont {Tengblad}}, \bibinfo {author} {\bibfnamefont
  {A.}~\bibnamefont {Turturica}}, \bibinfo {author} {\bibfnamefont {J.~M.}\
  \bibnamefont {Ud\'{\i}as}}, \bibinfo {author} {\bibfnamefont
  {P.}~\bibnamefont {Van~Duppen}}, \bibinfo {author} {\bibfnamefont
  {V.}~\bibnamefont {Vedia}}, \bibinfo {author} {\bibfnamefont
  {A.}~\bibnamefont {Villa-Abaunza}}, \bibinfo {author} {\bibfnamefont
  {S.}~\bibnamefont {Vi\~nals}}, \bibinfo {author} {\bibfnamefont {W.~B.}\
  \bibnamefont {Walters}}, \bibinfo {author} {\bibfnamefont {R.}~\bibnamefont
  {Wadsworth}},\ and\ \bibinfo {author} {\bibfnamefont {N.}~\bibnamefont
  {Warr}} (\bibinfo {collaboration} {IDS Collaboration}),\ }\href
  {https://doi.org/10.1103/PhysRevC.102.014328} {\bibfield  {journal} {\bibinfo
   {journal} {Phys. Rev. C}\ }\textbf {\bibinfo {volume} {102}},\ \bibinfo
  {pages} {014328} (\bibinfo {year} {2020})}\BibitemShut {NoStop}%
\bibitem [{\citenamefont {Jones}\ \emph {et~al.}(2010)\citenamefont {Jones},
  \citenamefont {Adekola}, \citenamefont {Bardayan}, \citenamefont {Blackmon},
  \citenamefont {Chae}, \citenamefont {Chipps}, \citenamefont {Cizewski},
  \citenamefont {Erikson}, \citenamefont {Harlin}, \citenamefont {Hatarik},
  \citenamefont {Kapler}, \citenamefont {Kozub}, \citenamefont {Liang},
  \citenamefont {Livesay}, \citenamefont {Ma}, \citenamefont {Moazen},
  \citenamefont {Nesaraja}, \citenamefont {Nunes}, \citenamefont {Pain},
  \citenamefont {Patterson}, \citenamefont {Shapira}, \citenamefont {Shriner},
  \citenamefont {Smith}, \citenamefont {Swan},\ and\ \citenamefont
  {Thomas}}]{kate}%
  \BibitemOpen
  \bibfield  {author} {\bibinfo {author} {\bibfnamefont {K.~L.}\ \bibnamefont
  {Jones}}, \bibinfo {author} {\bibfnamefont {A.~S.}\ \bibnamefont {Adekola}},
  \bibinfo {author} {\bibfnamefont {D.~W.}\ \bibnamefont {Bardayan}}, \bibinfo
  {author} {\bibfnamefont {J.~C.}\ \bibnamefont {Blackmon}}, \bibinfo {author}
  {\bibfnamefont {K.~Y.}\ \bibnamefont {Chae}}, \bibinfo {author}
  {\bibfnamefont {K.~A.}\ \bibnamefont {Chipps}}, \bibinfo {author}
  {\bibfnamefont {J.~A.}\ \bibnamefont {Cizewski}}, \bibinfo {author}
  {\bibfnamefont {L.}~\bibnamefont {Erikson}}, \bibinfo {author} {\bibfnamefont
  {C.}~\bibnamefont {Harlin}}, \bibinfo {author} {\bibfnamefont
  {R.}~\bibnamefont {Hatarik}}, \bibinfo {author} {\bibfnamefont
  {R.}~\bibnamefont {Kapler}}, \bibinfo {author} {\bibfnamefont {R.~L.}\
  \bibnamefont {Kozub}}, \bibinfo {author} {\bibfnamefont {J.~F.}\ \bibnamefont
  {Liang}}, \bibinfo {author} {\bibfnamefont {R.}~\bibnamefont {Livesay}},
  \bibinfo {author} {\bibfnamefont {Z.}~\bibnamefont {Ma}}, \bibinfo {author}
  {\bibfnamefont {B.~H.}\ \bibnamefont {Moazen}}, \bibinfo {author}
  {\bibfnamefont {C.~D.}\ \bibnamefont {Nesaraja}}, \bibinfo {author}
  {\bibfnamefont {F.~M.}\ \bibnamefont {Nunes}}, \bibinfo {author}
  {\bibfnamefont {S.~D.}\ \bibnamefont {Pain}}, \bibinfo {author}
  {\bibfnamefont {N.~P.}\ \bibnamefont {Patterson}}, \bibinfo {author}
  {\bibfnamefont {D.}~\bibnamefont {Shapira}}, \bibinfo {author} {\bibfnamefont
  {J.~F.}\ \bibnamefont {Shriner}}, \bibinfo {author} {\bibfnamefont {M.~S.}\
  \bibnamefont {Smith}}, \bibinfo {author} {\bibfnamefont {T.~P.}\ \bibnamefont
  {Swan}},\ and\ \bibinfo {author} {\bibfnamefont {J.~S.}\ \bibnamefont
  {Thomas}},\ }\href {https://doi.org/10.1038/nature09048} {\bibfield
  {journal} {\bibinfo  {journal} {Nature}\ }\textbf {\bibinfo {volume} {465}},\
  \bibinfo {pages} {454} (\bibinfo {year} {2010})}\BibitemShut {NoStop}%
\bibitem [{\citenamefont {Allmond}\ \emph {et~al.}(2014)\citenamefont
  {Allmond}, \citenamefont {Stuchbery}, \citenamefont {Beene}, \citenamefont
  {Galindo-Uribarri}, \citenamefont {Liang}, \citenamefont {Padilla-Rodal},
  \citenamefont {Radford}, \citenamefont {Varner}, \citenamefont {Ayres},
  \citenamefont {Batchelder}, \citenamefont {Bey}, \citenamefont {Bingham},
  \citenamefont {Howard}, \citenamefont {Jones}, \citenamefont {Manning},
  \citenamefont {Mueller}, \citenamefont {Nesaraja}, \citenamefont {Pain},
  \citenamefont {Peters}, \citenamefont {Ratkiewicz}, \citenamefont {Schmitt},
  \citenamefont {Shapira}, \citenamefont {Smith}, \citenamefont {Stone},
  \citenamefont {Stracener},\ and\ \citenamefont {Yu}}]{allmond}%
  \BibitemOpen
  \bibfield  {author} {\bibinfo {author} {\bibfnamefont {J.~M.}\ \bibnamefont
  {Allmond}}, \bibinfo {author} {\bibfnamefont {A.~E.}\ \bibnamefont
  {Stuchbery}}, \bibinfo {author} {\bibfnamefont {J.~R.}\ \bibnamefont
  {Beene}}, \bibinfo {author} {\bibfnamefont {A.}~\bibnamefont
  {Galindo-Uribarri}}, \bibinfo {author} {\bibfnamefont {J.~F.}\ \bibnamefont
  {Liang}}, \bibinfo {author} {\bibfnamefont {E.}~\bibnamefont
  {Padilla-Rodal}}, \bibinfo {author} {\bibfnamefont {D.~C.}\ \bibnamefont
  {Radford}}, \bibinfo {author} {\bibfnamefont {R.~L.}\ \bibnamefont {Varner}},
  \bibinfo {author} {\bibfnamefont {A.}~\bibnamefont {Ayres}}, \bibinfo
  {author} {\bibfnamefont {J.~C.}\ \bibnamefont {Batchelder}}, \bibinfo
  {author} {\bibfnamefont {A.}~\bibnamefont {Bey}}, \bibinfo {author}
  {\bibfnamefont {C.~R.}\ \bibnamefont {Bingham}}, \bibinfo {author}
  {\bibfnamefont {M.~E.}\ \bibnamefont {Howard}}, \bibinfo {author}
  {\bibfnamefont {K.~L.}\ \bibnamefont {Jones}}, \bibinfo {author}
  {\bibfnamefont {B.}~\bibnamefont {Manning}}, \bibinfo {author} {\bibfnamefont
  {P.~E.}\ \bibnamefont {Mueller}}, \bibinfo {author} {\bibfnamefont {C.~D.}\
  \bibnamefont {Nesaraja}}, \bibinfo {author} {\bibfnamefont {S.~D.}\
  \bibnamefont {Pain}}, \bibinfo {author} {\bibfnamefont {W.~A.}\ \bibnamefont
  {Peters}}, \bibinfo {author} {\bibfnamefont {A.}~\bibnamefont {Ratkiewicz}},
  \bibinfo {author} {\bibfnamefont {K.~T.}\ \bibnamefont {Schmitt}}, \bibinfo
  {author} {\bibfnamefont {D.}~\bibnamefont {Shapira}}, \bibinfo {author}
  {\bibfnamefont {M.~S.}\ \bibnamefont {Smith}}, \bibinfo {author}
  {\bibfnamefont {N.~J.}\ \bibnamefont {Stone}}, \bibinfo {author}
  {\bibfnamefont {D.~W.}\ \bibnamefont {Stracener}},\ and\ \bibinfo {author}
  {\bibfnamefont {C.-H.}\ \bibnamefont {Yu}},\ }\href
  {https://doi.org/10.1103/PhysRevLett.112.172701} {\bibfield  {journal}
  {\bibinfo  {journal} {Phys. Rev. Lett.}\ }\textbf {\bibinfo {volume} {112}},\
  \bibinfo {pages} {172701} (\bibinfo {year} {2014})}\BibitemShut {NoStop}%
\bibitem [{\citenamefont {Vaquero}\ \emph {et~al.}(2017)\citenamefont
  {Vaquero}, \citenamefont {Jungclaus}, \citenamefont {Doornenbal},
  \citenamefont {Wimmer}, \citenamefont {Gargano}, \citenamefont {Tostevin},
  \citenamefont {Chen}, \citenamefont {N\'acher}, \citenamefont {Sahin},
  \citenamefont {Shiga}, \citenamefont {Steppenbeck}, \citenamefont {Taniuchi},
  \citenamefont {Xu}, \citenamefont {Ando}, \citenamefont {Baba}, \citenamefont
  {Garrote}, \citenamefont {Franchoo}, \citenamefont {Hadynska-Klek},
  \citenamefont {Kusoglu}, \citenamefont {Liu}, \citenamefont {Lokotko},
  \citenamefont {Momiyama}, \citenamefont {Motobayashi}, \citenamefont
  {Nagamine}, \citenamefont {Nakatsuka}, \citenamefont {Niikura}, \citenamefont
  {Orlandi}, \citenamefont {Saito}, \citenamefont {Sakurai}, \citenamefont
  {S\"oderstr\"om}, \citenamefont {Tveten}, \citenamefont {Vajta},\ and\
  \citenamefont {Yalcinkaya}}]{vaquero17}%
  \BibitemOpen
  \bibfield  {author} {\bibinfo {author} {\bibfnamefont {V.}~\bibnamefont
  {Vaquero}}, \bibinfo {author} {\bibfnamefont {A.}~\bibnamefont {Jungclaus}},
  \bibinfo {author} {\bibfnamefont {P.}~\bibnamefont {Doornenbal}}, \bibinfo
  {author} {\bibfnamefont {K.}~\bibnamefont {Wimmer}}, \bibinfo {author}
  {\bibfnamefont {A.}~\bibnamefont {Gargano}}, \bibinfo {author} {\bibfnamefont
  {J.~A.}\ \bibnamefont {Tostevin}}, \bibinfo {author} {\bibfnamefont
  {S.}~\bibnamefont {Chen}}, \bibinfo {author} {\bibfnamefont {E.}~\bibnamefont
  {N\'acher}}, \bibinfo {author} {\bibfnamefont {E.}~\bibnamefont {Sahin}},
  \bibinfo {author} {\bibfnamefont {Y.}~\bibnamefont {Shiga}}, \bibinfo
  {author} {\bibfnamefont {D.}~\bibnamefont {Steppenbeck}}, \bibinfo {author}
  {\bibfnamefont {R.}~\bibnamefont {Taniuchi}}, \bibinfo {author}
  {\bibfnamefont {Z.~Y.}\ \bibnamefont {Xu}}, \bibinfo {author} {\bibfnamefont
  {T.}~\bibnamefont {Ando}}, \bibinfo {author} {\bibfnamefont {H.}~\bibnamefont
  {Baba}}, \bibinfo {author} {\bibfnamefont {F.~L.~B.}\ \bibnamefont
  {Garrote}}, \bibinfo {author} {\bibfnamefont {S.}~\bibnamefont {Franchoo}},
  \bibinfo {author} {\bibfnamefont {K.}~\bibnamefont {Hadynska-Klek}}, \bibinfo
  {author} {\bibfnamefont {A.}~\bibnamefont {Kusoglu}}, \bibinfo {author}
  {\bibfnamefont {J.}~\bibnamefont {Liu}}, \bibinfo {author} {\bibfnamefont
  {T.}~\bibnamefont {Lokotko}}, \bibinfo {author} {\bibfnamefont
  {S.}~\bibnamefont {Momiyama}}, \bibinfo {author} {\bibfnamefont
  {T.}~\bibnamefont {Motobayashi}}, \bibinfo {author} {\bibfnamefont
  {S.}~\bibnamefont {Nagamine}}, \bibinfo {author} {\bibfnamefont
  {N.}~\bibnamefont {Nakatsuka}}, \bibinfo {author} {\bibfnamefont
  {M.}~\bibnamefont {Niikura}}, \bibinfo {author} {\bibfnamefont
  {R.}~\bibnamefont {Orlandi}}, \bibinfo {author} {\bibfnamefont
  {T.}~\bibnamefont {Saito}}, \bibinfo {author} {\bibfnamefont
  {H.}~\bibnamefont {Sakurai}}, \bibinfo {author} {\bibfnamefont {P.~A.}\
  \bibnamefont {S\"oderstr\"om}}, \bibinfo {author} {\bibfnamefont {G.~M.}\
  \bibnamefont {Tveten}}, \bibinfo {author} {\bibfnamefont {Z.}~\bibnamefont
  {Vajta}},\ and\ \bibinfo {author} {\bibfnamefont {M.}~\bibnamefont
  {Yalcinkaya}},\ }\href {https://doi.org/10.1103/PhysRevLett.118.202502}
  {\bibfield  {journal} {\bibinfo  {journal} {Phys. Rev. Lett.}\ }\textbf
  {\bibinfo {volume} {118}},\ \bibinfo {pages} {202502} (\bibinfo {year}
  {2017})}\BibitemShut {NoStop}%
\bibitem [{\citenamefont {Catherall}\ \emph {et~al.}(2017)\citenamefont
  {Catherall}, \citenamefont {Andreazza}, \citenamefont {Breitenfeldt},
  \citenamefont {Dorsival}, \citenamefont {Focker}, \citenamefont {Gharsa},
  \citenamefont {J}, \citenamefont {Grenard}, \citenamefont {Locci},
  \citenamefont {Martins}, \citenamefont {Marzari}, \citenamefont {Schipper},
  \citenamefont {Shornikov},\ and\ \citenamefont {Stora}}]{isolde}%
  \BibitemOpen
  \bibfield  {author} {\bibinfo {author} {\bibfnamefont {R.}~\bibnamefont
  {Catherall}}, \bibinfo {author} {\bibfnamefont {W.}~\bibnamefont
  {Andreazza}}, \bibinfo {author} {\bibfnamefont {M.}~\bibnamefont
  {Breitenfeldt}}, \bibinfo {author} {\bibfnamefont {A.}~\bibnamefont
  {Dorsival}}, \bibinfo {author} {\bibfnamefont {G.~J.}\ \bibnamefont
  {Focker}}, \bibinfo {author} {\bibfnamefont {T.~P.}\ \bibnamefont {Gharsa}},
  \bibinfo {author} {\bibfnamefont {G.~T.}\ \bibnamefont {J}}, \bibinfo
  {author} {\bibfnamefont {J.-L.}\ \bibnamefont {Grenard}}, \bibinfo {author}
  {\bibfnamefont {F.}~\bibnamefont {Locci}}, \bibinfo {author} {\bibfnamefont
  {P.}~\bibnamefont {Martins}}, \bibinfo {author} {\bibfnamefont
  {S.}~\bibnamefont {Marzari}}, \bibinfo {author} {\bibfnamefont
  {J.}~\bibnamefont {Schipper}}, \bibinfo {author} {\bibfnamefont
  {A.}~\bibnamefont {Shornikov}},\ and\ \bibinfo {author} {\bibfnamefont
  {T.}~\bibnamefont {Stora}},\ }\href
  {https://doi.org/10.1088/1361-6471/aa7eba} {\bibfield  {journal} {\bibinfo
  {journal} {Journal of Physics G: Nuclear and Particle Physics}\ }\textbf
  {\bibinfo {volume} {44}},\ \bibinfo {pages} {094002} (\bibinfo {year}
  {2017})}\BibitemShut {NoStop}%
\bibitem [{\citenamefont {Fedosseev}\ \emph {et~al.}(2017)\citenamefont
  {Fedosseev}, \citenamefont {Chrysalidis}, \citenamefont {Goodacre},
  \citenamefont {Marsh}, \citenamefont {Rothe}, \citenamefont {Seiffert},\ and\
  \citenamefont {Wendt}}]{rilis}%
  \BibitemOpen
  \bibfield  {author} {\bibinfo {author} {\bibfnamefont {V.}~\bibnamefont
  {Fedosseev}}, \bibinfo {author} {\bibfnamefont {K.}~\bibnamefont
  {Chrysalidis}}, \bibinfo {author} {\bibfnamefont {T.~D.}\ \bibnamefont
  {Goodacre}}, \bibinfo {author} {\bibfnamefont {B.}~\bibnamefont {Marsh}},
  \bibinfo {author} {\bibfnamefont {S.}~\bibnamefont {Rothe}}, \bibinfo
  {author} {\bibfnamefont {C.}~\bibnamefont {Seiffert}},\ and\ \bibinfo
  {author} {\bibfnamefont {K.}~\bibnamefont {Wendt}},\ }\href
  {https://doi.org/10.1088/1361-6471/aa78e0} {\bibfield  {journal} {\bibinfo
  {journal} {Journal of Physics G: Nuclear and Particle Physics}\ }\textbf
  {\bibinfo {volume} {44}},\ \bibinfo {pages} {084006} (\bibinfo {year}
  {2017})}\BibitemShut {NoStop}%
\bibitem [{\citenamefont {Xu}\ \emph {et~al.}()\citenamefont {Xu} \emph
  {et~al.}}]{133In-prc}%
  \BibitemOpen
  \bibfield  {author} {\bibinfo {author} {\bibfnamefont {Z.~Y.}\ \bibnamefont
  {Xu}} \emph {et~al.},\ }\bibinfo {note} {submitted to Phys. Rev.
  C}\BibitemShut {NoStop}%
\bibitem [{\citenamefont {Duke}\ \emph {et~al.}(1970)\citenamefont {Duke},
  \citenamefont {Hansen}, \citenamefont {Nielsen},\ and\ \citenamefont
  {Rudstam}}]{sbeta}%
  \BibitemOpen
  \bibfield  {author} {\bibinfo {author} {\bibfnamefont {C.}~\bibnamefont
  {Duke}}, \bibinfo {author} {\bibfnamefont {P.}~\bibnamefont {Hansen}},
  \bibinfo {author} {\bibfnamefont {O.}~\bibnamefont {Nielsen}},\ and\ \bibinfo
  {author} {\bibfnamefont {G.}~\bibnamefont {Rudstam}},\ }\href
  {https://doi.org/https://doi.org/10.1016/0375-9474(70)90400-8} {\bibfield
  {journal} {\bibinfo  {journal} {Nuclear Physics A}\ }\textbf {\bibinfo
  {volume} {151}},\ \bibinfo {pages} {609} (\bibinfo {year}
  {1970})}\BibitemShut {NoStop}%
\bibitem [{\citenamefont {Orear}(1950)}]{fermi}%
  \BibitemOpen
  \bibfield  {author} {\bibinfo {author} {\bibfnamefont {J.}~\bibnamefont
  {Orear}},\ }\href@noop {} {\emph {\bibinfo {title} {{Nuclear physics, a
  course given by Enrico Fermi at the University of Chicago. Revised
  edition.}}}}\ (\bibinfo  {publisher} {University of Chicago Press},\ \bibinfo
  {address} {Chicago},\ \bibinfo {year} {1950})\BibitemShut {NoStop}%
\bibitem [{\citenamefont {Singh}\ \emph {et~al.}(1998)\citenamefont {Singh},
  \citenamefont {Rodriguez}, \citenamefont {Wong},\ and\ \citenamefont
  {Tuli}}]{logftreview}%
  \BibitemOpen
  \bibfield  {author} {\bibinfo {author} {\bibfnamefont {B.}~\bibnamefont
  {Singh}}, \bibinfo {author} {\bibfnamefont {J.}~\bibnamefont {Rodriguez}},
  \bibinfo {author} {\bibfnamefont {S.}~\bibnamefont {Wong}},\ and\ \bibinfo
  {author} {\bibfnamefont {J.}~\bibnamefont {Tuli}},\ }\href
  {https://doi.org/https://doi.org/10.1006/ndsh.1998.0015} {\bibfield
  {journal} {\bibinfo  {journal} {Nuclear Data Sheets}\ }\textbf {\bibinfo
  {volume} {84}},\ \bibinfo {pages} {487} (\bibinfo {year} {1998})}\BibitemShut
  {NoStop}%
\bibitem [{\citenamefont {Entem}\ and\ \citenamefont {Machleidt}(2003)}]{n3lo}%
  \BibitemOpen
  \bibfield  {author} {\bibinfo {author} {\bibfnamefont {D.~R.}\ \bibnamefont
  {Entem}}\ and\ \bibinfo {author} {\bibfnamefont {R.}~\bibnamefont
  {Machleidt}},\ }\href {https://doi.org/10.1103/PhysRevC.68.041001} {\bibfield
   {journal} {\bibinfo  {journal} {Phys. Rev. C}\ }\textbf {\bibinfo {volume}
  {68}},\ \bibinfo {pages} {041001} (\bibinfo {year} {2003})}\BibitemShut
  {NoStop}%
\bibitem [{\citenamefont {Wiringa}\ \emph {et~al.}(1995)\citenamefont
  {Wiringa}, \citenamefont {Stoks},\ and\ \citenamefont {Schiavilla}}]{v18}%
  \BibitemOpen
  \bibfield  {author} {\bibinfo {author} {\bibfnamefont {R.~B.}\ \bibnamefont
  {Wiringa}}, \bibinfo {author} {\bibfnamefont {V.~G.~J.}\ \bibnamefont
  {Stoks}},\ and\ \bibinfo {author} {\bibfnamefont {R.}~\bibnamefont
  {Schiavilla}},\ }\href {https://doi.org/10.1103/PhysRevC.51.38} {\bibfield
  {journal} {\bibinfo  {journal} {Phys. Rev. C}\ }\textbf {\bibinfo {volume}
  {51}},\ \bibinfo {pages} {38} (\bibinfo {year} {1995})}\BibitemShut {NoStop}%
\bibitem [{\citenamefont {Otsuka}\ \emph {et~al.}(2010)\citenamefont {Otsuka},
  \citenamefont {Suzuki}, \citenamefont {Honma}, \citenamefont {Utsuno},
  \citenamefont {Tsunoda}, \citenamefont {Tsukiyama},\ and\ \citenamefont
  {Hjorth-Jensen}}]{vmu}%
  \BibitemOpen
  \bibfield  {author} {\bibinfo {author} {\bibfnamefont {T.}~\bibnamefont
  {Otsuka}}, \bibinfo {author} {\bibfnamefont {T.}~\bibnamefont {Suzuki}},
  \bibinfo {author} {\bibfnamefont {M.}~\bibnamefont {Honma}}, \bibinfo
  {author} {\bibfnamefont {Y.}~\bibnamefont {Utsuno}}, \bibinfo {author}
  {\bibfnamefont {N.}~\bibnamefont {Tsunoda}}, \bibinfo {author} {\bibfnamefont
  {K.}~\bibnamefont {Tsukiyama}},\ and\ \bibinfo {author} {\bibfnamefont
  {M.}~\bibnamefont {Hjorth-Jensen}},\ }\href
  {https://doi.org/10.1103/PhysRevLett.104.012501} {\bibfield  {journal}
  {\bibinfo  {journal} {Phys. Rev. Lett.}\ }\textbf {\bibinfo {volume} {104}},\
  \bibinfo {pages} {012501} (\bibinfo {year} {2010})}\BibitemShut {NoStop}%
\bibitem [{\citenamefont {Bertsch}\ \emph {et~al.}(1977)\citenamefont
  {Bertsch}, \citenamefont {Borysowicz}, \citenamefont {McManus},\ and\
  \citenamefont {Love}}]{m3y}%
  \BibitemOpen
  \bibfield  {author} {\bibinfo {author} {\bibfnamefont {G.}~\bibnamefont
  {Bertsch}}, \bibinfo {author} {\bibfnamefont {J.}~\bibnamefont {Borysowicz}},
  \bibinfo {author} {\bibfnamefont {H.}~\bibnamefont {McManus}},\ and\ \bibinfo
  {author} {\bibfnamefont {W.}~\bibnamefont {Love}},\ }\href
  {https://doi.org/https://doi.org/10.1016/0375-9474(77)90392-X} {\bibfield
  {journal} {\bibinfo  {journal} {Nuclear Physics A}\ }\textbf {\bibinfo
  {volume} {284}},\ \bibinfo {pages} {399} (\bibinfo {year}
  {1977})}\BibitemShut {NoStop}%
\bibitem [{\citenamefont {Hjorth-Jensen}\ \emph {et~al.}(1995)\citenamefont
  {Hjorth-Jensen}, \citenamefont {Kuo},\ and\ \citenamefont {Osnes}}]{cens}%
  \BibitemOpen
  \bibfield  {author} {\bibinfo {author} {\bibfnamefont {M.}~\bibnamefont
  {Hjorth-Jensen}}, \bibinfo {author} {\bibfnamefont {T.~T.}\ \bibnamefont
  {Kuo}},\ and\ \bibinfo {author} {\bibfnamefont {E.}~\bibnamefont {Osnes}},\
  }\href {https://doi.org/https://doi.org/10.1016/0370-1573(95)00012-6}
  {\bibfield  {journal} {\bibinfo  {journal} {Physics Reports}\ }\textbf
  {\bibinfo {volume} {261}},\ \bibinfo {pages} {125} (\bibinfo {year}
  {1995})},\ \bibinfo {note} {code is available at
  \url{https://github.com/ManyBodyPhysics/CENS}}\BibitemShut {NoStop}%
\bibitem [{\citenamefont {Horoi}\ and\ \citenamefont {Brown}(2013)}]{jj77}%
  \BibitemOpen
  \bibfield  {author} {\bibinfo {author} {\bibfnamefont {M.}~\bibnamefont
  {Horoi}}\ and\ \bibinfo {author} {\bibfnamefont {B.~A.}\ \bibnamefont
  {Brown}},\ }\href {https://doi.org/10.1103/PhysRevLett.110.222502} {\bibfield
   {journal} {\bibinfo  {journal} {Phys. Rev. Lett.}\ }\textbf {\bibinfo
  {volume} {110}},\ \bibinfo {pages} {222502} (\bibinfo {year}
  {2013})}\BibitemShut {NoStop}%
\bibitem [{\citenamefont {Yoshida}\ \emph {et~al.}(2018)\citenamefont
  {Yoshida}, \citenamefont {Utsuno}, \citenamefont {Shimizu},\ and\
  \citenamefont {Otsuka}}]{yoshida}%
  \BibitemOpen
  \bibfield  {author} {\bibinfo {author} {\bibfnamefont {S.}~\bibnamefont
  {Yoshida}}, \bibinfo {author} {\bibfnamefont {Y.}~\bibnamefont {Utsuno}},
  \bibinfo {author} {\bibfnamefont {N.}~\bibnamefont {Shimizu}},\ and\ \bibinfo
  {author} {\bibfnamefont {T.}~\bibnamefont {Otsuka}},\ }\href
  {https://doi.org/10.1103/PhysRevC.97.054321} {\bibfield  {journal} {\bibinfo
  {journal} {Phys. Rev. C}\ }\textbf {\bibinfo {volume} {97}},\ \bibinfo
  {pages} {054321} (\bibinfo {year} {2018})}\BibitemShut {NoStop}%
\bibitem [{\citenamefont {Borzov}(2016)}]{borzov16}%
  \BibitemOpen
  \bibfield  {author} {\bibinfo {author} {\bibfnamefont {I.~N.}\ \bibnamefont
  {Borzov}},\ }\href {https://doi.org/10.1134/S1063778816060041} {\bibfield
  {journal} {\bibinfo  {journal} {Physics of Atomic Nuclei}\ }\textbf {\bibinfo
  {volume} {79}},\ \bibinfo {pages} {910} (\bibinfo {year} {2016})}\BibitemShut
  {NoStop}%
\bibitem [{\citenamefont {Sarriguren}\ \emph {et~al.}(2001)\citenamefont
  {Sarriguren}, \citenamefont {de~Guerra},\ and\ \citenamefont
  {Escuderos}}]{PedroNPA}%
  \BibitemOpen
  \bibfield  {author} {\bibinfo {author} {\bibfnamefont {P.}~\bibnamefont
  {Sarriguren}}, \bibinfo {author} {\bibfnamefont {E.~M.}\ \bibnamefont
  {de~Guerra}},\ and\ \bibinfo {author} {\bibfnamefont {A.}~\bibnamefont
  {Escuderos}},\ }\href
  {https://doi.org/https://doi.org/10.1016/S0375-9474(01)00565-6} {\bibfield
  {journal} {\bibinfo  {journal} {Nuclear Physics A}\ }\textbf {\bibinfo
  {volume} {691}},\ \bibinfo {pages} {631} (\bibinfo {year}
  {2001})}\BibitemShut {NoStop}%
\bibitem [{\citenamefont {Sarriguren}(2022)}]{PedroPriv}%
  \BibitemOpen
  \bibfield  {author} {\bibinfo {author} {\bibfnamefont {P.}~\bibnamefont
  {Sarriguren}},\ }\href@noop {} {}\bibinfo {howpublished} {Private
  communication} (\bibinfo {year} {2022})\BibitemShut {NoStop}%
\bibitem [{\citenamefont {Gysbers}\ \emph {et~al.}(2019)\citenamefont
  {Gysbers}, \citenamefont {Hagen}, \citenamefont {Holt}, \citenamefont
  {Jansen}, \citenamefont {Morris}, \citenamefont {Navr{\'a}til}, \citenamefont
  {Papenbrock}, \citenamefont {Quaglioni}, \citenamefont {Schwenk},
  \citenamefont {Stroberg},\ and\ \citenamefont {Wendt}}]{BGTnature}%
  \BibitemOpen
  \bibfield  {author} {\bibinfo {author} {\bibfnamefont {P.}~\bibnamefont
  {Gysbers}}, \bibinfo {author} {\bibfnamefont {G.}~\bibnamefont {Hagen}},
  \bibinfo {author} {\bibfnamefont {J.~D.}\ \bibnamefont {Holt}}, \bibinfo
  {author} {\bibfnamefont {G.~R.}\ \bibnamefont {Jansen}}, \bibinfo {author}
  {\bibfnamefont {T.~D.}\ \bibnamefont {Morris}}, \bibinfo {author}
  {\bibfnamefont {P.}~\bibnamefont {Navr{\'a}til}}, \bibinfo {author}
  {\bibfnamefont {T.}~\bibnamefont {Papenbrock}}, \bibinfo {author}
  {\bibfnamefont {S.}~\bibnamefont {Quaglioni}}, \bibinfo {author}
  {\bibfnamefont {A.}~\bibnamefont {Schwenk}}, \bibinfo {author} {\bibfnamefont
  {S.~R.}\ \bibnamefont {Stroberg}},\ and\ \bibinfo {author} {\bibfnamefont
  {K.~A.}\ \bibnamefont {Wendt}},\ }\href
  {https://doi.org/10.1038/s41567-019-0450-7} {\bibfield  {journal} {\bibinfo
  {journal} {Nature Physics}\ }\textbf {\bibinfo {volume} {15}},\ \bibinfo
  {pages} {428} (\bibinfo {year} {2019})}\BibitemShut {NoStop}%
\bibitem [{\citenamefont {Hinke}\ \emph {et~al.}(2012)\citenamefont {Hinke},
  \citenamefont {B{\"o}hmer}, \citenamefont {Boutachkov}, \citenamefont
  {Faestermann}, \citenamefont {Geissel}, \citenamefont {Gerl}, \citenamefont
  {Gernh{\"a}user}, \citenamefont {G{\'o}rska}, \citenamefont {Gottardo},
  \citenamefont {Grawe}, \citenamefont {Gr{\k e}bosz}, \citenamefont
  {Kr{\"u}cken}, \citenamefont {Kurz}, \citenamefont {Liu}, \citenamefont
  {Maier}, \citenamefont {Nowacki}, \citenamefont {Pietri}, \citenamefont
  {Podoly{\'a}k}, \citenamefont {Sieja}, \citenamefont {Steiger}, \citenamefont
  {Straub}, \citenamefont {Weick}, \citenamefont {Wollersheim}, \citenamefont
  {Woods}, \citenamefont {Al-Dahan}, \citenamefont {Alkhomashi}, \citenamefont
  {Ata{\c c}}, \citenamefont {Blazhev}, \citenamefont {Braun}, \citenamefont
  {{\v C}elikovi{\'c}}, \citenamefont {Davinson}, \citenamefont {Dillmann},
  \citenamefont {Domingo-Pardo}, \citenamefont {Doornenbal}, \citenamefont
  {de~France}, \citenamefont {Farrelly}, \citenamefont {Farinon}, \citenamefont
  {Goel}, \citenamefont {Habermann}, \citenamefont {Hoischen}, \citenamefont
  {Janik}, \citenamefont {Karny}, \citenamefont {Ka{\c s}ka{\c s}},
  \citenamefont {Kojouharov}, \citenamefont {Kr{\"o}ll}, \citenamefont
  {Litvinov}, \citenamefont {Myalski}, \citenamefont {Nebel}, \citenamefont
  {Nishimura}, \citenamefont {Nociforo}, \citenamefont {Nyberg}, \citenamefont
  {Parikh}, \citenamefont {Proch{\'a}zka}, \citenamefont {Regan}, \citenamefont
  {Rigollet}, \citenamefont {Schaffner}, \citenamefont {Scheidenberger},
  \citenamefont {Schwertel}, \citenamefont {S{\"o}derstr{\"o}m}, \citenamefont
  {Steer}, \citenamefont {Stolz},\ and\ \citenamefont {Strme{\v
  n}}}]{sn100nature}%
  \BibitemOpen
  \bibfield  {author} {\bibinfo {author} {\bibfnamefont {C.~B.}\ \bibnamefont
  {Hinke}}, \bibinfo {author} {\bibfnamefont {M.}~\bibnamefont {B{\"o}hmer}},
  \bibinfo {author} {\bibfnamefont {P.}~\bibnamefont {Boutachkov}}, \bibinfo
  {author} {\bibfnamefont {T.}~\bibnamefont {Faestermann}}, \bibinfo {author}
  {\bibfnamefont {H.}~\bibnamefont {Geissel}}, \bibinfo {author} {\bibfnamefont
  {J.}~\bibnamefont {Gerl}}, \bibinfo {author} {\bibfnamefont {R.}~\bibnamefont
  {Gernh{\"a}user}}, \bibinfo {author} {\bibfnamefont {M.}~\bibnamefont
  {G{\'o}rska}}, \bibinfo {author} {\bibfnamefont {A.}~\bibnamefont
  {Gottardo}}, \bibinfo {author} {\bibfnamefont {H.}~\bibnamefont {Grawe}},
  \bibinfo {author} {\bibfnamefont {J.~L.}\ \bibnamefont {Gr{\k e}bosz}},
  \bibinfo {author} {\bibfnamefont {R.}~\bibnamefont {Kr{\"u}cken}}, \bibinfo
  {author} {\bibfnamefont {N.}~\bibnamefont {Kurz}}, \bibinfo {author}
  {\bibfnamefont {Z.}~\bibnamefont {Liu}}, \bibinfo {author} {\bibfnamefont
  {L.}~\bibnamefont {Maier}}, \bibinfo {author} {\bibfnamefont
  {F.}~\bibnamefont {Nowacki}}, \bibinfo {author} {\bibfnamefont
  {S.}~\bibnamefont {Pietri}}, \bibinfo {author} {\bibfnamefont
  {Z.}~\bibnamefont {Podoly{\'a}k}}, \bibinfo {author} {\bibfnamefont
  {K.}~\bibnamefont {Sieja}}, \bibinfo {author} {\bibfnamefont
  {K.}~\bibnamefont {Steiger}}, \bibinfo {author} {\bibfnamefont
  {K.}~\bibnamefont {Straub}}, \bibinfo {author} {\bibfnamefont
  {H.}~\bibnamefont {Weick}}, \bibinfo {author} {\bibfnamefont {H.~J.}\
  \bibnamefont {Wollersheim}}, \bibinfo {author} {\bibfnamefont {P.~J.}\
  \bibnamefont {Woods}}, \bibinfo {author} {\bibfnamefont {N.}~\bibnamefont
  {Al-Dahan}}, \bibinfo {author} {\bibfnamefont {N.}~\bibnamefont
  {Alkhomashi}}, \bibinfo {author} {\bibfnamefont {A.}~\bibnamefont {Ata{\c
  c}}}, \bibinfo {author} {\bibfnamefont {A.}~\bibnamefont {Blazhev}}, \bibinfo
  {author} {\bibfnamefont {N.~F.}\ \bibnamefont {Braun}}, \bibinfo {author}
  {\bibfnamefont {I.~T.}\ \bibnamefont {{\v C}elikovi{\'c}}}, \bibinfo {author}
  {\bibfnamefont {T.}~\bibnamefont {Davinson}}, \bibinfo {author}
  {\bibfnamefont {I.}~\bibnamefont {Dillmann}}, \bibinfo {author}
  {\bibfnamefont {C.}~\bibnamefont {Domingo-Pardo}}, \bibinfo {author}
  {\bibfnamefont {P.~C.}\ \bibnamefont {Doornenbal}}, \bibinfo {author}
  {\bibfnamefont {G.}~\bibnamefont {de~France}}, \bibinfo {author}
  {\bibfnamefont {G.~F.}\ \bibnamefont {Farrelly}}, \bibinfo {author}
  {\bibfnamefont {F.}~\bibnamefont {Farinon}}, \bibinfo {author} {\bibfnamefont
  {N.}~\bibnamefont {Goel}}, \bibinfo {author} {\bibfnamefont {T.~C.}\
  \bibnamefont {Habermann}}, \bibinfo {author} {\bibfnamefont {R.}~\bibnamefont
  {Hoischen}}, \bibinfo {author} {\bibfnamefont {R.}~\bibnamefont {Janik}},
  \bibinfo {author} {\bibfnamefont {M.}~\bibnamefont {Karny}}, \bibinfo
  {author} {\bibfnamefont {A.}~\bibnamefont {Ka{\c s}ka{\c s}}}, \bibinfo
  {author} {\bibfnamefont {I.~M.}\ \bibnamefont {Kojouharov}}, \bibinfo
  {author} {\bibfnamefont {T.}~\bibnamefont {Kr{\"o}ll}}, \bibinfo {author}
  {\bibfnamefont {Y.}~\bibnamefont {Litvinov}}, \bibinfo {author}
  {\bibfnamefont {S.}~\bibnamefont {Myalski}}, \bibinfo {author} {\bibfnamefont
  {F.}~\bibnamefont {Nebel}}, \bibinfo {author} {\bibfnamefont
  {S.}~\bibnamefont {Nishimura}}, \bibinfo {author} {\bibfnamefont
  {C.}~\bibnamefont {Nociforo}}, \bibinfo {author} {\bibfnamefont
  {J.}~\bibnamefont {Nyberg}}, \bibinfo {author} {\bibfnamefont {A.~R.}\
  \bibnamefont {Parikh}}, \bibinfo {author} {\bibfnamefont {A.}~\bibnamefont
  {Proch{\'a}zka}}, \bibinfo {author} {\bibfnamefont {P.~H.}\ \bibnamefont
  {Regan}}, \bibinfo {author} {\bibfnamefont {C.}~\bibnamefont {Rigollet}},
  \bibinfo {author} {\bibfnamefont {H.}~\bibnamefont {Schaffner}}, \bibinfo
  {author} {\bibfnamefont {C.}~\bibnamefont {Scheidenberger}}, \bibinfo
  {author} {\bibfnamefont {S.}~\bibnamefont {Schwertel}}, \bibinfo {author}
  {\bibfnamefont {P.~A.}\ \bibnamefont {S{\"o}derstr{\"o}m}}, \bibinfo {author}
  {\bibfnamefont {S.~J.}\ \bibnamefont {Steer}}, \bibinfo {author}
  {\bibfnamefont {A.}~\bibnamefont {Stolz}},\ and\ \bibinfo {author}
  {\bibfnamefont {P.}~\bibnamefont {Strme{\v n}}},\ }\href
  {https://doi.org/10.1038/nature11116} {\bibfield  {journal} {\bibinfo
  {journal} {Nature}\ }\textbf {\bibinfo {volume} {486}},\ \bibinfo {pages}
  {341} (\bibinfo {year} {2012})}\BibitemShut {NoStop}%
\bibitem [{\citenamefont {Lubos}\ \emph {et~al.}(2019)\citenamefont {Lubos},
  \citenamefont {Park}, \citenamefont {Faestermann}, \citenamefont
  {Gernh\"auser}, \citenamefont {Kr\"ucken}, \citenamefont {Lewitowicz},
  \citenamefont {Nishimura}, \citenamefont {Sakurai}, \citenamefont {Ahn},
  \citenamefont {Baba}, \citenamefont {Blank}, \citenamefont {Blazhev},
  \citenamefont {Boutachkov}, \citenamefont {Browne}, \citenamefont {\ifmmode
  \check{C}\else \v{C}\fi{}elikovi\ifmmode~\acute{c}\else \'{c}\fi{}},
  \citenamefont {de~France}, \citenamefont {Doornenbal}, \citenamefont {Fang},
  \citenamefont {Fukuda}, \citenamefont {Giovinazzo}, \citenamefont {Goel},
  \citenamefont {G\'orska}, \citenamefont {Ilieva}, \citenamefont {Inabe},
  \citenamefont {Isobe}, \citenamefont {Jungclaus}, \citenamefont {Kameda},
  \citenamefont {Kim}, \citenamefont {Kojouharov}, \citenamefont {Kubo},
  \citenamefont {Kurz}, \citenamefont {Kwon}, \citenamefont {Lorusso},
  \citenamefont {Moschner}, \citenamefont {Murai}, \citenamefont {Nishizuka},
  \citenamefont {Patel}, \citenamefont {Rajabali}, \citenamefont {Rice},
  \citenamefont {Schaffner}, \citenamefont {Shimizu}, \citenamefont {Sinclair},
  \citenamefont {S\"oderstr\"om}, \citenamefont {Steiger}, \citenamefont
  {Sumikama}, \citenamefont {Suzuki}, \citenamefont {Takeda}, \citenamefont
  {Wang}, \citenamefont {Warr}, \citenamefont {Watanabe}, \citenamefont {Wu},\
  and\ \citenamefont {Xu}}]{sn100prl}%
  \BibitemOpen
  \bibfield  {author} {\bibinfo {author} {\bibfnamefont {D.}~\bibnamefont
  {Lubos}}, \bibinfo {author} {\bibfnamefont {J.}~\bibnamefont {Park}},
  \bibinfo {author} {\bibfnamefont {T.}~\bibnamefont {Faestermann}}, \bibinfo
  {author} {\bibfnamefont {R.}~\bibnamefont {Gernh\"auser}}, \bibinfo {author}
  {\bibfnamefont {R.}~\bibnamefont {Kr\"ucken}}, \bibinfo {author}
  {\bibfnamefont {M.}~\bibnamefont {Lewitowicz}}, \bibinfo {author}
  {\bibfnamefont {S.}~\bibnamefont {Nishimura}}, \bibinfo {author}
  {\bibfnamefont {H.}~\bibnamefont {Sakurai}}, \bibinfo {author} {\bibfnamefont
  {D.~S.}\ \bibnamefont {Ahn}}, \bibinfo {author} {\bibfnamefont
  {H.}~\bibnamefont {Baba}}, \bibinfo {author} {\bibfnamefont {B.}~\bibnamefont
  {Blank}}, \bibinfo {author} {\bibfnamefont {A.}~\bibnamefont {Blazhev}},
  \bibinfo {author} {\bibfnamefont {P.}~\bibnamefont {Boutachkov}}, \bibinfo
  {author} {\bibfnamefont {F.}~\bibnamefont {Browne}}, \bibinfo {author}
  {\bibfnamefont {I.}~\bibnamefont {\ifmmode \check{C}\else
  \v{C}\fi{}elikovi\ifmmode~\acute{c}\else \'{c}\fi{}}}, \bibinfo {author}
  {\bibfnamefont {G.}~\bibnamefont {de~France}}, \bibinfo {author}
  {\bibfnamefont {P.}~\bibnamefont {Doornenbal}}, \bibinfo {author}
  {\bibfnamefont {Y.}~\bibnamefont {Fang}}, \bibinfo {author} {\bibfnamefont
  {N.}~\bibnamefont {Fukuda}}, \bibinfo {author} {\bibfnamefont
  {J.}~\bibnamefont {Giovinazzo}}, \bibinfo {author} {\bibfnamefont
  {N.}~\bibnamefont {Goel}}, \bibinfo {author} {\bibfnamefont {M.}~\bibnamefont
  {G\'orska}}, \bibinfo {author} {\bibfnamefont {S.}~\bibnamefont {Ilieva}},
  \bibinfo {author} {\bibfnamefont {N.}~\bibnamefont {Inabe}}, \bibinfo
  {author} {\bibfnamefont {T.}~\bibnamefont {Isobe}}, \bibinfo {author}
  {\bibfnamefont {A.}~\bibnamefont {Jungclaus}}, \bibinfo {author}
  {\bibfnamefont {D.}~\bibnamefont {Kameda}}, \bibinfo {author} {\bibfnamefont
  {Y.~K.}\ \bibnamefont {Kim}}, \bibinfo {author} {\bibfnamefont
  {I.}~\bibnamefont {Kojouharov}}, \bibinfo {author} {\bibfnamefont
  {T.}~\bibnamefont {Kubo}}, \bibinfo {author} {\bibfnamefont {N.}~\bibnamefont
  {Kurz}}, \bibinfo {author} {\bibfnamefont {Y.~K.}\ \bibnamefont {Kwon}},
  \bibinfo {author} {\bibfnamefont {G.}~\bibnamefont {Lorusso}}, \bibinfo
  {author} {\bibfnamefont {K.}~\bibnamefont {Moschner}}, \bibinfo {author}
  {\bibfnamefont {D.}~\bibnamefont {Murai}}, \bibinfo {author} {\bibfnamefont
  {I.}~\bibnamefont {Nishizuka}}, \bibinfo {author} {\bibfnamefont
  {Z.}~\bibnamefont {Patel}}, \bibinfo {author} {\bibfnamefont {M.~M.}\
  \bibnamefont {Rajabali}}, \bibinfo {author} {\bibfnamefont {S.}~\bibnamefont
  {Rice}}, \bibinfo {author} {\bibfnamefont {H.}~\bibnamefont {Schaffner}},
  \bibinfo {author} {\bibfnamefont {Y.}~\bibnamefont {Shimizu}}, \bibinfo
  {author} {\bibfnamefont {L.}~\bibnamefont {Sinclair}}, \bibinfo {author}
  {\bibfnamefont {P.-A.}\ \bibnamefont {S\"oderstr\"om}}, \bibinfo {author}
  {\bibfnamefont {K.}~\bibnamefont {Steiger}}, \bibinfo {author} {\bibfnamefont
  {T.}~\bibnamefont {Sumikama}}, \bibinfo {author} {\bibfnamefont
  {H.}~\bibnamefont {Suzuki}}, \bibinfo {author} {\bibfnamefont
  {H.}~\bibnamefont {Takeda}}, \bibinfo {author} {\bibfnamefont
  {Z.}~\bibnamefont {Wang}}, \bibinfo {author} {\bibfnamefont {N.}~\bibnamefont
  {Warr}}, \bibinfo {author} {\bibfnamefont {H.}~\bibnamefont {Watanabe}},
  \bibinfo {author} {\bibfnamefont {J.}~\bibnamefont {Wu}},\ and\ \bibinfo
  {author} {\bibfnamefont {Z.}~\bibnamefont {Xu}},\ }\href
  {https://doi.org/10.1103/PhysRevLett.122.222502} {\bibfield  {journal}
  {\bibinfo  {journal} {Phys. Rev. Lett.}\ }\textbf {\bibinfo {volume} {122}},\
  \bibinfo {pages} {222502} (\bibinfo {year} {2019})}\BibitemShut {NoStop}%
\bibitem [{\citenamefont {Shimizu}\ \emph {et~al.}(2019)\citenamefont
  {Shimizu}, \citenamefont {Mizusaki}, \citenamefont {Utsuno},\ and\
  \citenamefont {Tsunoda}}]{kshell}%
  \BibitemOpen
  \bibfield  {author} {\bibinfo {author} {\bibfnamefont {N.}~\bibnamefont
  {Shimizu}}, \bibinfo {author} {\bibfnamefont {T.}~\bibnamefont {Mizusaki}},
  \bibinfo {author} {\bibfnamefont {Y.}~\bibnamefont {Utsuno}},\ and\ \bibinfo
  {author} {\bibfnamefont {Y.}~\bibnamefont {Tsunoda}},\ }\href
  {https://doi.org/https://doi.org/10.1016/j.cpc.2019.06.011} {\bibfield
  {journal} {\bibinfo  {journal} {Computer Physics Communications}\ }\textbf
  {\bibinfo {volume} {244}},\ \bibinfo {pages} {372} (\bibinfo {year}
  {2019})},\ \bibinfo {note} {code is available at
  \url{https://sites.google.com/a/cns.s.u-tokyo.ac.jp/kshell}}\BibitemShut
  {NoStop}%
\end{thebibliography}
%

\end{document}